              \documentstyle[amsfonts,amssymb,epsfig,12pt]{article}

\makeatletter
\renewcommand \thesection {\@arabic\c@section}
\renewcommand\thesubsection   {\thesection.\@arabic\c@subsection}
\renewcommand\thesubsubsection{\thesubsection .\@arabic\c@subsubsection}
\renewcommand\theparagraph    {\thesubsubsection.\@arabic\c@paragraph}
\renewcommand\section{\@startsection {section}{1}{\z@}%
                                   {-3.5ex \@plus -1ex \@minus -.2ex}%
                                   {1.9ex \@plus.2ex}%
                                   {\normalfont\large\bfseries\centering}}
\renewcommand\subsection{\@startsection{subsection}{2}{\z@}%
                                     {-2ex\@plus -1ex \@minus -.2ex}%
                                     {1.2ex \@plus .2ex}%
                                    {\normalfont\normalsize\bfseries\centering}}
\renewcommand\subsubsection{\@startsection{subsubsection}{3}{\z@}%
                                     {-2ex\@plus -1ex \@minus -.2ex}%
                                     {.5ex \@plus .2ex}%
                                     {\normalfont\normalsize\em}}
\renewcommand\paragraph{\@startsection{paragraph}{4}{\z@}%
                                    {3.25ex \@plus1ex \@minus.2ex}%
                                    {-1em}%
                                    {\normalfont\normalsize\em}}
\renewcommand\subparagraph{\@startsection{subparagraph}{5}{\parindent}%
                                       {3.25ex \@plus1ex \@minus .2ex}%
                                       {-1em}%
                                      {\normalfont\normalsize\em}}
\makeatother

\newcommand\abs[1]{{\left| #1 \right|}}
\newcommand\MnormSQ[1]{{{#1}^{\cdot{2}}}}

\newcounter{subequation}
	\newenvironment{subequation}%
	{\addtocounter{equation}{-1}%
	\stepcounter{subequation}%
	\begin{equation}}%
	{\end{equation}%
}

\newcommand{\beq}{\begin{equation}}
\newcommand{\eeq}{\end{equation}}
\newcommand{\bseq}{\begin{subequation}}
\newcommand{\eseq}{\end{subequation}}
\newcommand{\bea}{\begin{eqnarray}}
\newcommand{\eea}{\end{eqnarray}}

\newcommand{\refeq}[1]{(\ref{#1})}

\newcommand{\Id}{{\mathrm{Id}\,}}

\newcommand{\pmk}{{{z_k}}}

\newcommand{\plumi}{{\scriptstyle{(\pm)}}}
\newcommand{\veps}{\varepsilon} 

\newcommand{\QED}{$\quad$\textrm{Q.E.D.}}

\newcommand{\me}{m_{\mathrm{e}}}

\newcommand{\lC}{\lambda_{\mathrm{C}}}

\newcommand{\bulldif}[1]{{{#1}^{\!\!\!\stackrel{\bullet}{\phantom{.}}}}{}}

\newcommand{\dd}{{\mathrm{d}}}

\newcommand{\cA}{{\cal A}}

\newcommand{\cE}{{\cal E}}

\newcommand{\cH}{{\cal H}}
\newcommand{\cJ}{{\cal J}}

\newcommand{\cM}{{\cal M}}
\newcommand{\cN}{{\cal N}}
\newcommand{\cP}{{\cal P}}
\newcommand{\cQ}{{\cal Q}}

\newcommand{\cT}{{\cal T}}

\newcommand{\cX}{{\cal X}}
\newcommand{\cY}{{\cal Y}}

\newcommand{\MM}{{\Bbb M}}
\newcommand{\NN}{{\Bbb N}}

\newcommand{\RR}{{\Bbb R}}

\newcommand{\ZZ}{{\Bbb Z}}

\newcommand{\MTWtens}[1]{{\textbf{\textsf{#1}}}}

\newcommand{\MQ}{{\MTWtens{M}}}
\newcommand{\FQ}{{\MTWtens{F}}}

\newcommand{\dQ}{{\MTWtens{d}}}
\newcommand{\gQ}{{\MTWtens{g}}}

\newcommand{\MTWvec}[1]{{\mathbf{#1}}}

\newcommand{\AQ}{{\MTWvec{A}}}
\newcommand{\JQ}{{\MTWvec{J}}}
\newcommand{\PQ}{{\MTWvec{P}}}
\newcommand{\RQ}{{\MTWvec{R}}}

\newcommand{\eQ}{{\MTWvec{e}}}

\newcommand{\sQ}{{\MTWvec{s}}}
\newcommand{\tQ}{{\MTWvec{t}}}
\newcommand{\uQ}{{\MTWvec{u}}}

\newcommand{\wQ}{{\MTWvec{w}}}

\newcommand{\Beta}{{\mathrm{B}}}
\newcommand{\Eta}{{\mathrm{H}}}

\newcommand{\AlongH}{{\Big|_{_{\Eta}}\Big.}}

\newcommand{\AlongHk}{{\Big|_{_{\Eta_k}}\Big.}}
\newcommand{\alongH}{{\big|_{_{\Eta}}\big.}}
\newcommand{\alongHk}{{\big|_{_{\Eta_k}}\big.}}

\newcommand{\SPvec}[1]{{\textbf{\textsl{#1}}}}

\newcommand{\jV}{{\SPvec{j}}}
\newcommand{\sV}{{\SPvec{s}}}

\newcommand{\trg}{{\mathrm{tr}}_{\textsf{g}}}

\newcommand{\detg}{{\mathrm{det}}_{\textsf{g}}}

\newcommand{\dvol}{{\mathrm{d}^3}}

\newcommand{\Hodge}{{{}^\star}}

\newcommand{\haelfte}{{\textstyle{\frac{1}{2}}}}
\newcommand{\viertel}{{\textstyle{\frac{1}{4}}}}

\newtheorem{defn}{Definition}[section] % This defines the counter for
\newtheorem{Prop}[defn]{Proposition}

\begin{document}

\title{ ELECTROMAGNETIC FIELD THEORY \\
	WITHOUT DIVERGENCE PROBLEMS \\
	1. The Born Legacy}

\author{\textbf{MICHAEL K.-H. KIESSLING}\\
	Department of Mathematics\\
	Rutgers, The State University of New Jersey\\
	110 Frelinghuysen Rd., Piscataway, NJ 08854}
\date{}
\maketitle

\begin{abstract}
\noindent
Born's quest for the elusive divergence problem-free quantum theory of electromagnetism
led to the important discovery of the nonlinear Maxwell--Born--Infeld equations for the 
classical electromagnetic fields, the sources of which are classical point charges in motion. 
The law of motion for these point charges has however been missing, because
the Lorentz self-force in the relativistic Newtonian (formal) law of 
motion is ill-defined in magnitude and direction.
In the present paper it is shown that a relativistic Hamilton--Jacobi type 
law of point charge motion can be consistently coupled with the nonlinear 
Maxwell--Born--Infeld field equations to obtain a well-defined relativistic 
classical electrodynamics with point charges.
Curiously, while the point charges are spinless, the Pauli principle for bosons
can be incorporated.
Born's reasoning for calculating the value of his aether constant is re-assessed and 
found to be inconclusive.
\end{abstract} 

\textit{Electromagnetism}: electromagnetic fields, point charges;
\textit{Determinism}: Maxwell--Born--Infeld field equations,
Hamilton--Jacobi law of motion;
\textit{Permutability}: configuration space, Pauli principle.
\hrule
\smallskip
\noindent
Part I of two parts to appear in J. Stat. Phys. 
in honor of Elliott H. Lieb's 70th birthday;
received by JSP on June 29, 2003, accepted March 1, 2004.\\
\copyright{2004} The author.  Reproduction of this paper, in its
entirety and for noncommercial purposes only, is permitted.

\newpage
%%%%%%%%%%%%%%%%%%%%%%%%%%%%%%%%%%%%%%%%%%%%%%%%%%%%%%%%%%%%%%%%%%%%
%%%%%%%%%%%%%%%%%%%%%%%%%%%%%%%%%%%%%%%%%%%%%%%%%%%%%%%%%%%%%%%%%%%%
%%%%%%%%%%%%%%%%%%%%%%%%%%%%%%%%%%%%%%%%%%%%%%%%%%%%%%%%%%%%%%%%%%%%
%%%%%%%%%%%%%%%%%%%%%%%%%%%%%%%%%%%%%%%%%%%%%%%%%%%%%%%%%%%%%%%%%%%%
	\section{Introduction}
%%%%%%%%%%%%%%%%%%%%%%%%%%%%%%%%%%%%%%%%%%%%%%%%%%%%%%%%%%%%%%%%%%%%
%%%%%%%%%%%%%%%%%%%%%%%%%%%%%%%%%%%%%%%%%%%%%%%%%%%%%%%%%%%%%%%%%%%%
%%%%%%%%%%%%%%%%%%%%%%%%%%%%%%%%%%%%%%%%%%%%%%%%%%%%%%%%%%%%%%%%%%%%
%%%%%%%%%%%%%%%%%%%%%%%%%%%%%%%%%%%%%%%%%%%%%%%%%%%%%%%%%%%%%%%%%%%%

%
{\small{
% ``Quantum field theory deals with fields $\psi(x)$  that destroy and create particles at \hfill
% \  a spacetime point $x$. 
\ \ \ \ 
\ \ \ \ ``Earlier experience with classical electron theory provided a warning that a point\hfill

\ \ \ \   electron will have infinite electromagnetic self-mass; ... \hfill

\ \ \ \  Disappointingly this problem appeared with even greater severity in the early days\hfill

\ \ \ \  of quantum field theory, and although greatly ameliorated by subsequent improve- \hfill

\ \ \ \  ments in the theory, it remains with us to the present day.'' \hfill

{\hskip 8truecm Steven Weinberg (\cite{weinbergBOOKqftA}, p.31)}
}}
\smallskip

\noindent
  The making of quantum electrodynamics (QED) is generally recognized as
one of the biggest success stories of $20^{\mathrm{th}}$ century physics
	\cite{schweberBOOK}.
  The agreement of the measured values of the Lamb shift and the electron's $g$ factor 
with those computed perturbatively in QED is nothing less than spectacular
	\cite{Cottingham}.
  Yet, no physicist can be too happy about this feat, for 
QED's precision rests entirely on truncation.
  Without mathematical infrared (IR) and ultraviolet (UV) cutoffs QED is 
divergent at very large and very small length scales, and while
renormalization techniques allow these cutoffs to be removed in each order 
of perturbation theory
                \cite{weinberg},
the resulting series is still divergent 
		\cite{dysonC}
and must be truncated.
   Many experts have voiced their dissatisfaction with QED in this regard.\footnote{While R. Penrose files QED in his 
					``SUPERB'' category for its numerical precision, he remarks that 
				    ``[t]he theory as a whole does not have the compelling elegance or 
				    consistency of the earlier SUPERB theories, ...'' (p.153 in 
										    \cite{penroseEMPEROR}).
				    Peierls is more specific and insists that ``the use, as basic principle, of a 
				    semiconvergent series is unsatisfactory'' (p.92 
				                                              in \cite{peierlsBOOK}).
				    For Jost this suffices to declare that he ``cannot call QED a theory yet'' (p.93 
				                                                    in \cite{Jost}),
				    and Dirac's and Landau's radical opposition to QED in their later career is well documented in
				         \cite{schweberBOOK}.
				    The introductory quotation taken from 
										    \cite{weinbergBOOKqftA}
				    speaks for itself.}
   It is a difficult open question whether this state of affairs in electromagnetic theory can be overcome.

   For some three-plus decades it seemed that QED's divergence problems were merely a 
temporary nuisance that renormalization techniques would make go away eventually 
if we just keep working at it hard enough 
             \cite{glimmjaffeBOOK},
but the likelihood for this to come true seems very remote by now.
  While efforts to give mathematical sense to renormalized QED continue (e.g., 	
            \cite{conkreiA, conkreiB}), 
and progress is made in the rigorous control of non-perturbative low-energy 
(no pair creation / annihilation) approximations\footnote{The forthcoming monograph 
                                                             \cite{spohnBOOK}
							   summarizes the current state of affairs.}
            (e.g., \cite{BachFroehlichSigal, lieblossA, lieblossB}),
many physicists seem to have resigned to the view that QED, and for that matter
also the quantum field theory (QFT) of the standard model of the electromagnetic+weak+strong 
interactions, has to be relegated from the status of ``fundamental'' to that of ``effective''
	        \cite{weinbergBOOKqftA}.
  As effective theories, they may well require cutoffs\footnote{The jury is still out on whether a 
                       cutoff-free QFT of the strong interactions alone is feasible, as testified 
		       by the Clay Mathematics Institute's announcement of a millennium prize for the 
		       construction of a quantum Yang--Mills theory with compact gauge group on 3+1 
		       dimensional spacetime.}
to mimic the hypothetically 
regularizing effects of the omitted high-energy physics.
 By high-energy physics one means the realm of the truly fundamental theory, which 
aims at the unification of all (known) interactions, gravity  and 
the structure of spacetime included.
 It may sound fantastic that we should have to go to 
such length (in both senses of this phrase) as the Planck length
$\lambda_{\mathrm{Planck}} \approx 1.62 {\times}10^{-35} {\mathrm{m}}$
to truly understand a theory which was invented to understand effects that
live on scales between and about, say, the electron's classical radius
$\lambda_{\mathrm{classical}}\approx 2.82 {\times}10^{-15}{\mathrm{m}}$
and its Compton wave length 
$\lambda_{\mathrm{Compton}}\approx 3.86 {\times}10^{-13}{\mathrm{m}}$, 
yet if there is nothing ``rong'' with QED other than that it ignores all
non-electromagnetic effects, then this conclusion may be hard to avoid. 

%But perhaps something \emph{is} intrinsically wrong with QED, after all.

 However, almost from its beginnings, the dissenting opinion was expressed 
that (nowadays standard) QED, based as it is on the quantization of the classical 
Maxwell--Dirac field equations\footnote{Incidentally, the classical Maxwell--Dirac 
                                        equations are well-behaved, see 
                          		    \cite{flatoetal, bournaveas}.}
           \cite{BiBiBiBiBOOK, jauchrohrlichBOOK, weinbergBOOKqftA},
is not properly set up, that an intrinsically well-defined quantum theory 
of electromagnetism is yet to be found, and that an intrinsically well-defined 
classical (microscopic) theory of electromagnetism may serve as important  
stepping stone on our way to the quantum theory.
 Among the founding fathers of modern quantum theory this view was held most 
prominently by Dirac, Schr\"odinger, and by Born who assessed the situation 
in 1933 thus (cf. Weinberg's words in the introductory quotation):

{\small{
 ``The attempts to combine Maxwell's equations with the quantum theory (...) have \hfill

  not succeeded. One can see that the failure does not lie on the side of the quantum\hfill

  theory, but on the side of the field equations, which do not account for the existence \hfill

  of a radius of the electron (or its finite energy = mass).''
	{\hskip .5truecm Max Born}  \cite{BornA}
}}

\noindent
  Born surmised that the infinite self-energy problems that plague QED could be eliminated by 
quantizing nonlinear classical electromagnetic field equations which assign point charges a 
finite (classical) self-energy.
  To find suitable field equations, he recycled some earlier ideas of Gustaf Mie who, after 
his better known work on the scattering of electromagnetic waves in colloidal media
                \cite{MieSCATTERING},
had laid down a very general  Lagrangian framework for nonlinear classical electromagnetic field theories 
		\cite{MieFELDTHEORIE}
among which Mie hoped to find one in which electrons are smooth \emph{solitons}
--- a monumental series of papers which had not escaped  Born's attention
		\cite{BornsMIEpaper}
(see also 
		\cite{Hilbert, WeylBOOKrzm, PauliBOOK}).
   Born's paper 
 		\cite{BornA}
launched an alternate quest for QED, but despite an intense pursuit, 
by  Born and Infeld 
		\cite{BornA, BornInfeldA, BornInfeldB, BornB, BornInfeldC,  BornC},
Pryce 
                \cite{PryceB, PryceC, PryceD, PryceF},
Schr\"odinger
		\cite{ErwinEiHBiD, ErwinDUBLINa, ErwinDUBLINb, ErwinDUBLINc},
and Dirac
      \cite{diracBI},
a disappointed Born would near the end of his life concede that 
   ``[t]he adaption of these ideas to the principles of quantum theory and the 
     introduction of the spin has however met with no success'' 
           (\cite{BornD}, p.375).
 Ironically, only a year later the publications
          \cite{Boillat, Plebanski}
supplied a compelling new piece of evidence which suggested that Born and his prominent fellow 
dissenters were up to something after all, and which has suggested to us to pursue 
this matter a little further.

  To be specific, to Mie's requirements (i), that the field equations 
be covariant under the Poincar\'e group, and (ii), that they reduce to 
Maxwell's equations for the electromagnetic vacuum fields in the weak field limit, 
Born had added the postulates (iii), that the field equations be covariant under a Weyl 
(gauge) group, and (iv), that the electromagnetic field energy density 
surrounding a point charge be integrable. 
    Postulates (i)-(iv) do not uniquely identify the field equations, but in
             \cite{Boillat, Plebanski}
it was discovered that by adding to (i)--(iv) the reasonable physical requirement 
(v), that the speed of light [sic!] be independent
of the polarization of the wave fields,\footnote{The linear Maxwell--Lorentz equations for a point charge 
						 satisfy (i), (ii), (iii), (v), but not (iv).}
one arrives at a \emph{unique} one-parameter family of field equations, indeed the one proposed 
--- in ``one of those amusing cases of serendipity in theoretical physics''
                     (\cite{BiBiONE}, p.37) ---
by Born and Infeld
           \cite{BornInfeldA}.\footnote{While the unique characterization of the Maxwell--Born--Infeld field equations 
                                        in terms of (i)--(v) was apparently not known to Born and his contemporaries, 
					the fact that these field equations satisfy, beside items (i)-(iv), also
					item (v) is mentioned in passing also on p.102 in 
					\cite{ErwinDUBLINa} 
					as the absence of birefringence (double refraction).}
 In short, the Born--Infeld theory, which in essence replaces Maxwell's `pure aether'\footnote{Henceforth, 
                                                `aether' will be short for `electromagnetic vacuum,' 
                                                 and we will drop the quotes.}
by an aether with nontrivial polarizabilities to avoid the infinite electromagnetic self-contributions
to the mass of a point charge, does so in a distinctly unique way.

  Now, the  Lorentz program of electrodynamics 
           \cite{lorentzENCYCLOPb}
fails for point charges embedded in Maxwell's pure aether
because of their infinite electromagnetic self-masses and because their Lorentz self-force is 
``infinite in all directions.''\footnote{This (translated) phrase is borrowed from  a passage of Emil Wiechert's monumental paper
				       \cite{wiechertsBIGpaper} 
				       found on pp.41/42 in 
				       \cite{Jost};
				       it also happens to be the title of the published version of Freeman Dyson's 1985 Gifford 
				       lectures in Aberdeen \cite{dysonGIFFORD} (original title:``In praise of diversity''). 
				       (Both Wiechert and Dyson of course meant to indicate something more poetical.)}
 Unfortunately, the Maxwell--Born--Infeld field equations lead to a Lorentz self-force on a point charge which 
still is undefined in magnitude and direction.
  Many attempts have been made 
	\cite{BornB, BornInfeldB, PryceD, ErwinDUBLINb, diracBI, BiBiBiBiBOOK,
		Chernitskii, ChruscinskiA, ChruscinskiB}
to obtain a meaningful Newtonian equation of motion for the point charges, either
by \emph{imposing} the law of energy-momentum conservation or by regularization of the 
electromagnetic fields, yet upon reflection
(and close inspection) such attempts are found wanting.
   The main contribution of the present paper is a well-defined classical relativistic law of motion 
for spinless point charges which interact with the total classical electromagnetic Maxwell--Born--Infeld fields.
  This completes the consistent implementation of the notion of the point charge in the classical relativistic 
theory of electromagnetism,\footnote{The classical electromagnetic theory presented in this paper 
                                should not be mistaken as a merely mathematically fancier resurrection 
				of the so-called classical electron theory of Abraham and Lorentz
				     \cite{lorentzENCYCLOPb, abrahamBOOK, lorentzBOOKb, rohrlichBOOK}.
 			        In the present theory electrons are implemented as 
				\emph{point defects} in the electromagnetic potential field which
				cannot be transformed away, and which are characterized by 
				a Poincar\'e- and Weyl-invariant topological quantity that is identified 
				with the electric unit charge. 
				Thus, in this theory the electron is a \emph{true} point in the sense of Frenkel
				        \cite{Frenkel},
			        which remains the simplest notion compatible with the absence of empirical 
				evidence to the contrary.
				In contrast, in classical electron theory the electron was assumed to have an inner 
 				structure, to unlock the secrets of which was the  purpose of that theory (proven 
				by Einstein 
						     \cite{einsteinA}
				to be in vain). 
				Yet, classical electron theory has remained an interesting dynamical theory in its own right; 
see
				     \cite{KomechSpohn, KunzeSpohnA, BauerDuerr, AppKieAOP, AppKieLMP, spohnBOOK}.}
without any need for regularization or renormalization.\footnote{We add that since Dirac's paper
                               \cite{DiracA}, 
			  attempts to establish a classical Lorentz electrodynamics with point charges 
			  through (negative) infinite bare mass renormalization have continued. 
			  Recent rigorous contributions are 
			  \cite{bambusigalgani, bambusinoja, gitteletal, nojaposilicanoA}.}

 The gist of the matter is our observation, apparently not made elsewhere before, that the classical 
electromagnetic potentials for the solutions of the Maxwell--Born--Infeld field equations with point 
sources have just enough regularity so that a relativistic Hamilton--Jacobi theory can be put to work,
though not enough regularity for this Hamilton--Jacobi theory to reduce by differentiation to a 
(relativistic) Newtonian theory of motion driven by the total Lorentz force\footnote{Presumably 
                  this is as close as one can come to implementing the spirit of the Newtonian 
                  law of motion for point charges driven by the total Lorentz force, and yet 
		  not having quite the same.} 
 --- to which differentiation \emph{would} reduce it
\emph{if} the electromagnetic potentials \emph{were} differentiable, as functions of spacetime,
at the world-points of the point charges.
 In short, within the classical domain a relativistic Hamilton--Jacobi law of motion, i.e. 
a first-order (velocity) differential equation for the actual positions, proves to be the 
more fundamental notion; the relativistic Newtonian law, i.e. a second-order (acceleration) 
differential equation for the positions, is a derived concept with only approximate validity.
 
 What could seem to be merely a technical fine point that had been overlooked so far has, however,
conceptual consequences: since in classical Born--Infeld electrodynamics with point charges 
the electromagnetic potential $\AQ$ as a function on spacetime is not differentiable at the locations 
of the charges, the Hamilton--Jacobi phase function\footnote{In the physics literature the Hamilton--Jacobi phase function 
                                                             is usually denoted by $S\, (=\hbar \Phi)$.}
$\Phi$ cannot be eliminated and, hence, acquires significance in its own right.
 More to the point, what acquires significance is the guiding field on
configuration space which in particular guides the actual configuration 
of the point charges, and which is furnished by the gauge-invariant co-variant 
derivative\footnote{More precisely, it is the co-variant 
                    logarithmic derivative of $-ie^{i\Phi}$.}
of $\Phi$ w.r.t. a configuration space-indexed family of $\AQ$s.
 In this vein, $\Phi$ and the configuration-space indexed family of $\AQ$s become 
somewhat akin to the wave function in quantum mechanics, which certainly is a 
surprising departure from conventional wisdom about classical electromagnetic theory.

  Our theory allows us to re-assess the subtle issue of `Born's aether constant,' by which name 
we refer to the new dimensionless physical constant that enters the Born--Infeld law of the aether. 
 Reviving Abraham's ideas about the electromagnetic origin of the electron's inertia, a.k.a. mass,  
Born
       \cite{BornA}
proposed that the value of his aether constant be determined by identifying 
the empirical electron rest energy $\me c^2$ with the now finite electrostatic 
energy of a point charge at rest.
 Born and Infeld 
        \cite{BornInfeldA}
reinforced this thought by their erroneous result that the
field energy functional is the conserved energy quantity, which is not true in the 
presence of point charges.
 But even knowledge of the correct conserved energy functional is not in itself sufficient 
to determine the correct value of the aether constant.
 Interestingly enough, the true value of the aether constant may 
well be known only after the theory has been successfully quantized. 

 This brings the discussion back to Born's original motivation, namely to develop 
a consistent classical electrodynamics with point charges as a stepping stone in pursuit
of the elusive consistent quantum theory of electromagnetism with point charges.
 We here announce that we succeeded in the partial quantization of our theory (spin and photon are not
yet implemented), which we take as a major encouragement to pursue the full quantization, with 
spin and photon, in due course.
 The partially quantized theory will be presented in the sequel
         \cite{KiePapII}
to this paper.

 In the remainder of this paper, after we have stipulated the dimensionless units we use,
we first formulate the classical theory compactly in a natural dimensionless and 
manifestly Poincar\'e- and Weyl-(gauge-)covariant manner.
 In particular, the basic equations of the theory are listed and explained in section 3.2.
  In section 4, the more elaborate evolutionary formalism on the space-like standard 
foliation is extracted from the covariant formulation, the Cauchy problem is formulated
and the possibility to implement Pauli's principle discussed; also
the conservation laws are stated and their proof is sketched.
  Then, in section 5, for the benefit of the reader we recall the most important pertinent results 
about solutions with point charges that we could find in the literature; in particular, Born's field 
solution for a single point charge and Hoppe's field solution for an infinite crystal are listed; 
what seems to be known about charge-free solutions of the Maxwell--Born--Infeld field equations is 
collected in an appendix. 
 There we also add the new result that genuinely electromagnetic, subluminal 
charge-free soliton solutions of the Maxwell--Born--Infeld field equations can at most occur 
if the field strengths exceed a (huge) threshold, the rigorous proof of which we outline.
 In section 6 we illustrate the mathematical integrity of the Cauchy problem 
with point charges at hand of two examples. 
 In section 7 we assess Born's calculation of the value of the aether constant.
 The main part of the paper ends with a summary and outlook in section 8. 
%%%%%%%%%%%%%%%%%%%%%%%%%%%%%%%%%%%%%%%%%%%%%%%%%%%%%%%%%%%%%%%%%%%%
%%%%%%%%%%%%%%%%%%%%%%%%%%%%%%%%%%%%%%%%%%%%%%%%%%%%%%%%%%%%%%%%%%%%
%%%%%%%%%%%%%%%%%%%%%%%%%%%%%%%%%%%%%%%%%%%%%%%%%%%%%%%%%%%%%%%%%%%%
%%%%%%%%%%%%%%%%%%%%%%%%%%%%%%%%%%%%%%%%%%%%%%%%%%%%%%%%%%%%%%%%%%%%
	\section{The dimensionless formulation}
%%%%%%%%%%%%%%%%%%%%%%%%%%%%%%%%%%%%%%%%%%%%%%%%%%%%%%%%%%%%%%%%%%%%
%%%%%%%%%%%%%%%%%%%%%%%%%%%%%%%%%%%%%%%%%%%%%%%%%%%%%%%%%%%%%%%%%%%%
%%%%%%%%%%%%%%%%%%%%%%%%%%%%%%%%%%%%%%%%%%%%%%%%%%%%%%%%%%%%%%%%%%%%
%%%%%%%%%%%%%%%%%%%%%%%%%%%%%%%%%%%%%%%%%%%%%%%%%%%%%%%%%%%%%%%%%%%%
 The simplest presentation is achieved if the theory is written 
in an economical dimensionless manner.
 Moreover, in order to make the eventual quantization of the classical
theory that we undertake in our follow-up paper
        \cite{KiePapII}
as transparent as possible, we use the same units for the classical 
and for the quantum theory; in fact, this only helps in delineating
the range of validity of the classical theory in regard to the more
fundamental quantum theory. 
 Since quantum theory is more fundamental, we use those units that 
quantum theory itself suggests as most natural.
 As a consequence of this,
Sommerfeld's fine structure constant will necessarily make its appearance 
already in the classical theory. 
 However, its appearance in a
classical theory must not be misread  as meaning that classical theory
\emph{alone} would allow us to determine the value of Sommerfeld's fine structure constant.
 Clearly, units involving  $\hbar$ can be used only after the fact ($\hbar$),
which classical physics itself knows nothing about.
 After these words of warning, we now proceed and list the natural units.

%%%%%%%%%%%%%%%%%%%%%%%%%%%%%%%%%%%%%%%%%%%%%%%%%%%%%%%%%%%%%%%%%%%%
%%%%%%%%%%%%%%%%%%%%%%%%%%%%%%%%%%%%%%%%%%%%%%%%%%%%%%%%%%%%%%%%%%%%
%%%%%%%%%%%%%%%%%%%%%%%%%%%%%%%%%%%%%%%%%%%%%%%%%%%%%%%%%%%%%%%%%%%%
	\subsection{Natural physical units}
%%%%%%%%%%%%%%%%%%%%%%%%%%%%%%%%%%%%%%%%%%%%%%%%%%%%%%%%%%%%%%%%%%%%
%%%%%%%%%%%%%%%%%%%%%%%%%%%%%%%%%%%%%%%%%%%%%%%%%%%%%%%%%%%%%%%%%%%%
%%%%%%%%%%%%%%%%%%%%%%%%%%%%%%%%%%%%%%%%%%%%%%%%%%%%%%%%%%%%%%%%%%%%

  In its purest, genuinely electromagnetic setting the theory 
deals exclusively with positively and negatively charged electrons 
(though spinless, for now) and electromagnetic fields in spacetime,
which in turn might be flat and passive or curved and dynamical. 
 Therefore the arguably most natural dimensionless formulation is 
obtained with the following conversion factors between Gaussian 
and dimensionless units: $\hbar$ (Planck's constant divided by $2{\pi}$) 
for both the unit of action and the magnitude of angular momentum, 
$e$ (elementary charge) for the unit of charge, 
$\me$ (electron rest mass) for the unit of mass, 
$c$ (speed of light \textit{in vacuo}) for the unit of speed.
  Thus, length and time are both referred to in the same units, for which 
the Compton wave length of the electron $\lC = {\hbar}/{\me}c$ is used
to convert the dimensionless unit of length and time.
  Accordingly, the unit magnitude of the electromagnetic 
fields is to be converted by a factor $e/\lC^2$, 
while the natural unit for the magnitude  of momentum and the 
energy are converted, respectively, by factors ${\me}c$ and ${\me}c^2$.

%%%%%%%%%%%%%%%%%%%%%%%%%%%%%%%%%%%%%%%%%%%%%%%%%%%%%%%%%%%%%%%%%%%%
%%%%%%%%%%%%%%%%%%%%%%%%%%%%%%%%%%%%%%%%%%%%%%%%%%%%%%%%%%%%%%%%%%%%
%%%%%%%%%%%%%%%%%%%%%%%%%%%%%%%%%%%%%%%%%%%%%%%%%%%%%%%%%%%%%%%%%%%%
	\subsection{The universal parameters}
%%%%%%%%%%%%%%%%%%%%%%%%%%%%%%%%%%%%%%%%%%%%%%%%%%%%%%%%%%%%%%%%%%%%
%%%%%%%%%%%%%%%%%%%%%%%%%%%%%%%%%%%%%%%%%%%%%%%%%%%%%%%%%%%%%%%%%%%%
%%%%%%%%%%%%%%%%%%%%%%%%%%%%%%%%%%%%%%%%%%%%%%%%%%%%%%%%%%%%%%%%%%%%

  In the general-relativistic spacetime, the genuinely electromagnetic theory 
features precisely three universal dimensionless constants,\footnote{Other dimensionless parameters, like the 
                                                                     numbers of positive and negative electrons, 
								     are better thought of as part of the initial data, 
								     which means they are non-universal parameters.}
$\alpha$, $\beta$, and $\gamma$, but only $\alpha$ and $\beta$ figure in the 
special-relativistic spacetime.

 The constant $\alpha$ is Sommerfeld's fine structure constant, i.e.
\beq
\alpha 
\equiv 
\frac{e^2}{{\hbar}c}
\approx \frac{1}{137.036}
\,.
\label{eq:SOMMERFELDconstant}
\eeq
 Strictly speaking, since our set of dynamical equations is new, we have to 
vindicate the identification \refeq{eq:SOMMERFELDconstant}. 
  The value of $\alpha$ can be determined by demanding that in the non-relativistic 
radiation-free limit of the {classical} theory the well-known law of motion 
for the electrons obtains (in our units). 
 This classical dynamical vindication of \refeq{eq:SOMMERFELDconstant}
we will carry out non-rigorously.
  In our follow-up paper 
        \cite{KiePapII}, 
which deals with the quantum theory, we will 
carry out a rigorous study of the hydrogen spectrum which confirms our 
result \refeq{eq:SOMMERFELDconstant} for $\alpha$.

 The parameter $\beta$ is Born's aether constant,\footnote{Born \cite{BornA} originally introduced the dimensional parameter 
                                                     $a$, but subsequently switched to use $b\, \equiv a^{-1}$.	Our 
                                                     dimensionless $\beta^2\, \propto a$ (the reason for the `square' is
						     simply to avoid having to write some awkward `square roots' later on).}
which enters through the Born--Infeld aether law.
 Its value is a rather subtle issue.
 Born's result for $\beta$ is
          \cite{BornA}
\beq
{\beta}\big|_{_{\mathrm{Born}}}
=
     {\textstyle{\frac{1}{6}}}
\Beta\left({\textstyle{\frac{1}{4},\frac{1}{4}}}\right){\alpha}
\approx 
1.2361{\alpha}
\,,
\label{eq:BORNconstant}
\eeq
but as we will explain in this paper, \refeq{eq:BORNconstant} must be taken with a grain of salt.
 A study of the hydrogen spectrum in our follow-up paper 
   \cite{KiePapII}
will supply  upper bounds on $\beta$ 
compatible with \refeq{eq:BORNconstant}, but as long as spin and the photon have not been incorporated 
in the theory, the value for $\beta$ remains tentative.

 Finally, $\gamma$ is the electron's `gravitational fine structure constant,' 
given by
\beq
\gamma
\equiv 
\frac{G\me^2}{{\hbar}c}
\approx 1.75 {\times}10^{-45}
\,
\label{eq:gravitationalSOMMERFELDconstant}
\eeq
and obtained by rewriting Einstein's field equations in our dimensionless units;
here, $G$ is Newton's constant of universal gravitation. 
  We will focus mostly on the genuinely electromagnetic theory
in the special-relativistic limit $\gamma\downarrow 0$, choosing Minkowski spacetime 
as solution for the vacuum Einstein equations.

 Some non-genuine yet electromagnetically important phenomena, such as nuclei charged with $z$  
electric units and having considerably bigger masses than electrons due to their strong interactions, 
can be readily accommodated\footnote{Particles which do not interact at all electromagnetically can be 
                                    readily accommodated, too.}
at the price of introducing additional dimensionless (effective) parameters into the theory, viz. the very 
$z$'s and the relative masses of the nuclei. 
  We will outline the modifications in a separate subsection.
 These modifications should be derivable from some deeper theory 
with fewer parameters which ultimately would be truly universal.
%%%%%%%%%%%%%%%%%%%%%%%%%%%%%%%%%%%%%%%%%%%%%%%%%%%%%%%%%%%%%%%%%%%%
%%%%%%%%%%%%%%%%%%%%%%%%%%%%%%%%%%%%%%%%%%%%%%%%%%%%%%%%%%%%%%%%%%%%
%%%%%%%%%%%%%%%%%%%%%%%%%%%%%%%%%%%%%%%%%%%%%%%%%%%%%%%%%%%%%%%%%%%%
%%%%%%%%%%%%%%%%%%%%%%%%%%%%%%%%%%%%%%%%%%%%%%%%%%%%%%%%%%%%%%%%%%%%
	\section{The  covariant formalism}
%%%%%%%%%%%%%%%%%%%%%%%%%%%%%%%%%%%%%%%%%%%%%%%%%%%%%%%%%%%%%%%%%%%%
%%%%%%%%%%%%%%%%%%%%%%%%%%%%%%%%%%%%%%%%%%%%%%%%%%%%%%%%%%%%%%%%%%%%
%%%%%%%%%%%%%%%%%%%%%%%%%%%%%%%%%%%%%%%%%%%%%%%%%%%%%%%%%%%%%%%%%%%%
%%%%%%%%%%%%%%%%%%%%%%%%%%%%%%%%%%%%%%%%%%%%%%%%%%%%%%%%%%%%%%%%%%%%
%
%%%%%%%%%%%%%%%%%%%%%%%%%%%%%%%%%%%%%%%%%%%%%%%%%%%%%%%%%%%%%%%%%%%%
%%%%%%%%%%%%%%%%%%%%%%%%%%%%%%%%%%%%%%%%%%%%%%%%%%%%%%%%%%%%%%%%%%%%
%%%%%%%%%%%%%%%%%%%%%%%%%%%%%%%%%%%%%%%%%%%%%%%%%%%%%%%%%%%%%%%%%%%%
	{\subsection{The Minkowski spacetime}}
%%%%%%%%%%%%%%%%%%%%%%%%%%%%%%%%%%%%%%%%%%%%%%%%%%%%%%%%%%%%%%%%%%%%
%%%%%%%%%%%%%%%%%%%%%%%%%%%%%%%%%%%%%%%%%%%%%%%%%%%%%%%%%%%%%%%%%%%%
%%%%%%%%%%%%%%%%%%%%%%%%%%%%%%%%%%%%%%%%%%%%%%%%%%%%%%%%%%%%%%%%%%%%

%%%%%%%%%%%%%%%%%%%%%%%%%%%%%%%%%%%%%%%%%%%%%%%%%%%%%%%%%%%%%%%%%%%%
%%%%%%%%%%%%%%%%%%%%%%%%%%%%%%%%%%%%%%%%%%%%%%%%%%%%%%%%%%%%%%%%%%%%
	{\subsubsection{The manifold and its tangent space}}
%%%%%%%%%%%%%%%%%%%%%%%%%%%%%%%%%%%%%%%%%%%%%%%%%%%%%%%%%%%%%%%%%%%%
%%%%%%%%%%%%%%%%%%%%%%%%%%%%%%%%%%%%%%%%%%%%%%%%%%%%%%%%%%%%%%%%%%%%

	Except in a brief subsection, in this paper the arena for the 
electromagnetic phenomena is the passive Minkowski spacetime $\MM^4$, 
a pseudo-Riemannian flat, oriented $4$-manifold with Lorentzian metric 
tensor $\gQ$ of signature\footnote{Here we follow mathematical conventions as in
                                   \cite{hawkingellisBOOK};
				   in the physics literature this corresponds to
				   saying that the signature of $\gQ$ is $(-,+,+,+)$.}
$+2$. 
	The points $\varpi$ in Minkowski spacetime are called 
world-points (or sometimes, events).

	The  geometry of Minkowski spacetime is defined through 
the Poincar\'e group acting on $\MM^4$, consisting of the isometries with 
respect to $\gQ$, i.e. the spacetime translations, rotations, and reflections,
		\cite{streaterwightmanBOOK, barutBOOKa}.
	Hence, we can choose any particular world-point $\varpi_0\in{\MM}^{4}$
and identify any other world-point $\varpi\in{\MM}^{4}$ with the 
world-vector $\MTWvec{w}\in {\mathbb T}_{\varpi_0}({\MM}^{4})$ 
by stipulating that $\MTWvec{w}$ is the oriented chord from $\varpi_0$ 
to $\varpi$.
	In this way ${\MM}^{4}$ becomes identified with 
${\mathbb T}_{\varpi_0}({\MM}^{4})$, its tangent space at $\varpi_0$.

        A {Lorentz frame} is a basis for ${\mathbb T}_{\varpi_0}({\MM}^{4})$, 
viz. any tetrad $\{\eQ_0,\eQ_1,\eQ_2,\eQ_3\}$ of fixed unit world-vectors 
in ${\mathbb T}_\varpi({\MM}^{4})$ that satisfy the elementary 
inner product rules
\begin{equation}
\gQ(\eQ_\mu,\eQ_\nu) 
\equiv 
\eQ_\mu\cdot \eQ_\nu 
        = 
        \left\{
        \begin{array}{rl}
                -1 & \mathrm{for}\quad \mu =   \nu = 0 \\
                 1 & \mathrm{for}\quad \mu =   \nu > 0 \\
                 0 & \mathrm{for}\quad \mu\neq \nu     
        \end{array}
        \right.
        \, .
        \label{eq:ipr} %inner product rules
\end{equation}
	With respect to such a Lorentz frame any real-valued vector 
$\wQ \in {\mathbb T}_{\varpi_0}({\MM}^{4})$ has the representation
(Einstein summation convention understood)
\begin{equation}
        \wQ
        =       
	%        \sum_{\mu=0}^3 
        w^\mu \eQ_\mu
        \label{eq:vEXPAND} % four vector tetrad expansion 
\end{equation}
with $w^0 = - \wQ \cdot \MTWvec{e}_0$ and
$w^\mu = \wQ\cdot \MTWvec{e}_\mu$ for $\mu= 1,2,3$.
	The identification $\wQ\cong (w^0,w^1,w^2,w^3)\in \RR^{1,3}$, 
where $\RR^{1,3}$ is the set of ordered real $4$-tuples 
equipped with Minkowski metric, provides a global coordinatization
for ${\MM}^{4}$.

	World-vectors $\wQ$ are classified into space-like, 
light-like, and time-like according as $\MnormSQ{\wQ} >0$, $\MnormSQ{\wQ} =0$, 
or $\MnormSQ{\wQ} <0$, where $\MnormSQ{\wQ}\equiv \wQ\cdot\wQ$.
        A space-like world-vector is connected via a continuous orbit 
of the Lorentz group with a non-zero world-vector $\sQ$ satisfying 
$\sQ\cdot \MTWvec{e}_0=0$, a time-like one with a non-zero world-vector 
$\tQ$ satisfying $\tQ\cdot \MTWvec{e}_\mu=0$ for all $\mu=1,2,3$.
        The light-like world-vectors form the light-(double)cone 
in ${\mathbb T}_{\varpi_0}({\MM}^{4})$, which separates the space-like 
from the time-like world-vectors in ${\mathbb T}_{\varpi_0}({\MM}^{4})$.

\smallskip
%%%%%%%%%%%%%%%%%%%%%%%%%%%%%%%%%%%%%%%%%%%%%%%%%%%%%%%%%%%%%%%%%%%%
%%%%%%%%%%%%%%%%%%%%%%%%%%%%%%%%%%%%%%%%%%%%%%%%%%%%%%%%%%%%%%%%%%%%
	{\subsubsection{World-lines and point-histories in spacetime}}
%%%%%%%%%%%%%%%%%%%%%%%%%%%%%%%%%%%%%%%%%%%%%%%%%%%%%%%%%%%%%%%%%%%%
%%%%%%%%%%%%%%%%%%%%%%%%%%%%%%%%%%%%%%%%%%%%%%%%%%%%%%%%%%%%%%%%%%%%

  One-dimensional geometrical objects in the Minkowski spacetime, 
viz. continuous curves in spacetime, are called \emph{world-lines}. 
  A time-like future-oriented world-line is called a 
\emph{point-history}, or just history for short, and denoted $\Eta$.

  We shall assume that histories are  of class $C^1$, in fact $C^{1,\alpha}$.
  In that case, at each event on a history there is a unique future-oriented
unit tangent vector to the history.
  Its metric dual is a co-vector $\uQ$, satisfying $\uQ\cdot\uQ =-1$ and $u_0 >0$. 

  By  $\varpi_p\Eta\varpi_f$ we denote a truncated point history between two 
events, $\varpi_p$ and $\varpi_f$, where $\varpi_f$ lies in the future 
of $\varpi_p$.
  By integrating the co-vector $\uQ$ along a truncated history, we obtain the 
invariant \emph{proper-time span} $\cT[\varpi_p\Eta\varpi_f]$ of the truncated history, 
$
\cT[\varpi_p\Eta\varpi_f] = \int_{\varpi_p\Eta\varpi_f} \uQ\, .
$
  In particular, by choosing any convenient event $\varpi_0$ on a history 
as reference world-point to which is assigned the proper-time Null, 
we can then assign to each event $\varpi$ on that history a proper-time 
$\tau\in\RR$ relative to $\varpi_0$ defined to be the proper-time span 
$\cT[\varpi_0\Eta\varpi]$ ($\cT[\varpi\Eta\varpi_0)]$) if $\varpi$ lies in the future 
(past) of $\varpi_0$.
        Since the  history is time-like future-oriented, its proper-time 
assignment is one-to-one onto and order-preserving. 
Hence, proper-time
in turn can serve as a natural parameterization for the  history, i.e.
a  history is given by a $C^{1,\alpha}$ mapping 
$\eta:\, \tau\mapsto \varpi = \eta(\tau)\in {\MM}^{4}$.
        Similarly, the metric dual of the unit tangent vector at each world-point $\varpi$ on the history
is then a co-vector-valued $C^\alpha$ map $\tau\mapsto \uQ(\tau)$.        

\smallskip
%%%%%%%%%%%%%%%%%%%%%%%%%%%%%%%%%%%%%%%%%%%%%%%%%%%%%%%%%%%%%%%%%%%%
%%%%%%%%%%%%%%%%%%%%%%%%%%%%%%%%%%%%%%%%%%%%%%%%%%%%%%%%%%%%%%%%%%%%
	{\subsubsection{Space-like slices}}
%%%%%%%%%%%%%%%%%%%%%%%%%%%%%%%%%%%%%%%%%%%%%%%%%%%%%%%%%%%%%%%%%%%%
%%%%%%%%%%%%%%%%%%%%%%%%%%%%%%%%%%%%%%%%%%%%%%%%%%%%%%%%%%%%%%%%%%%%

	Dual as a concept to point-histories in spacetime
are space-like slices of spacetime. 
	A space-like slice $\Sigma\subset{\MM}^{4}$ 
is a three-dimensional simply connected hypersurface in $\MM^4$
with a time-like normal vector at every point in $\Sigma$.
	Topologically, $\Sigma$ is homeomorphic to $\RR^3$. 
	Without loss of generality one can assume that  
$\Sigma_t =\{\varpi: T(\varpi) =t\}$ is the boundary of a 
level set of a differentiable function $T:\MM^4\to\RR$ which has the properties that
ran$(T)=\RR$ and that the metric dual to $\dQ T(\varpi)$ is time-like for all $\varpi$ 
in $\MM^4$; here $\dQ$ denotes E. Cartan's exterior derivative on ${\MM}^4$.

\smallskip
%%%%%%%%%%%%%%%%%%%%%%%%%%%%%%%%%%%%%%%%%%%%%%%%%%%%%%%%%%%%%%%%%%%%
%%%%%%%%%%%%%%%%%%%%%%%%%%%%%%%%%%%%%%%%%%%%%%%%%%%%%%%%%%%%%%%%%%%%
	{\subsubsection{Space-like foliations}}
%%%%%%%%%%%%%%%%%%%%%%%%%%%%%%%%%%%%%%%%%%%%%%%%%%%%%%%%%%%%%%%%%%%%
%%%%%%%%%%%%%%%%%%%%%%%%%%%%%%%%%%%%%%%%%%%%%%%%%%%%%%%%%%%%%%%%%%%%

	The light-cone structure of ${\MM}^{4}$ allows one to
foliate this manifold with space-like leafs.
	The intersection of any two such space-like hypersurfaces 
$\Sigma_{t_1}$ and $\Sigma_{t_2}$ with $t_1\neq t_2$ is empty, and
each $\varpi$ is in precisely one $\Sigma_t$ for some suitable 
$t\in {\mathrm{ran}}(T)$. 
	As a consequence, $\bigcup_{t\in{\RR}}\Sigma_t$ can 
be identified with all of spacetime, and this identification 
constitutes the {foliation} of spacetime (generated by $T$).

	We can coordinatize all leafs $\Sigma_t$, $t\neq 0$, with the
coordinates on $\Sigma_0$ by following the integral curves of the metric
dual of $\dQ{T}$ 
to their piercing through points in $\Sigma_0$. 
	Equipped with such a global parameterization of a 
foliation, Minkowski spacetime is always diffeomorphic 
to the product manifold ${\mathbb R}\times \Sigma_0$. 
	If $(s^1,s^2,s^3)$ are arbitrary coordinates on the space-like
$\Sigma_0$, we can coordinatize all of $\MM^4$ by 
$\varpi\cong(s^0,s^1,s^2,s^3)$, with $s^0=t$. 
	The spacetime metric of $\MM^4$ with respect to the
foliation generated by $T$ in these coordinates is given by the
world-line element
\begin{equation}
	\dd s\,{}^2  
= 
	- \ell^2(\varpi)\, \dd t\,{}^2  
	+ \sum_{1\leq m,n\leq 3}g_{mn}(\varpi) \dd s^m \dd s^n,
\label{eq:lapseMETRIC}
\end{equation}
where $\ell = (- \dQ T \cdot \dQ T)^{-1/2}$ 
is the \emph{lapse function}, 
and where the $g_{mn}(\varpi)$ are the components,
with respect to an arbitrary space frame of $\Sigma_t$, of the 
\emph{first fundamental form} $g$ of the leaf at $\varpi$, which 
is induced on $\Sigma_t$ by the Minkowski metric on $\MM^4$.

\medskip
%%%%%%%%%%%%%%%%%%%%%%%%%%%%%%%%%%%%%%%%%%%%%%%%%%%%%%%%%%%%%%%%
%%%%%%%%%%%%%%%%%%%%%%%%%%%%%%%%%%%%%%%%%%%%%%%%%%%%%%%%%%%%%%%%
%%%%%%%%%%%%%%%%%%%%%%%%%%%%%%%%%%%%%%%%%%%%%%%%%%%%%%%%%%%%%%%%
        \subsection{The electromagnetic spacetime}
%%%%%%%%%%%%%%%%%%%%%%%%%%%%%%%%%%%%%%%%%%%%%%%%%%%%%%%%%%%%%%%%
%%%%%%%%%%%%%%%%%%%%%%%%%%%%%%%%%%%%%%%%%%%%%%%%%%%%%%%%%%%%%%%%
%%%%%%%%%%%%%%%%%%%%%%%%%%%%%%%%%%%%%%%%%%%%%%%%%%%%%%%%%%%%%%%%

  We are now ready to define classical \emph{electromagnetic (flat) spacetime}
as Minkowski spacetime equipped with a classical electromagnetic structure, 
consisting of a two-form (field tensor) which is defined almost everywhere, 
the exception being $N$ time-like oriented complete world-lines (the histories 
of the point charges).
  Any such electromagnetic structure satisfies the Maxwell--Born--Infeld laws
of the electromagnetic field with point sources, which we first state and 
then assess.

%%%%%%%%%%%%%%%%%%%%%%%%%%%%%%%%%%%%%%%%%%%%%%%%%%%%%%%%%%%%%%%%
%%%%%%%%%%%%%%%%%%%%%%%%%%%%%%%%%%%%%%%%%%%%%%%%%%%%%%%%%%%%%%%%
\subsubsection{The Maxwell--Born--Infeld field laws with point sources}
%%%%%%%%%%%%%%%%%%%%%%%%%%%%%%%%%%%%%%%%%%%%%%%%%%%%%%%%%%%%%%%%
%%%%%%%%%%%%%%%%%%%%%%%%%%%%%%%%%%%%%%%%%%%%%%%%%%%%%%%%%%%%%%%%

 Let $\bigcup_k\Eta_k$ denote the set of $N$ point histories with which $\MM^4$ is
threaded. 
 Then on $\MM^4\backslash\bigcup_k\Eta_k$ there is stipulated to exist a 
two-form $\FQ$, called Faraday's electromagnetic field tensor.
 The postulate that $\FQ$ is closed on $\MM^4$, i.e.
\begin{equation}
        \dQ{\FQ} 
=
\MTWvec{0}
\, 
\label{eq:GEOmaxwEQhom}
\end{equation}
 in the sense of distributions, is known as the \emph{Faraday--Maxwell law} 
(cf. 
		\cite{WheeleretalBOOK}).

 To relate the two-form $\FQ$ to the point histories, $\FQ$ is mapped to another 
two-form $\MQ$, called Maxwell's electromagnetic displacement tensor. 
 The map is effected by the \emph{Born and Infeld law of the aether},
\beq
-\Hodge\MQ 
= 
\frac{\FQ - \beta^4  {}\Hodge\big(\FQ\wedge\FQ \big){}\Hodge\FQ} 
     {\sqrt{1 - \beta^4 {}\Hodge\big(\FQ\wedge \Hodge\FQ \big)
              - \beta^8 \left({}\Hodge \big(\FQ\wedge\FQ \big)\right)^2 }}
\,,
\label{eq:GEOaetherLAWborninfeld}
\eeq
where $\Hodge\FQ$ (etc.) is the Hodge dual of $\FQ$ (etc.).
 The rather complicated  law \refeq{eq:GEOaetherLAWborninfeld}, 
reduces to the linear law  $\MQ = \Hodge\FQ$ of Maxwell's pure aether
in the limit\footnote{More precisely, since $\beta$ is a universal physical 
                      constant which has to be assigned a nonzero value to avoid the infinite self-energies
		      of the classical Lorentz electrodynamics, the proper way of putting it is to say 
		      that \refeq{eq:GEOaetherLAWborninfeld} reduces to Maxwell's linear law of the 
		      aether whenever the magnitude of $\FQ$ is sufficiently small; i.e., when
		      $\beta^4 |{}\Hodge\big(\FQ\wedge \Hodge\FQ \big)| 
		       + \beta^8 \left({}\Hodge \big(\FQ\wedge\FQ \big)\right)^2\ll 1$, 
		      then $\MQ \sim  \Hodge\FQ  \quad (\mathrm{weak\ field\ limit})$.}
$\beta \downarrow 0$. 

 The distributional exterior derivative of the two-form $\MQ$ now gives the
electromagnetic current density,
\begin{equation}
\dQ\MQ 
= 
4\pi \JQ 
\,.
\label{eq:GEOmaxwEQinh}
\end{equation}
  Equation \refeq{eq:GEOmaxwEQinh} will be called the \emph{Amp\'ere--Coulomb--Maxwell} 
law; cf. 
		\cite{WheeleretalBOOK}. 

    Equation \refeq{eq:GEOmaxwEQinh} implies that the electromagnetic current density $\JQ$
is a closed three-form, i.e. its exterior derivative on ${\MM}^{4}$, in the sense of distributions, 
vanishes,
\begin{equation}
  	\dQ\JQ 
=
\MTWvec{0}
\,.
\label{eq:GEOconservechargeLAW}
\end{equation}
   Equation \refeq{eq:GEOconservechargeLAW} is known as the \emph{law of the conservation 
of electric charge}. 

   In the genuinely electromagnetic setting of the theory, all charges
are either positive or negative unit point charges, representing electrons of either 
variety.
   For a system of $N\geq 0$ electric unit point charges the electromagnetic current 
density $\JQ(\varpi)$ at $\varpi$  is given by the familiar expression
                 \cite{WheeleretalBOOK, ThirringBOOKa}, 
\begin{equation}
         \JQ(\varpi) 
=
	\sum_{k\in\cN} \int_{-\infty}^{+\infty}
		\pmk\Hodge\uQ_k(\tau)\delta_{\eta_k(\tau)}\big(\varpi\big)\,\dd\tau
\, ,
\label{eq:GEOptchargecurrent}
\end{equation}
cf. also 
                 \cite{weinbergBOOKart, jacksonBOOK}.
    Here, $\delta_{\eta(\tau)}\big(\,.\,\big)$ 
is the Dirac measure on $\MM^4$ concentrated at $\eta(\tau)$ 
at proper-time $\tau$, $\Hodge\uQ_k(\tau)$ is the Hodge dual of 
the future-oriented Minkowski-velocity co-vector $\uQ_k(\tau)$,
and $\pmk$ is the sign of the $k^{\mathrm{th}}$ point charge;\footnote{This formula for $\JQ$ reveals very nicely that
                                                                       the `sign of the point charge' $\pmk$ is 
								       actually an artifact of our insistence (psychologically 
								       inherited from how we perceive the world around and in us) 
								       that the $\uQ_k$  be future oriented.
								       Geometrically it is much more natural to absorb $\pmk$ 
								       into $\uQ_k$ and simply distinguish future- and 
								       past-oriented $\uQ_k$s, as Wheeler has suggested.}
furthermore, $\cN\subset\NN\cup\{0\}$ is the set of $N$ indices.
 We set $\cN\equiv\emptyset$ if $N=0$, in which case
$\sum_{k\in\emptyset} (...) \equiv 0$, so
that the charge-free situation is included in \refeq{eq:GEOptchargecurrent}.

 The manifestly covariant character of the Maxwell--Born--Infeld laws 
makes it plain that, whichever particular electromagnetic structure satisfying these
laws may be the actually realized one in Minkowski spacetime, any two 
Lorentz observers of this electromagnetic spacetime would necessarily
conclude that they see their respective Lorentz frame manifestations of 
the \emph{same} electromagnetic spacetime, whatever their relative state of 
uniform motion with respect to each other might be.

%%%%%%%%%%%%%%%%%%%%%%%%%%%%%%%%%%%%%%%%%%%%%%%%%%%%%%%%%%%%%%%%
%%%%%%%%%%%%%%%%%%%%%%%%%%%%%%%%%%%%%%%%%%%%%%%%%%%%%%%%%%%%%%%%
\subsubsection{Limitations of the Maxwell--Born--Infeld field laws with point sources}
%%%%%%%%%%%%%%%%%%%%%%%%%%%%%%%%%%%%%%%%%%%%%%%%%%%%%%%%%%%%%%%%
%%%%%%%%%%%%%%%%%%%%%%%%%%%%%%%%%%%%%%%%%%%%%%%%%%%%%%%%%%%%%%%%
 
 For any particular electromagnetic spacetime as defined above, a-priori knowledge of 
$\FQ$ would in principle allow us to compute from it the remaining electromagnetic 
quantities $\MQ$ and $\JQ$, the latter giving $\bigcup_k\Eta_k$ and the $\uQ_k(\tau)$. 
 Of course, we have no complete a-priori knowledge of the `actual' $\FQ$ (i.e., `actual'
to the extent that curvature and non-electromagnetic effects can be neglected), but the 
partial knowledge we do have of our local electromagnetic spacetime indicates much more 
order in the world than demanded by the Maxwell--Born--Infeld field laws alone.
 In particular, even granted we would have complete knowledge of the past and present
part of the electromagnetic spacetime with respect to some space-like hypersurface of some 
standard Lorentz frame (and also granted we would  know $\FQ$ at space-like infinity), 
the  Maxwell--Born--Infeld laws would not allow us to continue that information uniquely 
into the future to extend the knowledge of the electromagnetic spacetime, not to mention
completing the knowledge; unless, that is, $\JQ$ would vanish identically. 
 But since a basic hypothesis of Born--Infeld's, and also our, theory is that there are
point charges, in order to be able to uniquely continue such data of the electromagnetic 
spacetime we need to supply a mathematically well-defined law for the point histories, 
and not just any such law but one which accounts for the observed regularities of point 
charge motions. 
 In particular, the law for the point histories has to reduce to the known asymptotic law 
in the limit of gently accelerated, slowly moving, far separated point charges, which is
Newton's law of motion equipped with the Coulomb law for the  forces between the point charges;
we also know that magnetic effects can be taken into account very accurately (in the same 
regime of motion) by replacing the Coulomb force on a particle with the Lorentz force due 
to all other particles. 

 Unfortunately, the formal expression of Newton's law of motion equipped with the 
total Lorentz force involves $\FQ$ at the location of the point charges, but $\FQ$ inherits 
from $\MQ$ (recall \refeq{eq:GEOmaxwEQinh}) those locations as singular points; and while in 
contrast to the singularities of $\MQ$ those of $\FQ$ are very mild (finite discontinuities) 
when $\beta \neq 0$, the tensor field $\FQ$ cannot be extended continuously into those points. 
 This renders the Newtonian law of motion with the total electromagnetic fields
an only formal arrangement of symbols without proper meaning. 
 We need a crucial new insight.

%%%%%%%%%%%%%%%%%%%%%%%%%%%%%%%%%%%%%%%%%%%%%%%%%%%%%%%%%%%%%%%%%%%%
%%%%%%%%%%%%%%%%%%%%%%%%%%%%%%%%%%%%%%%%%%%%%%%%%%%%%%%%%%%%%%%%%%%%
	\subsection{A first-order law for the histories of the point-charges}
%%%%%%%%%%%%%%%%%%%%%%%%%%%%%%%%%%%%%%%%%%%%%%%%%%%%%%%%%%%%%%%%%%%%
%%%%%%%%%%%%%%%%%%%%%%%%%%%%%%%%%%%%%%%%%%%%%%%%%%%%%%%%%%%%%%%%%%%%

 To guide our search for the law of motion, we argue heuristically as follows.
  First of all, we recall that \refeq{eq:GEOmaxwEQhom} implies that, except at the locations 
of the point charges where it is necessarily undefined, an otherwise differentiable field tensor 
$\FQ$ is an exact two-form on $\MM^4\backslash\bigcup_k\Eta_k$,  i.e.
\beq
\FQ=\dQ\AQ
\,,
\label{eq:FdA}
\eeq
where $\AQ$, the \emph{electromagnetic potential}, is a one-form on ${\MM}^{4}\backslash\bigcup_k\Eta_k$.
 Recall also 
that such an $\AQ$ is not uniquely defined by \refeq{eq:FdA}, for adding an exact one-form 
to $\AQ$, i.e.
\beq
\AQ
\to
\AQ+\dQ\Upsilon,
\label{eq:GEOgaugetrA}
\eeq
where $\Upsilon:\MM^4\to\RR$ is any zero-form, obviously leaves  $\FQ$ unchanged. 
 The interesting point now is that, since the singularities of such an $\FQ$ are mild, namely finite jump 
discontinuities, we can assume that the one-form $\AQ$ can be Lipschitz-continuously extended into the 
locations of the point charges, hence onto all of $\MM^4$, and \refeq{eq:FdA} is then to be understood 
in the sense of distributions. 
 Actually, what we have just been saying is rigorously known to be true only for a point charge in arbitrary 
uniform motion, but it is very much expected to be true also for accelerated point charges, 
for in the immediate vicinity of the accelerated point charge its field should look like the
co-moving stationary field of a charge in uniform motion with same instantaneous velocity, 
while the acceleration makes only a \emph{relatively small} correction to the fields.
 Although the intuition behind this expectation is borrowed from the experience with the 
Li\'enard--Wiechert solutions 
	       \cite{lienard, wiechert}
to the linear Maxwell--Lorentz equations with point charge source, \emph{in this respect} 
the nonlinear Maxwell--Born--Infeld field equations should behave just like 
the linear Maxwell--Lorentz equations;\footnote{Clearly, a rigorous proof of this assertion 
                                                needs to be supplied at some point.} 
unless, that is, shocks would form.
 Fortunately, shock formation is ruled out by the complete linear degeneracy of the linearized 
Maxwell--Born--Infeld field equations
	      \cite{Boillat, Plebanski}.
 It follows that, if all what we just said is indeed rigorously true for any electromagnetic 
spacetime satisfying the Maxwell--Born--Infeld field laws with point charges, then the 
familiar canonical momentum of the $k$-th point charge,
\begin{equation}
    \PQ_k(\tau) 
= \uQ_k(\tau) + z_k \alpha\AQ(\eta(\tau)),
\label{eq:canonicalMOMENTUM}
\end{equation}
is well-defined along its point history $\Eta_k$ in any such an electromagnetic spacetime.
 In any event, \refeq{eq:canonicalMOMENTUM} is certainly well-defined for those parts of an 
electromagnetic spacetime satisfying the Maxwell--Born--Infeld field laws with point charges 
for which what we said is true. 
 After having stressed this, we will now take \refeq{eq:canonicalMOMENTUM} as our
point of departure for constructing a well-defined \emph{first-order} law for the 
point histories.

 To prepare the ground for the statement of the law of motion, at this point it is helpful
to recall the co-variant formulation of the \emph{test particles} Hamilton--Jacobi theory
of Newton's law of motion for point charges in  \emph{given} nice (read: at least Lip\-schitz continuous) 
electromagnetic fields on $\MM^4$ so that the electromagnetic potentials are differentiable.
 Next we upgrade this test particles  Hamilton--Jacobi formalism to nice
electromagnetic fields which are to be \emph{solved for alongside} the Hamilton--Jacobi equation. 
 While this formulation is inadequate for our situation in which the electromagnetic
fields are not Lipschitz continuous at the locations of the actual point charges, a brief
analysis of why the `upgraded test particle Hamilton--Jacobi formulation' fails
will right away lead us to the remedy, which we call the \emph{proper particles} 
Hamilton--Jacobi formulation.

%%%%%%%%%%%%%%%%%%%%%%%%%%%%%%%%%%%%%%%%%%%%%%%%%%%%%%%%%%%%%%%%%%%%
	\subsubsection{Test particles Hamilton--Jacobi theory}
%%%%%%%%%%%%%%%%%%%%%%%%%%%%%%%%%%%%%%%%%%%%%%%%%%%%%%%%%%%%%%%%%%%%

 In this vein, we temporarily pretend that $\AQ$ in \refeq{eq:canonicalMOMENTUM} were not
the electromagnetic one-form for the total electromagnetic Maxwell--Born--Infeld fields 
created by the point charges, but instead just a smoothly differentiable electromagnetic 
potential field on $\MM^4$ for some given, nice electromagnetic fields. 
 Then the familiar\footnote{Perhaps it will become explicitly obvious
           only in the space and time decomposition, and after synchronization w.r.t. Lorentz 
	   time, that these co-variant equations describe the familiar $N$ test particles 
	   Hamilton--Jacobi formalism.
           The manifest simplicity of the co-variant formalism should leave no doubt, however, 
	   that this is the natural language for relativistic Hamilton--Jacobi theory.}
{test particles} Hamilton--Jacobi theory sets 
$\PQ_k(\tau) = (\dQ_k \widetilde\Phi) (\eta_1(\tau),...,\eta_N(\tau))$ for 
$\widetilde\Phi(\varpi_1,...,\varpi_N)$ a scalar field on world configuration space $\MM^{4N}$, so
that \refeq{eq:canonicalMOMENTUM} becomes
 \beq
\uQ_k(\tau)
=
      {\dQ}_k \widetilde\Phi(\varpi_1,...,\varpi_N)
-\pmk {\alpha}\AQ(\varpi_k) \Big|_{\{\varpi_n = \eta_n(\tau)\}}\Big.
\,,
\label{eq:GEOhamjacLAWtestPARTICLES}
\eeq
which needs to be solved for all $k$, with initial data for the $\eta_k(0)$ supplied.
 Interestingly, the scalar field $\widetilde\Phi$ in turn is implicitly determined by 
\refeq{eq:GEOhamjacLAWtestPARTICLES}, too, for notice that the r.h.s. of \refeq{eq:GEOhamjacLAWtestPARTICLES} 
has to produce a future-oriented co-vector $\uQ_k$ satisfying $\Hodge\left(\uQ_k\wedge\Hodge\uQ_k\right) = 1$.
 But since the world-line for $\uQ_k$ is not known a-priori, the r.h.s. of \refeq{eq:GEOhamjacLAWtestPARTICLES} 
can only do what it is supposed to do if $\widetilde\Phi$ on $\MM^{4N}$ satisfies the $N$ equations 
\beq
\Hodge
\bigl( \bigl( 
{\dQ}_k \widetilde{\Phi}(\varpi_1,...,\varpi_N)
-\pmk {\alpha}\AQ_k(\varpi_k)
\bigr)\wedge
\Hodge \bigl( 
{\dQ}_k \widetilde{\Phi}(\varpi_1,...,\varpi_N)
-\pmk {\alpha}\AQ(\varpi_k)
\bigr)\bigr)
 = 1
\,.
\label{eq:GEOhamjacPDEtestPARTICLES}
\eeq
 Each equation \refeq{eq:GEOhamjacPDEtestPARTICLES} has a double root, of which the future-oriented 
one is to be chosen.
 This system of equations is consistent if the $N$ world-points $\varpi_k$ are distributed over a 
space-like hypersurface in $\MM^4$.
 However, since in given fields each test particle moves as if it were the only one around, we may 
separate variables as 
$\widetilde{\Phi}(\varpi_1,...,\varpi_N) = \sum_{k=1}^N\widetilde\Phi^{(k)}(\varpi_k)$, 
where $\widetilde\Phi^{(k)}(\varpi_k)$ is a genuine function on the $k^{\mathrm{th}}$ component of 
$\MM^{4N}$.
 In that case $\dQ_k\widetilde{\Phi}(\varpi_1,...,\varpi_k,...,\varpi_N) = \dQ_k\widetilde{\Phi}^{(k)}(\varpi_k)$,
and the $N$ equations \refeq{eq:GEOhamjacLAWtestPARTICLES} become individual equations for the $\widetilde{\Phi}^{(k)}$,
and no compatibility issue arises.
 By differentiating \refeq{eq:GEOhamjacLAWtestPARTICLES} with respect to (proper) time and
using \refeq{eq:GEOhamjacPDEtestPARTICLES} and \refeq{eq:FdA}, we recover Newton's law of
motion for the `actual test particles' which are accelerated by the Lorentz force of the 
given electromagnetic fields.
 
 Notice that the test particles Hamilton--Jacobi equation \refeq{eq:GEOhamjacPDEtestPARTICLES} 
in given fields does not depend on the motion of the actual test point charges.
 Equation \refeq{eq:GEOhamjacPDEtestPARTICLES} is solved autonomously with initial 
data for $\widetilde\Phi$ compatible with the initial 
velocities of the $N$ actual test particles at their starting positions.
 Subsequently the guiding equation \refeq{eq:GEOhamjacLAWtestPARTICLES} is solved to obtain
the actual motions.

%%%%%%%%%%%%%%%%%%%%%%%%%%%%%%%%%%%%%%%%%%%%%%%%%%%%%%%%%%%%%%%%%%%%
	\subsubsection{Upgraded test particles Hamilton--Jacobi theory}
%%%%%%%%%%%%%%%%%%%%%%%%%%%%%%%%%%%%%%%%%%%%%%%%%%%%%%%%%%%%%%%%%%%%

 Less well-known is the fact that, if the actual electromagnetic fields are not a-priori given 
but nevertheless remain nice  at the locations of the point 
charges, then the test particles Hamilton--Jacobi formalism can be adapted to the computation 
of the motion of the point charges alongside the unknown fields which they generate. 
 We stress that this excludes fields satisfying the Maxwell--Lorentz and Maxwell--Born--Infeld 
field equations with point charge sources, but there are plenty of other constitutive relations 
between $\FQ$ and $\MQ$ which, when used in place of the Born--Infeld aether laws 
\refeq{eq:GEOaetherLAWborninfeld}, ensure that $\FQ$ is at least Lipschitz continuous at the 
locations of the point charges. 
 For instance, Infeld 
       \cite{InfeldA, InfeldB}
has proposed nonlinear aether laws that would seem to qualify in this respect (however, the
(say) Maxwell--Infeld field equations produce birefringence and other unwanted  phenomena 
which seem to disqualify them in these respects). 
 The main difference, as compared to the previous case of test particle motion in given fields, is that 
now the equations \refeq{eq:GEOhamjacLAWtestPARTICLES} and \refeq{eq:GEOhamjacPDEtestPARTICLES} are 
to be solved simultaneously, i.e. now the Hamilton--Jacobi partial differential equation 
\refeq{eq:GEOhamjacPDEtestPARTICLES} does depend on the actual motion.
 At the end of the day, almost all streamlines of the velocity field in generic configuration space 
represent motions of test particles in the actual field; however, precisely one such streamline 
coincides with the path of the actual configuration that provides the sources for the actual field, 
so that this particular `test particles' motion is not a test particles motion at all.
 To distinguish this situation from the one with given fields it may be appropriate, 
perhaps, to speak now of an `upgraded test particles Hamilton--Jacobi theory.' 

 We remark that as in the test particles Hamilton--Jacobi theory, also here we may separate 
variables through $\widetilde{\Phi}(\varpi_1,...,\varpi_N) = \sum_{k=1}^N\widetilde\Phi^{(k)}(\varpi_k)$. 
 Again the $N$ equations \refeq{eq:GEOhamjacLAWtestPARTICLES} become individual equations for 
the $\widetilde{\Phi}^{(k)}$, but now they are still coupled through the unknown 
electromagnetic potential which the $N$ actual motions generate. 
 That this is general enough to obtain all the actual motions from upgraded
test particles Hamilton--Jacobi theory follows again from the simple fact
that differentiating \refeq{eq:GEOhamjacLAWtestPARTICLES} with respect to proper time, and
using \refeq{eq:GEOhamjacPDEtestPARTICLES} and \refeq{eq:FdA}, yields Newton's law for the 
actual motions of the particles, but now accelerated by the Lorentz force in their self-generated
electromagnetic fields, to be computed alongside the actual motions.
 
%%%%%%%%%%%%%%%%%%%%%%%%%%%%%%%%%%%%%%%%%%%%%%%%%%%%%%%%%%%%%%%%%%%%
	\subsubsection{Limitations of the upgraded test particles Hamilton--Jacobi theory}
%%%%%%%%%%%%%%%%%%%%%%%%%%%%%%%%%%%%%%%%%%%%%%%%%%%%%%%%%%%%%%%%%%%%

 Unfortunately, the upgraded test particles Hamilton--Jacobi formalism fails to give 
a well-defined law of motion when coupled to the electromagnetic potentials for the total 
electromagnetic fields that satisfy the Maxwell--Born--Infeld field equations with point sources. 
 The reason is simply that at the end of the day only one of the test particles' trajectories in 
configuration space can coincide with the actual particles' trajectory, while the remaining 
trajectories --- and that means almost all --- indeed represent just  motions of test particles
which have no influence on the field that determines their motion. 
  But this means that almost all these test particles configurations do not sample the 
actual $\AQ$ (it is understood that the `actual' $\AQ$ requires the stipulation of a gauge)
the way the configuration of the $N$ actual particles does, the exception occurring when 
the test particles' world-configuration coincides with the actual one.
 As a consequence of this, the actual $\AQ$ in \refeq{eq:GEOhamjacPDEtestPARTICLES} which acts
as a field on the $k$-th copy of $\MM^4$ is exactly as non-differentiable at the actual 
particles' world-points as the actual $\AQ$ is as a field on Minkowski spacetime itself. 
 Now $\widetilde\Phi$ in \refeq{eq:GEOhamjacPDEtestPARTICLES} inherits this non-differentiability 
from the actual $\AQ$, and that simply means that \refeq{eq:GEOhamjacLAWtestPARTICLES} ceases to be 
well-defined precisely at the actual positions where it is needed.

 Fortunately, the above analysis of why the upgraded test particles Hamilton--Jacobi formalism fails 
with the electromagnetic potentials of the total Maxwell--Born--Infeld fields with 
point sources hints clearly at what is needed.
 Namely, instead of using a formalism in which almost all the generic configurations represent 
virtual test particles configurations which are not the sources for the field they sample, 
and which is the actual field, we should be working with a formalism in which each generic 
configuration represents the point sources for the very field that it samples, and which 
therefore will generally not be the actual one.
 We will call this the \emph{proper many-body} Hamilton--Jacobi theory.
 Curiously, the pertinent Hamilton--Jacobi literature is silent about such a 
formalism,\footnote{Barut
                           \cite{barutBOOKa} % p.58(middle); p.73(above sect.4)
		   seems quite pessimistic about the prospects of achieving a relativistic 
		   proper many-body Hamilton--Jacobi theory and proposes as alternative 
		   the Fokker--Schwarzschild--Tetrode-/-Feynman--Wheeler electrodynamics
		           \cite{wheelerfeynman}; 
		   ch. VI in
                           \cite{barutBOOKa}.
		   That theory does not pose an initial value problem and, hence, departs
		   ultra-radically from conventional modes of thought about physics; 
		   for rigorous results on two-particle scattering, see
                          \cite{bauer}.}
even though --- with hindsight --- it could have been developed long ago!

%%%%%%%%%%%%%%%%%%%%%%%%%%%%%%%%%%%%%%%%%%%%%%%%%%%%%%%%%%%%%%%%%%%%
	\subsubsection{Proper many-particles Hamilton--Jacobi theory}
%%%%%%%%%%%%%%%%%%%%%%%%%%%%%%%%%%%%%%%%%%%%%%%%%%%%%%%%%%%%%%%%%%%%

 With the help of the geometrical language of forms, a relativistically covariant 
proper many-body Hamilton--Jacobi theory of motion can be set up as follows.
 Once again, that a Hamilton--Jacobi structure is behind this formulation may not be
any more obvious than for the (plain or upgraded) test particles versions,
but it will be unveiled in our sections on the evolutionary formalism
in space and time decomposition.

 According to what we wrote in our last subsection, the proper many-particles Hamilton--Jacobi 
formalism does not work with the a-priori unknown actual fields on $\MM^4$ but with an 
(also a-priori unknown) family of fields on $\MM^4$ which is indexed by $N$ world-points;
hence instead of the actual  $\JQ$, $\MQ$ (whence, $\FQ$) and $\AQ$ on $\MM^4$ 
we consider fields $^\sharp\JQ$, $^\sharp\MQ$, $^\sharp\FQ$ and $^\sharp\AQ$ 
with arguments $\in \MM^4\times\MM^{4N}_{\neq}$, where $\MM^{4N}_{\neq}$ is 
the configuration space of $N$ ordered world-points with co-incidence points removed.
 These $^\sharp$fields are of course to be associated with the actual fields in
a natural way, and the obvious requirement which seems to suggest itself is to 
demand that, when the generic configuration point $(\varpi_1,...,\varpi_N)\in\MM^{4N}_{\neq}$ 
in the argument of a $^\sharp$field is replaced by an actual configuration point
$(\eta_1(\tau),...,\eta_N(\tau))\in\MM^{4N}_{\neq}$, then this so ``conditioned'' 
$^\sharp$field becomes the corresponding actual field on $\MM^4$ in the remaining 
world-point variable $\varpi\in\MM^4$, i.e. 
$^\sharp\AQ(\varpi,\eta_1(\tau),...,\eta_N(\tau))
 = \AQ(\varpi)$, etc. 
 However, to ask for this much is actually asking just a little too much, for 
by re-parametrization of the point histories we can arrange that 
$(\eta_1(\tau),...,\eta_N(\tau))$ is any configuration point that can be 
formed by picking one arbitrary point from each of the $N$ actual point histories
without changing $\AQ(\varpi)$ (etc.). 
 This in turn would imply that the individual directional derivatives of 
$^\sharp\AQ(\varpi,\varpi_1,...,\varpi_N)$ along each of the $N$ actual 
point histories would have to vanish.
 But all we really need to ask for is that for any $N+1$ world-points 
$(\varpi,\varpi_1,...,\varpi_N)$ picked from any leaf of a foliation we 
have $^\sharp\AQ(\varpi,\eta_1(\tau),...,\eta_N(\tau)) = \AQ(\varpi)$, etc. 
 This also fixes in a canonical way the field equations for the $^\sharp$fields
on $\MM^4\times\MM^{4N}_{\neq}$ synchronized w.r.t. the foliation.
 Since we need a foliation for this, we postpone the presentation of these
$^\sharp$field equations until we work out the formalism in standard foliation. 

 Next we form $N$ fields $\widetilde\AQ_k$ on synchronized $\MM^{4N}_{\neq}$,
this time by conditioning $^\sharp\AQ$ by successively replacing the world-point
$\varpi$ in its first argument with the $N$ components $\varpi_k$ of the generic 
configuration point; more precisely, we define
$\widetilde\AQ_k(\varpi_1,...,\varpi_N) \equiv {}^\sharp\AQ(\varpi_k,\varpi_1,...,\varpi_N)$
for each $k=1,...,N$.
 Since we may think of ${}^\sharp\AQ(\varpi,\varpi_1,...,\varpi_N)$ essentially as 
the electromagnetic potential field for $N$ point sources at $(\varpi_1,...,\varpi_N)$,
the field $\widetilde\AQ_k(\varpi_1,...,\varpi_N)$ samples 
${}^\sharp\AQ(\varpi,\varpi_1,...,\varpi_N)$ exactly at its $k$-th source point. 
 
 We now formulate a law of motion for the actual configuration point in 
synchronized $\MM^{4N}_{\neq}$ 
of the $N$ point charges in $\MM^4$ which is similar in appearance to 
\refeq{eq:GEOhamjacLAWtestPARTICLES}.
 Namely, we replace $\AQ(\varpi_k)$ in \refeq{eq:GEOhamjacLAWtestPARTICLES} by 
$\widetilde\AQ_k(\varpi_1,...,\varpi_N)$ and obtain the system of first order 
guiding laws 
\beq
\uQ_k(\tau)
=
      {\dQ}_k \widetilde\Phi(\varpi_1,...,\varpi_N)
-\pmk {\alpha}\widetilde\AQ_k(\varpi_1,...,\varpi_N) \Big|_{\{\varpi_n = \eta_n(\tau)\}}\Big.
\,
\label{eq:GEOhamjacLAW}
\eeq
for the actual world-points $\varpi_k = \eta_k(\tau)$ on the point-histories 
$\Eta_k$, with $k=1,...,N$.
 Since \refeq{eq:GEOhamjacLAW} must produce a future-oriented  co-vector $\uQ_k$ 
satisfying $\Hodge\left(\uQ_k\wedge\Hodge\uQ_k\right) = 1$, the phase function
$\widetilde\Phi$ on synchronized $\MM^{4N}_{\neq}$ must satisfy the $N$ equations
\beq
\Hodge
\Bigl( \Bigl( 
{\dQ}_k \widetilde{\Phi}
-\pmk {\alpha}\widetilde\AQ_k
\Bigr)\wedge
\Hodge \Bigl( 
{\dQ}_k \widetilde{\Phi}
-\pmk {\alpha}\widetilde\AQ_k
\Bigr)\Bigr)
 = 1
\,,
\label{eq:GEOhamjacPDE}
\eeq
of each of which the future-oriented root is to be chosen.
 These $N$ equations are to be understood  w.r.t. the same foliation as the 
$^\sharp$fields, for otherwise the $N$ equations would in general over-determine 
$\widetilde\Phi$ when $N>1$.
 While \refeq{eq:GEOhamjacLAW} and \refeq{eq:GEOhamjacPDE} are
manifestly Poincar\'e-invariant, the foliation is not but takes a status akin 
to the world lines. 
 We add that \refeq{eq:GEOhamjacLAW} and \refeq{eq:GEOhamjacPDE} are also
manifestly gauge (Weyl-)invariant. 
 Namely, since $^\sharp\AQ$ inherits its gauge transformation from 
\refeq{eq:GEOgaugetrA}, a gauge transformation becomes
\bea
\widetilde\AQ_k(\varpi_1,...,\varpi_N) 
\!\!\!
&\to&
\!\!\!
\widetilde\AQ_k(\varpi_1,...,\varpi_N) +(\dQ\Upsilon)(\varpi_k) 
\label{eq:GEOgaugetrAk}
\\
	\widetilde\Phi(\varpi_1,...,\varpi_N) 
\!\!\!
&\to&
\!\!\!
	\widetilde\Phi(\varpi_1,...,\varpi_N) +  \sum_k \pmk \alpha\Upsilon(\varpi_k) 
\,,
\label{eq:GEOgaugetrPHI}
\eea
and this obviously leaves $\FQ$, $^\sharp\FQ$, and the $\uQ_k$ invariant. 

 We remark that while in general $\widetilde\AQ_k(\varpi_1,...,\varpi_N)$
is a genuine function of all $\varpi_\ell$, $\ell=1,...,N$, in a wide variety of 
``decoherent'' situations we may assume that $\widetilde\AQ_k$ depends only 
on $\varpi_k$, and in that case we can separate variables 
and demand that 
$\widetilde{\Phi}(\varpi_1,...,\varpi_N) = \sum_{k=1}^N\widetilde\Phi^{(k)}(\varpi_k)$, 
where $\widetilde\Phi^{(k)}(\varpi_k)$ is a genuine function on the $k^{\mathrm{th}}$ 
component of synchronized $\MM^{4N}_{\neq}$, i.e. $\widetilde\Phi^{(k)}(\varpi_k)$ is not indexed 
by the $N-1$ other component world-points. 
 In that case we have 
$\dQ_k\widetilde{\Phi}(\varpi_1,...,\varpi_k,...,\varpi_N) = \dQ_k\widetilde{\Phi}^{(k)}(\varpi_k)$,
and the $N$ equations \refeq{eq:GEOhamjacLAW} become individual equations for the $\widetilde{\Phi}^{(k)}$,
though still coupled through the family of electromagnetic potentials.
 This decoherent situation is closely reminiscent of the upgraded test particles formulation,
and yet different.

 To summarize, proper many-body Hamilton--Jacobi theory is formulated with the help of 
a field $^\sharp\AQ$ on synchronized $\MM^4\times\MM^{4N}_{\neq}$.
 The equation for $^\sharp\AQ$ is induced by the requirement that insertion of the actual 
configuration in place of the generic one produces the actual $\AQ$ for the actual 
Maxwell--Born--Infeld fields, which concept is well-defined w.r.t. some foliation.
 The field $^\sharp\AQ$ induces $N$ fields $\widetilde\AQ_k$ on synchronized $\MM^{4N}_{\neq}$,
with the help of which an $N$ Minkowski-velocities field on synchronized $\MM^{4N}_{\neq}$ can be
defined, the $k$-th component of which is given by the r.h.s. of \refeq{eq:GEOhamjacLAW},
at generic world configuration points. 
 The requirement that the components are Minkowski-velocity co-vectors gives the $N$
equations \refeq{eq:GEOhamjacPDE}, defined w.r.t. the same foliation with respect to 
which the equation for $^\sharp\AQ$ is formulated.
 The actual dynamics of the $N$ point charges is formulated as the motion of a single 
point in synchronized $\MM^{4N}_{\neq}$ which moves according to a system of covariant guiding laws, 
the $k$-th component of which is  \refeq{eq:GEOhamjacLAW}. 

 We end by emphasizing one more time that the fields $\widetilde\Phi$ and 
$^\sharp\AQ$, understood with respect to a foliation, are elevated from the status of 
mathematical auxiliary fields (which is what they would be if the Newtonian law of 
motion would be well-defined) to fields that have a certain physical significance 
in their own right. 
 More precisely, it is the gauge-invariant linear combinations of $\dQ_k\widetilde\Phi$ 
and $\widetilde\AQ_k$ on the r.h.s. of \refeq{eq:GEOhamjacLAW} which have physical significance. 
 Even if $\widetilde\Phi$ separates into a sum of $N$ functions $\widetilde\Phi^{(k)}$ of
the $N$ individual generic world-points, the $\widetilde\Phi^{(k)}$ have significance only 
at the world-points of the $k^{\mathrm{th}}$ particle.
 The  domain for $\widetilde\Phi^{(k)}$ is the set of 
all world-points that are a-priori available to the 
$k^{\mathrm{th}}$ particle, which  is the $k^{\mathrm{th}}$ copy  $\MM^4_k$ of $\MM^4$.

%%%%%%%%%%%%%%%%%%%%%%%%%%%%%%%%%%%%%%%%%%%%%%%%%%%%%%%%%%%%%%%%%%%%
%%%%%%%%%%%%%%%%%%%%%%%%%%%%%%%%%%%%%%%%%%%%%%%%%%%%%%%%%%%%%%%%%%%%
%%%%%%%%%%%%%%%%%%%%%%%%%%%%%%%%%%%%%%%%%%%%%%%%%%%%%%%%%%%%%%%%%%%%
  \subsection{Extensions to non-genuinely electromagnetic systems (I)}
	\label{sect:NongenuineEdynI}
%%%%%%%%%%%%%%%%%%%%%%%%%%%%%%%%%%%%%%%%%%%%%%%%%%%%%%%%%%%%%%%%%%%%
%%%%%%%%%%%%%%%%%%%%%%%%%%%%%%%%%%%%%%%%%%%%%%%%%%%%%%%%%%%%%%%%%%%%
%%%%%%%%%%%%%%%%%%%%%%%%%%%%%%%%%%%%%%%%%%%%%%%%%%%%%%%%%%%%%%%%%%%%

    While in its genuine setting the theory is only about positive and 
negative electrons and the electromagnetic fields, 
so-called \emph{external charges} of physical importance are readily accommodated,
as are electromagnetically non-interacting particles.
  Since the theory aims at providing the classical stepping stone on our 
way to a more fundamental physical theory, only extensions necessary to mimic
such microscopically relevant situations will be 
considered.\footnote{A derivation of such extensions should be possible in terms of the
                     more fundamental theory.}
   Macroscopic reformulations of the theory with continuum 
charge and current densities, while also possible, are \emph{not} 
necessary and should be {derivable} from the microscopic theory as expounded 
here, to the extent that it is correct, with the help of some law of large numbers.

  The most important special case of this generalization is a typical system
of many negative electrons and many positive nuclei.
  Therefore, consider a system of $N$ positive, neutral, or negative point charges, 
carrying (integer) multiples of the unit charge, and possibly more massive than
the electrons.
   Multiple-units charges are accommodated by simply letting 
$z_k\in\ZZ$ now denote the charge of the $k^{\mathrm{th}}$ particle (note 
that the particle may be neutral).
    If it is desired, the finite size of the nuclei can also be modeled 
with form factors  that have to be put into the theory by hand, but 
for the sake of simplicity we will not do this here (see 
   \cite{AppKieAOP, AppKieLMP} 
for how to treat extended, spinning charge distribution covariantly in Lorentz electrodynamics.)
   Heavier masses than the electron's can simply be accommodated by 
numerical parameters $\kappa_k$, expressing the ratio of the electron's 
to the $k^{\mathrm{th}}$ particle's rest mass, which enter the covariant 
guiding law for the $k^{\mathrm{th}}$ history as multipliers of the coupling 
constant,
\beq
\uQ_k
=
{\dQ}_k\widetilde\Phi -  \kappa_k z_k {\alpha}\widetilde\AQ_k
\,.
\label{eq:GEOhamjacLAWz}
\eeq
 Note that $\kappa_k\leq 1$ if $|z_k| = 1$, and $0< \kappa_k \ll 1$ if $|z_k|\neq 1$,
thereby reducing the influence of the electromagnetic  connection on the point history.
%\newpage
%%%%%%%%%%%%%%%%%%%%%%%%%%%%%%%%%%%%%%%%%%%%%%%%%%%%%%%%%%%%%%%%%%%%
%%%%%%%%%%%%%%%%%%%%%%%%%%%%%%%%%%%%%%%%%%%%%%%%%%%%%%%%%%%%%%%%%%%%
%%%%%%%%%%%%%%%%%%%%%%%%%%%%%%%%%%%%%%%%%%%%%%%%%%%%%%%%%%%%%%%%%%%%
%%%%%%%%%%%%%%%%%%%%%%%%%%%%%%%%%%%%%%%%%%%%%%%%%%%%%%%%%%%%%%%%%%%%
        \section{The evolutionary formalism}
%%%%%%%%%%%%%%%%%%%%%%%%%%%%%%%%%%%%%%%%%%%%%%%%%%%%%%%%%%%%%%%%%%%%
%%%%%%%%%%%%%%%%%%%%%%%%%%%%%%%%%%%%%%%%%%%%%%%%%%%%%%%%%%%%%%%%%%%%
%%%%%%%%%%%%%%%%%%%%%%%%%%%%%%%%%%%%%%%%%%%%%%%%%%%%%%%%%%%%%%%%%%%%
%%%%%%%%%%%%%%%%%%%%%%%%%%%%%%%%%%%%%%%%%%%%%%%%%%%%%%%%%%%%%%%%%%%%
%
%%%%%%%%%%%%%%%%%%%%%%%%%%%%%%%%%%%%%%%%%%%%%%%%%%%%%%%%%%%%%%%%%%%%
%%%%%%%%%%%%%%%%%%%%%%%%%%%%%%%%%%%%%%%%%%%%%%%%%%%%%%%%%%%%%%%%%%%%
%%%%%%%%%%%%%%%%%%%%%%%%%%%%%%%%%%%%%%%%%%%%%%%%%%%%%%%%%%%%%%%%%%%%
      {\subsection{Space and time; electricity and magnetism}}
%%%%%%%%%%%%%%%%%%%%%%%%%%%%%%%%%%%%%%%%%%%%%%%%%%%%%%%%%%%%%%%%%%%%
%%%%%%%%%%%%%%%%%%%%%%%%%%%%%%%%%%%%%%%%%%%%%%%%%%%%%%%%%%%%%%%%%%%%
%%%%%%%%%%%%%%%%%%%%%%%%%%%%%%%%%%%%%%%%%%%%%%%%%%%%%%%%%%%%%%%%%%%%
Space-like foliations of spacetime have to satisfy 
             \cite{hawkingellisBOOK, christodoulouklainermanBOOK}
the first and second variation equations for the evolution of the 
first and second fundamental forms, which is constrained by the 
Gauss and Gauss--Codacci 
equations.\footnote{Constructing suitable foliations will be of 
               prime importance when formulating the Cauchy problem
	       for the general-relativistic extension of our special-relativistic 
	       electromagnetic theory, which we already anticipated can be done.
	       As to the importance of selecting good foliations for the Cauchy 
	       problem of the general-relativistic vacuum Einstein equations, see
		     \cite{christodoulouklainermanBOOK}.} 
  However, in the following the theory will be formulated in the familiar space and time 
splitting of the standard foliation, to which we turn
next.
%%%%%%%%%%%%%%%%%%%%%%%%%%%%%%%%%%%%%%%%%%%%%%%%%%%%%%%%%%%%%%%%%%%%
%%%%%%%%%%%%%%%%%%%%%%%%%%%%%%%%%%%%%%%%%%%%%%%%%%%%%%%%%%%%%%%%%%%%
	{\subsubsection{The standard foliation}}
%%%%%%%%%%%%%%%%%%%%%%%%%%%%%%%%%%%%%%%%%%%%%%%%%%%%%%%%%%%%%%%%%%%%
%%%%%%%%%%%%%%%%%%%%%%%%%%%%%%%%%%%%%%%%%%%%%%%%%%%%%%%%%%%%%%%%%%%%
	The simplest foliation of $\MM^4$, called 
the {standard foliation}, consists of affine flat space slices.
	It is generated by the function $T(\varpi)\equiv-\eQ_0\cdot\wQ$, 
which has a constant time-like four-gradient, $\dQ T(\varpi) = -\gQ(\eQ_0)$.
	Here, world-vectors are in ${\mathbb T}_{\varpi_0}({\MM}^{4})$ 
for $\varpi_0 \in \Sigma_0\subset \MM^4$, with $\eQ_0$ the constant
time-like unit normal to $\Sigma_0$.
	Since $\Sigma_0$ is then identifiable with its own tangent
space, $\Sigma_0\cong\RR^3$, we can globally coordinatize $\Sigma_0$
by the restriction to $\Sigma_0$ of the global coordinatization of $\MM^4$ 
w.r.t. the Lorentz frame $\{\eQ_0,\eQ_1,\eQ_2,\eQ_3\}$ at $\varpi_0$.
	We identify $\varpi\cong(s^0,s^1,s^2,s^3)$.
	In these coordinates, we have $g = {\mathrm{diag}}\,(1,1,1)$ 
and the spacetime metric is given by the world-line element 
\begin{equation}
	(\dd s){}^2 
= 
	- (\dd s^0){}^2 
	+ \sum_{1 \leq m\leq 3} (\dd s^m){}^2.
\end{equation}
   If we want to emphasize 
the `time' and `space' coordinatization of $\varpi$ we write 
$\varpi\cong(t,\sV)$, where $t\in\RR$ denotes an `instant of time' and 
$\sV\in \Sigma_t$ a `space point,'  the bold font used for $\sV$ indicating
that, in the standard foliation, $\Sigma_t \equiv \RR^3$ is a vector space. 

%\smallskip
%%%%%%%%%%%%%%%%%%%%%%%%%%%%%%%%%%%%%%%%%%%%%%%%%%%%%%%%%%%%%%%%
%%%%%%%%%%%%%%%%%%%%%%%%%%%%%%%%%%%%%%%%%%%%%%%%%%%%%%%%%%%%%%%%
%%%%%%%%%%%%%%%%%%%%%%%%%%%%%%%%%%%%%%%%%%%%%%%%%%%%%%%%%%%%%%%%
        \subsubsection{Electric and magnetic decomposition of electromagnetism} 
        \label{sec:EandM}
%%%%%%%%%%%%%%%%%%%%%%%%%%%%%%%%%%%%%%%%%%%%%%%%%%%%%%%%%%%%%%%%
%%%%%%%%%%%%%%%%%%%%%%%%%%%%%%%%%%%%%%%%%%%%%%%%%%%%%%%%%%%%%%%%
%%%%%%%%%%%%%%%%%%%%%%%%%%%%%%%%%%%%%%%%%%%%%%%%%%%%%%%%%%%%%%%%

      In a local basis of one-forms dual to a
Lorentz frame $\{\eQ_0,\eQ_1,\eQ_2,\eQ_3\}$ at
$\varpi\in\MM^4$, the various electromagnetic $p$ forms read:
$\AQ = A_\mu\dQ{s}^\mu$ and
$\FQ =  F_{\mu\nu}\dQ{s}^\mu\otimes\dQ{s}^\nu$,
with $F_{\mu\nu} = - F_{\nu\mu}$.
        Written with a wedge product this becomes
$\FQ = \haelfte F_{\mu\nu}\dQ{s}^\mu\wedge\dQ{s}^\nu$.
                Similarly, $\MQ = M_{\mu\nu}\dQ{s}^\mu\otimes\dQ{s}^\nu$,
with $M_{\mu\nu} = - M_{\nu\mu}$; hence, 
$\MQ = \haelfte M_{\mu\nu}\dQ{s}^\mu\wedge\dQ{s}^\nu$.
   Finally, 
$\JQ = \frac{1}{3!} j^\kappa \veps_{\kappa\lambda\mu\nu}\dQ{s^\lambda}\wedge\dQ{s^\mu} \wedge\dQ{s^\nu}$,
where $\veps_{\kappa\lambda\mu\nu}$ is the totally antisymmetric
symbol with $\veps_{0123}=1$.
        The Hodge dual of $\JQ$ is the one-form
$\Hodge\JQ = j_\kappa\dQ{s^\kappa}$.

  The electric and magnetic decomposition of the electromagnetic connection $\AQ$
w.r.t. the space and time decomposition gives the electric potential 
 $A = A^0$ and the magnetic vector-potential ${\SPvec{A}} = (A^1,A^2,A^3)$.
	The decomposition of the Faraday tensor $\FQ$ 
gives us $\FQ = - E_k\dQ{t}\wedge\dQ{s^k} + B_1 \dQ{s^2}\wedge\dQ{s^3}
+ B_2 \dQ{s^3}\wedge\dQ{s^1} + B_3 \dQ{s^1}\wedge\dQ{s^2}$, where
${\SPvec{E}} = (E^1,E^2,E^3)$
and 
${\SPvec{B}} = (B^1,B^2,B^3)$
are the electric, respectively magnetic induction, fields; cf. 
          \cite{WheeleretalBOOK}.
        Similarly, the decomposition of the Maxwell tensor $\MQ$ gives us
$\MQ = H_k\dQ{t}\wedge\dQ{s^k} + D_1 \dQ{s^2}\wedge\dQ{s^3}
+ D_2 \dQ{s^3}\wedge\dQ{s^1} + D_3 \dQ{s^1}\wedge\dQ{s^2}$, where
${\SPvec{D}} = (D^1,D^2,D^3)$
and 
${\SPvec{H}}
 = (H^1,H^2,H^3)$
are the electric displacement, respectively the magnetic, vector fields.
	The decomposition of the electromagnetic current density three-form
gives the charge density $j = j^0$
and the current vector-density $\jV = (j^1,j^2,j^3)$.
%\smallskip
%%%%%%%%%%%%%%%%%%%%%%%%%%%%%%%%%%%%%%%%%%%%%%%%%%%%%%%%%%%%%%%%%%%%
%%%%%%%%%%%%%%%%%%%%%%%%%%%%%%%%%%%%%%%%%%%%%%%%%%%%%%%%%%%%%%%%%%%%
%%%%%%%%%%%%%%%%%%%%%%%%%%%%%%%%%%%%%%%%%%%%%%%%%%%%%%%%%%%%%%%%%%%%
  \subsection{The space and time decomposition of the actual field equations}
%%%%%%%%%%%%%%%%%%%%%%%%%%%%%%%%%%%%%%%%%%%%%%%%%%%%%%%%%%%%%%%%%%%%
%%%%%%%%%%%%%%%%%%%%%%%%%%%%%%%%%%%%%%%%%%%%%%%%%%%%%%%%%%%%%%%%%%%%
%%%%%%%%%%%%%%%%%%%%%%%%%%%%%%%%%%%%%%%%%%%%%%%%%%%%%%%%%%%%%%%%%%%%
	We now list the geometrical equations in their space and time
decomposition, which brings out the evolutionary and the constraint 
aspects of the theory.
  In principle, the evolution and constraint equations for 
the space-like foliations of $\MM^4$ will  have to be listed 
among the fundamental equations for the theory.
  However, since for our Minkowski spacetime we will
not require nonstandard foliations, we relegate the
discussion of the foliation equations to a later paper.
  Having stressed this, we now move on with our development of the
evolutionary formalism in the standard foliation.
  In the following, partial derivative w.r.t. time is denoted by the symbol $\partial$, 
while $\nabla$ denotes the usual space gradient operator w.r.t. the standard foliation.

%%%%%%%%%%%%%%%%%%%%%%%%%%%%%%%%%%%%%%%%%%%%%%%%%%%%%%%%%%%%%%%%%%%%
%%%%%%%%%%%%%%%%%%%%%%%%%%%%%%%%%%%%%%%%%%%%%%%%%%%%%%%%%%%%%%%%%%%%
  \subsubsection{The continuity equation for systems of electric point charges}
%%%%%%%%%%%%%%%%%%%%%%%%%%%%%%%%%%%%%%%%%%%%%%%%%%%%%%%%%%%%%%%%%%%%
%%%%%%%%%%%%%%%%%%%%%%%%%%%%%%%%%%%%%%%%%%%%%%%%%%%%%%%%%%%%%%%%%%%%

  It is readily shown (e.g.,
		\cite{weinbergBOOKart})
that the space and time decomposition of \refeq{eq:GEOptchargecurrent}
into the microscopic charge `density' $j(t,{\SPvec{s}})$ 
and current `vector-density' ${\SPvec{j}}(t,{\SPvec{s}})$ for a system 
of $N^+\! \geq \! 0$ positive and $N^-\! \geq \! 0$ negative unit point charges reads
\bea
        j (t,{\SPvec{s}})
\!\!\!&=&\!\!\!
\sum_{k\in\cN} \pmk\delta_{{\SPvec{r}}_{k}(t)}({\SPvec{s}})
\label{eq:FOLIpointSOURCESrho}
\\
&&
\nonumber
\\
        {\SPvec{j}}(t,{\SPvec{s}})
\!\!\!&=&\!\!\!
\sum_{k\in\cN}
\pmk\delta_{{\SPvec{r}}_{k}(t)}({\SPvec{s}})\bulldif{\SPvec{r}}{_k}(t)
\, ,
\label{eq:FOLIpointSOURCESjjj}
\eea
where $\delta_{{\SPvec{r}}_{k}(t)}({\SPvec{s}})$ denotes the 
Dirac probability distribution of ${\SPvec{s}}\in\RR^3$ concentrated 
at the position ${\SPvec{s}} = {\SPvec{r}}_k(t)\in {\RR}^3$,
and moving with the linear velocity 
$\frac{\dd{\SPvec{s}}}{\dd{t}} = \bulldif{\SPvec{r}}{_k}(t)$, 
at foliation time $t$.
  These measure-valued expressions for $j$ and ${\SPvec{j}}$ satisfy, in a weak sense, 
the familiar continuity equation 
\beq
{\partial}j + {\nabla}\cdot {\SPvec{j}}
= 
0,
\label{eq:FOLIchargecontinuityEQ}
\eeq
which expresses the law of charge conservation \refeq{eq:GEOconservechargeLAW} in
space and time decomposition.
  
\smallskip
%%%%%%%%%%%%%%%%%%%%%%%%%%%%%%%%%%%%%%%%%%%%%%%%%%%%%%%%%%%%%%%%%%%%
%%%%%%%%%%%%%%%%%%%%%%%%%%%%%%%%%%%%%%%%%%%%%%%%%%%%%%%%%%%%%%%%%%%%
	\subsubsection{Maxwell's electromagnetic field equations}
%%%%%%%%%%%%%%%%%%%%%%%%%%%%%%%%%%%%%%%%%%%%%%%%%%%%%%%%%%%%%%%%%%%%
%%%%%%%%%%%%%%%%%%%%%%%%%%%%%%%%%%%%%%%%%%%%%%%%%%%%%%%%%%%%%%%%%%%%

 We now address the spacetime between the particle world-lines.
	The standard space and time decompositions of the Faraday--Maxwell law
and the Amp\'ere--Coulomb--Maxwell law yield the general Maxwell equations
of the classical electromagnetic fields, familiar from Maxwell's theory of
electromagnetic fields in a medium with nontrivial dielectric and magnetic 
permeability properties and listed in many textbooks, e.g. in
		\cite{panofskyphillipsBOOK, landaulifshitzBOOK, jacksonBOOK}.
 Here of course the `medium' is the nonlinear aether itself.
 The dynamical variables are the field vectors ${\SPvec{B}}$ and ${\SPvec{D}}$ 
at each space point ${\SPvec{s}}$, the locations of the point charges excepted.

 The evolution of ${\SPvec{B}}$ is governed by
\beq
	{\partial}{\SPvec{B}}
=
        - {\nabla}\times {\SPvec{E}} 
\, ,
\label{eq:FOLIeB} 
\eeq
and constrained by
\beq
        {\nabla}\cdot  {\SPvec{B}}
=
        0
\, ,
\label{eq:FOLIcB}
\eeq
while the evolution of ${\SPvec{D}}$ is governed by
\beq
        {\partial}{\SPvec{D}}
=
        {\nabla}\times{\SPvec{H}}
-
	4\pi {\SPvec{j}}
\,,
\label{eq:FOLIeD}
\eeq
and constrained by
\beq
        {\nabla}\cdot {\SPvec{D}}
=
        4 \pi j 
\,.
\label{eq:FOLIcD}
\eeq

%%%%%%%%%%%%%%%%%%%%%%%%%%%%%%%%%%%%%%%%%%%%%%%%%%%%%%%%%%%%%%%%%%%%
%%%%%%%%%%%%%%%%%%%%%%%%%%%%%%%%%%%%%%%%%%%%%%%%%%%%%%%%%%%%%%%%%%%%
	\subsubsection{Born and Infeld's electromagnetic aether laws}
%%%%%%%%%%%%%%%%%%%%%%%%%%%%%%%%%%%%%%%%%%%%%%%%%%%%%%%%%%%%%%%%%%%%
%%%%%%%%%%%%%%%%%%%%%%%%%%%%%%%%%%%%%%%%%%%%%%%%%%%%%%%%%%%%%%%%%%%%

	The electric and magnetic decomposition of 
the aether laws of Born and Infeld
		\cite{BornInfeldB}
defines the field strengths 
${\SPvec{E}}$  and ${\SPvec{H}}$
locally in terms of the natural evolutionary variables
electric displacement 
${\SPvec{D}}$
and the magnetic induction
${\SPvec{B}}$, 
\bea
&&
{\SPvec{E}} 
= 
\frac{
{\SPvec{D}}-\beta^4{\SPvec{B}}\times({\SPvec{B}}\times{\SPvec{D}})
}{
\sqrt{    1 
	+ \beta^4(|{\SPvec{B}}|^2 + |{\SPvec{D}}|^2) 
	+ \beta^8|{\SPvec{B}}\times {\SPvec{D}}|^2
      }}
\label{eq:FOLIeqEofBD}
\\
&&
{\SPvec{H}} 
 = 
\frac{
{\SPvec{B}}-\beta^4{\SPvec{D}}\times({\SPvec{D}}\times {\SPvec{B}})
     }{
\sqrt{	  1 
	+ \beta^4(|{\SPvec{B}}|^2 + |{\SPvec{D}}|^2) 
	+ \beta^8|{\SPvec{B}}\times {\SPvec{D}}|^2
      }}
\label{eq:FOLIeqHofBD}
\, ,
\eea
for $\beta\in (0,\infty)$. 
   The notion of weak-field limit is intrinsically well-defined for the Born--Infeld aether laws, 
in which limit they reduce to Maxwell's pure aether laws
\bea
&&
{\SPvec{E}} 
\sim
{\SPvec{D}}
\, \qquad (\mathrm{weak\ field\ limit})
\\
&&
{\SPvec{H}} 
 \sim
{\SPvec{B}}
\, \qquad (\mathrm{weak\ field\ limit})\, .
\eea

  We remark that in the formal limit $\beta\to 0$, the 
Born--Infeld aether laws yield precisely Maxwell's laws of the pure aether,
${\SPvec{E}} = {\SPvec{D}}$ and ${\SPvec{H}} = {\SPvec{B}}$. 
 On the other hand, the limit $\beta\to \infty$ of the Born--Infeld aether laws 
yields \emph{ultra}\footnote{This coinage was proposed in \cite{BiBiONE}.}
Born--Infeld laws
\bea
&&
{\SPvec{E}} 
= 
\frac{{\SPvec{B}}\times{\SPvec{D}}}
     {|{\SPvec{B}}\times {\SPvec{D}}|}
\times{\SPvec{B}}
\label{eq:FOLIeqEofBDultra}
\\
&&
{\SPvec{H}} 
 = 
\frac{{\SPvec{D}}\times{\SPvec{B}}}
     {|{\SPvec{D}}\times {\SPvec{B}}|}
\times{\SPvec{D}}
\label{eq:FOLIeqHofBDultra}
\, ,
\eea
unless  ${\SPvec{B}}\times {\SPvec{D}} = \SPvec{0}$, in which case we find
${\SPvec{E}} = \SPvec{0}$ and  ${\SPvec{H}} = \SPvec{0}$.
 Completing the source-free Maxwell field equations with either the Maxwell laws of the 
pure aether or with the ultra Born--Infeld laws of the aether result in field theories
which are invariant under the full conformal group of Minkowski space. 
 The source-free Maxwell field equations for the Maxwell laws of the pure aether
have of course been exhaustively studied, e.g. 
 		\cite{panofskyphillipsBOOK, landaulifshitzBOOK, jacksonBOOK}.
  The source-free Maxwell field equations for the ultra Born--Infeld laws of the aether 
are studied in 
   \cite{BiBiONE, BiBiTWO, Brenier}.

%%%%%%%%%%%%%%%%%%%%%%%%%%%%%%%%%%%%%%%%%%%%%%%%%%%%%%%%%%%%%%%%%%%%
%%%%%%%%%%%%%%%%%%%%%%%%%%%%%%%%%%%%%%%%%%%%%%%%%%%%%%%%%%%%%%%%%%%%
	\subsubsection{The laws for the electric and magnetic potentials}
%%%%%%%%%%%%%%%%%%%%%%%%%%%%%%%%%%%%%%%%%%%%%%%%%%%%%%%%%%%%%%%%%%%%
%%%%%%%%%%%%%%%%%%%%%%%%%%%%%%%%%%%%%%%%%%%%%%%%%%%%%%%%%%%%%%%%%%%%

  While $\FQ$ is defined as $\FQ=\dQ\AQ$, this equation at the same time 
gives the status of a primitive dynamical field variable to the magnetic 
vector potential $\SPvec{A}$, with the electric field strength ${\SPvec{E}}$ 
and the magnetic induction ${\SPvec{B}}$ acting as source terms in the evolution, 
respectively constraint equations for $\SPvec{A}$.
  Thus, the evolution of ${\SPvec{A}}$ is governed by
\beq
	{\partial}{\SPvec{A}}
=
        - {\nabla}A - {\SPvec{E}}
\label{eq:FOLIeAmagn}
\eeq
and constrained by
\beq
        {\nabla}\times  {\SPvec{A}}
=
          {\SPvec{B}}\, .
\label{eq:FOLIcAmagn}
\eeq
  
  The electric potential $A$ on the other hand acquires a dynamical 
status only within a Poincar\'e-invariant gauge. 
  For instance, in the V.Lorenz--H.A.Lorentz gauge
   \cite{hawkingellisBOOK, jacksonokun, jacksonB}
the evolution of ${A}$ is governed by
\bea
{\partial} A = - {\nabla} \cdot  {\SPvec{A}} 
\label{eq:FOLIeAelec}
\eea
and not constrained by any other equation.
  However, while in the Lorenz--Lorentz gauge the
Maxwell--Lorentz equations with prescribed (point) sources 
are equivalent to a decoupled set of non-homogeneous wave equations for 
$A,{\SPvec{A}}$ that are readily solved by the Li\'enard--Wiechert potentials
    \cite{lienard, wiechert},
this gauge achieves no such simplification 
for the Maxwell--Born--Infeld equations with (point) sources.
  Hence, while for the sake of concreteness we shall work in
the Lorenz--Lorentz gauge one might as well look for a more convenient
Poincar\'e-invariant gauge to work in.

%%%%%%%%%%%%%%%%%%%%%%%%%%%%%%%%%%%%%%%%%%%%%%%%%%%%%%%%%%%%%%%%%%%%
%%%%%%%%%%%%%%%%%%%%%%%%%%%%%%%%%%%%%%%%%%%%%%%%%%%%%%%%%%%%%%%%%%%%
%%%%%%%%%%%%%%%%%%%%%%%%%%%%%%%%%%%%%%%%%%%%%%%%%%%%%%%%%%%%%%%%%%%%
	\subsection{The $t$-synchronized proper many-body Hamilton--Jacobi formalism}
%%%%%%%%%%%%%%%%%%%%%%%%%%%%%%%%%%%%%%%%%%%%%%%%%%%%%%%%%%%%%%%%%%%%
%%%%%%%%%%%%%%%%%%%%%%%%%%%%%%%%%%%%%%%%%%%%%%%%%%%%%%%%%%%%%%%%%%%%
%%%%%%%%%%%%%%%%%%%%%%%%%%%%%%%%%%%%%%%%%%%%%%%%%%%%%%%%%%%%%%%%%%%%

%%%%%%%%%%%%%%%%%%%%%%%%%%%%%%%%%%%%%%%%%%%%%%%%%%%%%%%%%%%%%%%%%%%%
%%%%%%%%%%%%%%%%%%%%%%%%%%%%%%%%%%%%%%%%%%%%%%%%%%%%%%%%%%%%%%%%%%%%
	\subsubsection{$t$-synchronized ordered configurations}
%%%%%%%%%%%%%%%%%%%%%%%%%%%%%%%%%%%%%%%%%%%%%%%%%%%%%%%%%%%%%%%%%%%%
%%%%%%%%%%%%%%%%%%%%%%%%%%%%%%%%%%%%%%%%%%%%%%%%%%%%%%%%%%%%%%%%%%%%

 We begin by defining the (foliation time) $t$-synchronized configuration 
space of $N$ ordered space-points, which consists of ordered $3N$-tupels 
${\SPvec{S}}\equiv ({\SPvec{s}}_1,..., {\SPvec{s}}_N)\in \RR^{3N}_{\neq}$;
here, $\RR^{3N}_{\neq} \equiv
{\RR}^{3N}\backslash \{\SPvec{S}:\SPvec{s}_k=\SPvec{s}_l \ for some\ k\neq{l}\}$, 
i.e.  coincidence points are removed.
  The $N$ individual point charge trajectories 
$t\mapsto{\SPvec{s}}\alongHk =  {\SPvec{r}}_k(t) \in \RR^{3}$ now correspond 
to a single trajectory $t\mapsto\SPvec{S}= \SPvec{R}(t)\in \RR^{3N}_{\neq}$ 
in this ordered configuration space, where
$\SPvec{R}(t) = ({\SPvec{r}}_1(t),..., {\SPvec{r}}_N(t))$.
 Our goal is a Hamilton--Jacobi law of evolution such that
$\dot{\SPvec{R}}(t) = {\SPvec{V}}(t,{\SPvec{R}(t)})$ for a velocity
flow field ${\SPvec{V}}(t,{\SPvec{S}})
\equiv ({\SPvec{v}}_1,..., {\SPvec{v}}_N)(t,{\SPvec{S}})$
on $\RR^{3N}_{\neq}$ obtained with the phase 
function $\Phi(t,{\SPvec{S}})$ on $\RR^{3N}_{\neq}$ satisfying 
a Hamilton--Jacobi PDE.

%%%%%%%%%%%%%%%%%%%%%%%%%%%%%%%%%%%%%%%%%%%%%%%%%%%%%%%%%%%%%%%%%%%%
%%%%%%%%%%%%%%%%%%%%%%%%%%%%%%%%%%%%%%%%%%%%%%%%%%%%%%%%%%%%%%%%%%%%
	\subsubsection{The $t$-synchronized equations of the space and time decomposition of the $^\sharp$fields}
%%%%%%%%%%%%%%%%%%%%%%%%%%%%%%%%%%%%%%%%%%%%%%%%%%%%%%%%%%%%%%%%%%%%
%%%%%%%%%%%%%%%%%%%%%%%%%%%%%%%%%%%%%%%%%%%%%%%%%%%%%%%%%%%%%%%%%%%%

 The space and time decomposition of 
$^\sharp\AQ(\varpi,\varpi_1,...,\varpi_N)$ into components 
$^\sharp{A}$ 
and 
$^\sharp{\SPvec{A}}$
and subsequent  $t$-synchronization gives us the fields
\beq
{{A}}^\sharp(t,\SPvec{s},\SPvec{S})
\equiv
{}^\sharp{{A}}(t,\SPvec{s}, t_1,\SPvec{s}_1,...,t_k,\SPvec{s}_k,...,t_N,\SPvec{s}_N)
\big|_{t_1=t_2=...=t_N=t}\big.,
\eeq
\beq
{\SPvec{A}}^\sharp(t,\SPvec{s},\SPvec{S})
\equiv
{}^\sharp{\SPvec{A}}(t,\SPvec{s}, t_1,\SPvec{s}_1,...,t_k,\SPvec{s}_k,...,t_N,\SPvec{s}_N)
\big|_{t_1=t_2=...=t_N=t}\big.
\eeq
on $\RR^{3(N+1)}$ (etc. for the other $^\sharp$fields).
 As stipulated earlier, by conditioning with the actual configuration we 
want to obtain the actual fields on $\MM^4$ (in Lorentz gauge, say), 
i.e. ${A}^\sharp (t,\SPvec{s},\SPvec{R}(t)) = A(t,\SPvec{s})$ and 
${\SPvec{A}}^\sharp (t,\SPvec{s},\SPvec{R}(t))= {\SPvec{A}}(t,\SPvec{s})$ (etc.).
 This canonically fixes the equations for the $t$-synchronized space and time 
decomposition of the $^\sharp$fields. 
 Namely, ${{A}}^\sharp(t,\SPvec{s},\SPvec{S})$, ${\SPvec{A}}^\sharp(t,\SPvec{s},\SPvec{S})$, 
and ${\SPvec{D}}^\sharp(t,\SPvec{s},\SPvec{S})$ satisfy the evolution equations
\beq
\partial{{A}}^\sharp(t,\SPvec{s},\SPvec{S})
=
- \SPvec{V}(t,\SPvec{S})\cdot \nabla_{\!\!\SPvec{S}} 
{A}^\sharp(t,\SPvec{s},\SPvec{S})
- \nabla\cdot \SPvec{A}^\sharp(t,\SPvec{s},\SPvec{S}) ,
\label{eq:FOLIeqAscalSHARP}
\eeq
\beq
\partial {\SPvec{A}}^\sharp(t,\SPvec{s},\SPvec{S})
=
- \SPvec{V}(t,\SPvec{S})\cdot\nabla_{\!\!\SPvec{S}} 
 {\SPvec{A}}^\sharp(t,\SPvec{s},\SPvec{S})
- \nabla A^\sharp(t,\SPvec{s},\SPvec{S}) -  {\SPvec{E}}^\sharp(t,\SPvec{s},\SPvec{S}),
\label{eq:FOLIeqAvectSHARP}
\eeq
\beq
\partial {\SPvec{D}}^\sharp(t,\SPvec{s},\SPvec{S})
=
- \SPvec{V}(t,\SPvec{S})\cdot\nabla_{\!\!\SPvec{S}} 
 {\SPvec{D}}^\sharp(t,\SPvec{s},\SPvec{S})
+ \nabla \times {\SPvec{H}}^\sharp(t,\SPvec{s},\SPvec{S}) 
- 4\pi {\SPvec{j}}^\sharp(t,{\SPvec{s}},{\SPvec{S}}),
\label{eq:FOLIeqDvectSHARP}
\eeq
where 
$\SPvec{V}(t,\SPvec{S})\cdot\nabla_{\!\!\SPvec{S}} \equiv \sum_{k=1}^N \SPvec{v}_k(t,\SPvec{S})\cdot\nabla_k$;
furthermore, ${\SPvec{D}}^\sharp(t,\SPvec{s},\SPvec{S})$ obeys the constraint equation
\beq
\nabla\cdot {\SPvec{D}}^\sharp(t,\SPvec{s},\SPvec{S})
=
 4\pi {{j}}^\sharp(t,{\SPvec{s}},{\SPvec{S}}),
\label{eq:FOLIeqDvectSHARPconstraint}
\eeq
where\footnote{Notice that $j^\sharp$ is actually $t$-independent. 
               Notice furthermore that 
	       $\nabla_{\SPvec{s}_k}\delta_{{\SPvec{s}}_{k}}({\SPvec{s}}) 
	       = - \nabla_{\SPvec{s}}\delta_{{\SPvec{s}}_{k}}({\SPvec{s}})$.
	       Hence, the $^\sharp$field re-formulation of the continuity equation 
	       for the charge conservation (in spacetime), 
	       $\partial{{j}}^\sharp(t,\SPvec{s},\SPvec{S})
	       = - \SPvec{V}(t,\SPvec{S})\cdot \nabla_{\!\!\SPvec{S}} 
	       {j}^\sharp(t,\SPvec{s},\SPvec{S})
	       - \nabla\cdot \SPvec{j}^\sharp(t,\SPvec{s},\SPvec{S})$, 
	       is an identity, not an independent equation.}
\bea
        j^\sharp (t,{\SPvec{s}},{\SPvec{S}})
\!\!\!&=&\!\!\!
{\textstyle\sum_{k\in\cN}} \pmk\delta_{{\SPvec{s}}_{k}}({\SPvec{s}}),
\label{eq:FOLIpointSOURCESrhoSHARP}
\\
&&
\nonumber
\\
        {\SPvec{j}}^\sharp(t,{\SPvec{s}},{\SPvec{S}})
\!\!\!&=&\!\!\!
{\textstyle\sum_{k\in\cN}}
\pmk\delta_{{\SPvec{s}}_{k}}({\SPvec{s}}){\SPvec{v}}{_k}(t,{\SPvec{S}})
\, .
\label{eq:FOLIpointSOURCESjjjSHARP}
\eea
 The fields 
${\SPvec{E}}^\sharp(t,\SPvec{s},\SPvec{S})$ and 
${\SPvec{H}}^\sharp(t,\SPvec{s},\SPvec{S})$ in 
\refeq{eq:FOLIeqAvectSHARP}, \refeq{eq:FOLIeqDvectSHARP}
are defined in terms of 
${\SPvec{D}}^\sharp(t,\SPvec{s},\SPvec{S})$ 
and 
${\SPvec{B}}^\sharp(t,\SPvec{s},\SPvec{S})$ 
in precisely the same manner as the actual fields 
${\SPvec{E}}(t,\SPvec{s})$ and ${\SPvec{H}}(t,\SPvec{s})$ 
are defined in terms of 
${\SPvec{D}}(t,\SPvec{s})$ and ${\SPvec{B}}(t,\SPvec{s})$ 
through the Born--Infeld aether laws 
\refeq{eq:FOLIeqEofBD}, \refeq{eq:FOLIeqHofBD}, 
while 
${\SPvec{B}}^\sharp(t,\SPvec{s},\SPvec{S})$ 
in turn is defined in terms of 
${\SPvec{A}}^\sharp(t,\SPvec{s},\SPvec{S})$ 
in precisely the same manner as the actual 
${\SPvec{B}}(t,\SPvec{s})$ 
is defined in terms of the actual
${\SPvec{A}}(t,\SPvec{s})$ 
in 
\refeq{eq:FOLIcAmagn}.

 It is straightforward to verify that by substituting the actual configuration 
$\SPvec{R}(t)$ for the generic $\SPvec{S}$ in the $t$-synchronized $^\sharp$fields 
satisfying the above equations, we obtain the actual electromagnetic potentials, 
fields, and charge-current densities satisfying the Maxwell--Born--Infeld field 
equations (in Lorentz--Lorenz gauge). 

%%%%%%%%%%%%%%%%%%%%%%%%%%%%%%%%%%%%%%%%%%%%%%%%%%%%%%%%%%%%%%%%%%%%
%%%%%%%%%%%%%%%%%%%%%%%%%%%%%%%%%%%%%%%%%%%%%%%%%%%%%%%%%%%%%%%%%%%%
	\subsubsection{The $t$-synchronized Hamilton--Jacobi equations}
%%%%%%%%%%%%%%%%%%%%%%%%%%%%%%%%%%%%%%%%%%%%%%%%%%%%%%%%%%%%%%%%%%%%
%%%%%%%%%%%%%%%%%%%%%%%%%%%%%%%%%%%%%%%%%%%%%%%%%%%%%%%%%%%%%%%%%%%%

  By conditioning ${\SPvec{A}}^\sharp(t,\SPvec{s},\SPvec{S})$
and ${A}^\sharp(t,\SPvec{s},\SPvec{S})$ with $\SPvec{s}=\SPvec{s}_k$ 
for each $k=1,...,N$, we now obtain the $t$-synchronized $\widetilde{A}_k$ 
and $\widetilde{\SPvec{A}}_k$ (etc.) fields on $\RR^{3N}_{\neq}$, for 
which we write 
\beq
{A}_k(t,\SPvec{S})
\equiv
\widetilde{A}_k
(t_1,{\SPvec{s}}_1,...,t_N,{\SPvec{s}}_N)\Big|_{t_1=t_2=...=t_N=t}\Big.
\eeq
and
\beq
\SPvec{A}_k(t,\SPvec{S})
\equiv
\widetilde{\SPvec{A}}_k
(t_1,{\SPvec{s}}_1,...,t_N,{\SPvec{s}}_N)\Big|_{t_1=t_2=...=t_N=t}\Big. .
\eeq
    We shall now show that the space and time decomposition of the geometric 
law for each point-history \refeq{eq:GEOhamjacLAW} is equivalent
to a Hamilton--Jacobi law of point charge motion. 
   We consider the genuinely electromagnetic setting first.
  The non-genuinely electromagnetic extensions are given 
subsequently.

%\newpage

{\textit{A single positive or negative unit point charge}}

  When there is only a single point charge, in space and time decomposition 
(no synchronization necessary now; i.e. $t_1=t$ automatically), we can 
immediately identify
$\widetilde{{A}}_1(\varpi_1) \equiv A_1(t,{\SPvec{s}_1})$
and
$\widetilde{\SPvec{A}}_1(\varpi_1) \equiv \SPvec{A}_1(t,{\SPvec{s}_1})$;
furthermore, we clearly have
$\widetilde\Phi_1 (\varpi_1)\equiv\Phi(t,{\SPvec{s}_1})$, 
where $\Phi : \RR^{1,3} \to{\RR}$
is the scalar {phase function associated with the point-history}
of the only electron in that world.
 In terms of $\Phi$ we now define a Minkowski velocity field on $\MM^4_1$, 
given by the system of equations
\bea
&& {{u}}{}(t,{\SPvec{s}_1})
=
-{\partial} \Phi(t,{\SPvec{s}_1}) 
-\plumi {\alpha}{A}_1(t,{\SPvec{s}_1})\, ,
\label{eq:FOLIvelofieldLAWtSINGLE}
\\
&&{\SPvec{u}}{}(t,{\SPvec{s}_1})
=
\ {\nabla}_1 \Phi(t,{\SPvec{s}_1}) 
-\plumi {\alpha} {\SPvec{A}}_1(t,{\SPvec{s}_1}) \,.
\label{eq:FOLIvelofieldLAWsSINGLE}
\eea
  The condition that the l.h.s. be a Minkowski velocity vector for each value of its arguments
gives the space and time decomposition of the geometrical law \refeq{eq:GEOhamjacPDE} for the phase function 
$\Phi$,
\beq
-\left({\partial} \Phi + \plumi {\alpha}{A}_1\right)^2
+
  {|{\nabla}_1 \Phi -\plumi {\alpha} {\SPvec{A}_1}|}^2
= 
- 1\, ,
\label{eq:HamJacONEeEQsquare}
\eeq
and the future-orientation of $\uQ$ selects
the following root of \refeq{eq:HamJacONEeEQsquare}, 
\beq 
 {\partial} \Phi
= 
-  \sqrt{1 +{|{\nabla}_1 \Phi -\plumi {\alpha} {\SPvec{A}_1}|}^2} -\plumi {\alpha}{A}_1.
\label{eq:HamJacONEeEQfuture}
\eeq
 The appearance of \refeq{eq:HamJacONEeEQsquare}, and of \refeq{eq:HamJacONEeEQfuture},
is familiar from the relativistic Hamilton--Jacobi equation for a
single point charge interacting with external electromagnetic fields
               \cite{landaulifshitzBOOK};
cf. our subsection on test particles Hamilton--Jacobi theory.
 Appearances are however misleading, for here the fields are the total fields.

 The geometric law  \refeq{eq:GEOhamjacLAW} for the actual history $\Eta$ of the single 
negative or positive point electron is then the system of guiding equations
\bea
&&\ \textstyle{\frac{\dd{t}}{\dd{\tau}}\alongH}
=
{{u}}{}(t,{\SPvec{s}_1})\alongH
\\
&&\textstyle{\frac{\dd{\SPvec{s}_1}}{\dd{\tau}}\alongH}
=
{\SPvec{u}}{}(t,{\SPvec{s}_1})\alongH
\label{eq:FOLIguidingLAWsSINGLE}
\eea
  We finally eliminate $\tau$ in favor of our foliation time $t$ to get 
the equation for $t\mapsto\SPvec{s}_1$.
 Thus, $\tau\mapsto (t,{\SPvec{s}_1}) = \left(w_1(\tau),{\SPvec{w}_1}(\tau)\right)$ 
is replaced by 
$t\mapsto(t,{\SPvec{s}_1})  = (r_1(t),{\SPvec{r}_1}(t))$, whence
$r(\,.\,)\equiv {\mathrm{id}}(\,.\,)$
(we would not need to keep the index $_1$, but for the sake of clarity we do).
	The change of the derivatives is done according 
to the familiar relativistic formula
$
{\dd{\tau}} 
=
{\sqrt{1- |\bulldif{\SPvec{r}}_1|{}^2}}\,{\dd{t}} 
$,
where $\bulldif{\SPvec{r}}_1(t)$ is the conventional velocity of the particle at 
${\SPvec{s}_1} = {\SPvec{r}_1}(t)$ in the standard-foliation space  $\RR^3$.
  Hence, for a single point charge coupled to the electromagnetic Maxwell--Born--Infeld 
fields, 
we can finally rewrite the space-part of the guiding equation into
\beq
	\frac{\dd{\SPvec{s}_1}}{\dd{t}}\AlongH
=
\frac{{\nabla}_1 \Phi(t,{\SPvec{s}_1}) - \plumi \alpha {\SPvec{A}}_1(t,{\SPvec{s}_1})}
  {\sqrt{1 +{|{\nabla}_1 \Phi(t,{\SPvec{s}_1}) -\plumi \alpha {\SPvec{A}}_1(t,{\SPvec{s}_1})|}{}^2} }\AlongH
\,.
\label{eq:STANDARDhamjacLAWofMOTION}
\eeq

 Lastly, we note that substitution of the actual position $\SPvec{r}_1(t)$ for the 
generic $\SPvec{s}_1$ in ${A}_1(t,\SPvec{s}_1)$ and $\SPvec{A}_1(t,\SPvec{s}_1)$
gives the actual potential fields evaluated at the location of the single point charge, 
i.e. we have 
${A}_1(t,\SPvec{r}_1(t)) = A(t,\SPvec{r}_1(t))$ 
and 
$\SPvec{A}_1(t,\SPvec{r}_1(t))  = \SPvec{A}(t,\SPvec{r}_1(t))$. 
 Thus, for the actual solution $t\mapsto {\SPvec{s}_1} = {\SPvec{r}_1}(t)$ 
of \refeq{eq:STANDARDhamjacLAWofMOTION}, we find the identity
\beq
\bulldif{\SPvec{r}}_1(t)
=
\frac{{\nabla}_1 \Phi(t,{\SPvec{r}_1}(t)) - \plumi \alpha {\SPvec{A}}(t,{\SPvec{r}_1}(t))}
  {\sqrt{1 +{|{\nabla}_1 \Phi(t,{\SPvec{r}_1}(t)) -\plumi \alpha {\SPvec{A}}(t,{\SPvec{r}_1}(t))|}{}^2} }
\,,
\label{eq:STANDARDhamjacCANONICALmomentumID}
\eeq
which is precisely \refeq{eq:canonicalMOMENTUM},  our point of departure.
 We have come full circle.

{\textit{Many positive and negative unit point charges}}

        For $N$ (negative and positive) unit point charges, 
space and time decomposition gives
$\widetilde\Phi(\varpi_1,...,\varpi_N)
 = \widetilde\Phi(t_1,\SPvec{s}_1,...,t_N,\SPvec{s}_N)$, 
and synchronization gives us
the \emph{$t$-synchronized configuration space phase function}
$\Phi(t,\SPvec{S})\equiv
 \widetilde\Phi(t,\SPvec{s}_1,...,t,\SPvec{s}_N)$,
with the help of which we now define a Minkowski velocities field on $\RR\times\RR^{3N}_{\neq}$, 
the $k$-th component of which is given by the system of equations
\bea
&&{u_k}{}(t,{\SPvec{S}})
=
-{\partial}_k\widetilde\Phi(...,t,\SPvec{s}_k,...)
-\pmk{\alpha}{A}_k(t,\SPvec{S})
\label{eq:FOLIvelofieldLAWtMANY}
\\
&&{\SPvec{u}_k}{}(t,{\SPvec{S}})
=\ 
{\nabla}_k\widetilde\Phi(...,t,\SPvec{s}_k,...)
- \pmk{\alpha} {\SPvec{A}}_k(t,\SPvec{S})
\,.
\label{eq:FOLIvelofieldLAWsMANY}
\eea
  The constraint that the left-hand sides of \refeq{eq:FOLIvelofieldLAWtMANY}
and \refeq{eq:FOLIvelofieldLAWsMANY} are the components in the $k^{\mathrm{th}}$ 
copy $\MM^4_k$ of $\MM^4$ of a future-oriented Minkowski-velocities vector field 
on configuration space gives a Hamilton--Jacobi type partial 
differential equation on $\RR\times\RR^{3N}_{\neq}$, 
\beq
 {\partial}_k \widetilde\Phi(...,t,\SPvec{s}_k,...)
= 
-  \sqrt{1 +{|{\nabla}_k \widetilde\Phi(...,t,\SPvec{s}_k,...) 
-  \pmk {\alpha} {\SPvec{A}}_k(t,\SPvec{S})|}^2}
-  \pmk {\alpha}{A}_k(t,\SPvec{S})\,.
\label{eq:HamJacEQkFUTURE}
\eeq
  Using now
$
{\dd{\tau}} 
=
{\sqrt{1- |\bulldif{\SPvec{r}}_k|{}^2}}\,{\dd{t}} 
$
for each $k$ gives us the Hamilton--Jacobi type guiding equation for 
the actual motion of the $k^{\mathrm{th}}$ point-charge along $\Eta_k$,
\beq
	\frac{\dd{\SPvec{s}_k}}{\dd{t}}\AlongHk
=
\frac{{\nabla}_k \widetilde\Phi(...,t,{\SPvec{s}_k},...)  
- \pmk \alpha
{\SPvec{A}}_k\left(t,{\SPvec{S}}\right) }
{\sqrt{1 +
	{|{\nabla}_k \widetilde\Phi(...,t,{\SPvec{s}_k},...)   
	-\pmk \alpha {\SPvec{A}}_k\left(t,{\SPvec{S}}\right)|}{}^2} }\AlongHk\,.
\label{eq:FOLIguidingLAWkMANY}
\eeq
 Finally, since substitution of the actual configuration $\SPvec{R}(t)$ for the 
generic $\SPvec{S}$ in ${A}_k(t,\SPvec{S})$ and $\SPvec{A}_k(t,\SPvec{S})$
gives the actual potential fields evaluated at the location of the $k$-th 
point charge, i.e. since
${A}_k(t,\SPvec{R}(t)) = \widetilde{A}_k(t,\SPvec{r}_1(t),...,t,\SPvec{r}_N(t))
                       = A(t,\SPvec{r}_k(t))$ 
and 
$\SPvec{A}_k(t,\SPvec{R}(t)) =
\widetilde{\SPvec{A}}_k(t,\SPvec{r}_1(t),...,t,\SPvec{r}_N(t))
	     = \SPvec{A}(t,\SPvec{r}_k(t))$, 
\refeq{eq:FOLIguidingLAWkMANY} turns into the identity
\beq
\bulldif{\SPvec{r}}_k(t)
=
\frac{{\nabla}_k \widetilde\Phi(...,t,{\SPvec{r}_k}(t),...)  
- \pmk \alpha
{\SPvec{A}}\left(t,{\SPvec{r}_k}(t)\right) }
{\sqrt{1 +
	{|{\nabla}_k \widetilde\Phi(...,t,{\SPvec{r}_k}(t),...)   
	-\pmk \alpha {\SPvec{A}}\left(t,{\SPvec{r}}_k(t)\right)|}{}^2} }\,,
\label{eq:FOLIcanonicalMOMENTUMkID}
\eeq
which is \refeq{eq:canonicalMOMENTUM}.
 Once again we have completed the loop back to our point of departure, though not quite.
 We still have to state the results in terms of $\Phi(t,\SPvec{S})$ satisfying a
single Hamilton--Jacobi PDE. 
 But, clearly, $\Phi(t,\SPvec{S})$ satisfies
\beq
 {\partial} \Phi 
= 
-\sum_{k\in\cN} 
     \left(
           \sqrt{1 +{|{\nabla}_k \Phi - \pmk {\alpha}{\SPvec{A}}_k|}^2}
	   + \pmk {\alpha} A_k
     \right),
\label{eq:configSPACEhamjacEQmany}
\eeq
and \refeq{eq:FOLIguidingLAWkMANY} can be rephrased as 
\beq
	\frac{\dd{\SPvec{s}}_k}{\dd{t}}\AlongHk
=
\frac{{\nabla}_k \Phi(t,{\SPvec{S}})  
- \pmk \alpha
{\SPvec{A}}_k\left(t,{\SPvec{S}}\right) }
{\sqrt{1 +
	{|{\nabla}_k \Phi(t,{\SPvec{S}})   
	-\pmk \alpha {\SPvec{A}}_k\left(t,{\SPvec{S}}\right)|}{}^2}}\AlongHk
\,.
\label{eq:configSPACEguidingEQmany}
\eeq

  We finally remark that in the decoherent case variables can even be separated in 
$\Phi$ as $\Phi =\sum_k \Phi^{(k)}$, so that the $k^{\mathrm{th}}$ point-history 
has associated with it an individual scalar phase function $\Phi^{(k)}: \RR^{1,3}_k\to{\RR}$.

%%%%%%%%%%%%%%%%%%%%%%%%%%%%%%%%%%%%%%%%%%%%%%%%%%%%%%%%%%%%%%%%%%%%
  \textit{Extensions to non-genuinely electromagnetic systems (II)}
	\label{sect:NongenuineEdynII}
%%%%%%%%%%%%%%%%%%%%%%%%%%%%%%%%%%%%%%%%%%%%%%%%%%%%%%%%%%%%%%%%%%%%

	We here briefly summarize how the extensions to 
non-genuinely electromagnetic settings affect the equations of the
evolutionary formulation of the theory.
  The effect that these modified formulas have on the conservation laws 
and variational principles can then readily be worked out.

   In the fairly general microscopic setting presented in subsection
\ref{sect:NongenuineEdynI}, the change of value space for the $z_k$ 
from $\{-1,1\}$ to $\ZZ$ does not entail any change in the formulas.
  However, the $k^{\mathrm{th}}$ component of 
configuration space trajectory ${\SPvec{S}} = {\SPvec{R}}(t)$, 
where $\SPvec{S} \equiv ({\SPvec{s}}_1,...,{\SPvec{s}}_{N})$,
now satisfies the Hamilton--Jacobi guiding law
\beq
	\frac{\dd{\SPvec{s}_k}}{\dd{t}}\AlongHk
=
\frac{{\nabla}_k \Phi(t,{\SPvec{S}})  
- \kappa_k \pmk   \alpha
{\SPvec{A}}_k\left(t,{\SPvec{S}}\right) }
{\sqrt{1 +
	{|{\nabla}_k \Phi(t,{\SPvec{S}})   
	-\kappa_k \pmk  \alpha {\SPvec{A}}_k\left(t,{\SPvec{S}}\right)|}{}^2} }\AlongHk\,.
\label{eq:allVELO}
\eeq
where again $\kappa_k \leq 1$ is the ratio of the electron's to $k^{\mathrm{th}}$ particle's 
rest mass, and the subscripts at the operators indicate once again with respect to which of 
the particle co-ordinates to take the derivative.
 The Hamilton--Jacobi equation on $t$-synchronized $3N$-dimensional configuration space becomes
\beq
 {\partial} \Phi (t,\SPvec{S})
=
- \sum_{k\in\cN} 
\Bigl(
      \sqrt{1 +{|{\nabla}_k \Phi(t,\SPvec{S}) -\kappa_k z_k {\alpha} {\SPvec{A}}_k(t,\SPvec{S})|}^2} 
      + \kappa_k  z_k{\alpha}A_k(t,\SPvec{S})
\Bigr)
\,.
\eeq

  While it is obvious from the generality of the above formulas that the electromagnetic 
effects of particles and antiparticles are modeled on an equal footing, 
the primary application of the theory should be to normal matter modeled
by $N^+$ positive point charges of various sorts, and $N^-$ negative 
unit point charges, representing a system of nuclei and electrons.
  The most sensible application should be to high-temperature plasma, but to the extent 
that quantum effects are negligible, applications to fluids, solids, and even molecules 
and atoms may be possible.
  For $N^- >1$ we may wonder whether we can even impose the Pauli principle for the electrons already
at the classical level; we come to this point in a moment.
  
%%%%%%%%%%%%%%%%%%%%%%%%%%%%%%%%%%%%%%%%%%%%%%%%%%%%%%%%%%%%%%%%%%%%
{\textit{The Born--Oppenheimer approximation}}
%%%%%%%%%%%%%%%%%%%%%%%%%%%%%%%%%%%%%%%%%%%%%%%%%%%%%%%%%%%%%%%%%%%%

  We remark that  the Born--Oppenheimer approximation, 
in which the nuclei are assumed to be infinitely massive, can also easily be implemented
by formally letting $\kappa_k\downarrow{0}$ for the positive charges representing nuclei
in our model.
  Inspection of \refeq{eq:allVELO} reveals that $\kappa_k\downarrow{0}$ gives  for each nucleus
the equation of a free straight motion, as it should be the case. 
  This simplifies the dynamical problem for the charges to solving a reduced Hamilton--Jacobi
equation for the $N^-$ electrons, which is coupled to the nuclei only 
via the electromagnetic potentials. 
 Furthermore, if there is only one nucleus, then by performing a Lorentz boost and a translation 
one can assume that the nucleus is permanently at rest at the origin.

%%%%%%%%%%%%%%%%%%%%%%%%%%%%%%%%%%%%%%%%%%%%%%%%%%%%%%%%%%%%%%%%%%%%
%%%%%%%%%%%%%%%%%%%%%%%%%%%%%%%%%%%%%%%%%%%%%%%%%%%%%%%%%%%%%%%%%%%%
	\subsubsection{Gauge invariance (Part II)}
%%%%%%%%%%%%%%%%%%%%%%%%%%%%%%%%%%%%%%%%%%%%%%%%%%%%%%%%%%%%%%%%%%%%
%%%%%%%%%%%%%%%%%%%%%%%%%%%%%%%%%%%%%%%%%%%%%%%%%%%%%%%%%%%%%%%%%%%%

  Since the $A^\sharp$ and $\SPvec{A}^\sharp$ inherit the gauge transformations
from $A$ and $\SPvec{A}$, we see that 
equations \refeq{eq:FOLIeAmagn}, \refeq{eq:FOLIcAmagn},
\refeq{eq:FOLIeAelec} for the electric and magnetic potentials,
and equations \refeq{eq:configSPACEhamjacEQmany}, 
\refeq{eq:configSPACEguidingEQmany}
for the point charge motions, are invariant under gauge transformations
\bea
  \Phi(t,{\SPvec{S}})
 \!\!\!\!\!\!\!\! && 
\to \Phi(t,{\SPvec{S}})  + \alpha {\textstyle{\sum_{k\in\cN}}} \pmk \Upsilon(t,{\SPvec{s}}_k),
\\
  A_k(t,{\SPvec{S}}) 
 \!\!\!\!  \!\!\!\! 
&& 
\to A_k(t,{\SPvec{S}}) - {\partial} \Upsilon(t,{\SPvec{s}_k}),
\\
    {\SPvec{A}}_k (t,{\SPvec{S}})
\!\!\!\!  \!\!\!\! 
&& 
\to  {\SPvec{A}}_k(t,{\SPvec{S}}) + ({\nabla} \Upsilon)(t,{\SPvec{s}_k}),
\label{eq:GEOgaugetrAPHIA}
\eea
with any relativistic scalar field $\Upsilon: {\RR}^{1,3}\to{\RR}$ 
satisfying the wave equation
\bea
{\partial}^2 \Upsilon = {\nabla^2}  \Upsilon 
\label{eq:waveEQgamma}
\eea
with $\nabla^2 = \Delta$, the Laplacian on $\RR^3$.
  Since a (sufficiently regular) solution of  \refeq{eq:waveEQgamma} 
in $\RR_+\times \RR^3$ is uniquely determined by the initial data for $\Upsilon$ and its 
time derivative $\partial\Upsilon$, the gauge freedom that is left concerns the initial 
conditions of $\Phi,A,{\SPvec{A}}$. 

%%%%%%%%%%%%%%%%%%%%%%%%%%%%%%%%%%%%%%%%%%%%%%%%%%%%%%%%%%%%%%%%%%%%
%%%%%%%%%%%%%%%%%%%%%%%%%%%%%%%%%%%%%%%%%%%%%%%%%%%%%%%%%%%%%%%%%%%%
%%%%%%%%%%%%%%%%%%%%%%%%%%%%%%%%%%%%%%%%%%%%%%%%%%%%%%%%%%%%%%%%%%%%
	\subsection{The Cauchy problem for the physical state}
%%%%%%%%%%%%%%%%%%%%%%%%%%%%%%%%%%%%%%%%%%%%%%%%%%%%%%%%%%%%%%%%%%%%
%%%%%%%%%%%%%%%%%%%%%%%%%%%%%%%%%%%%%%%%%%%%%%%%%%%%%%%%%%%%%%%%%%%%
%%%%%%%%%%%%%%%%%%%%%%%%%%%%%%%%%%%%%%%%%%%%%%%%%%%%%%%%%%%%%%%%%%%%

  After having recast the geometrical laws of electromagnetism into the 
evolutionary formalism w.r.t. the standard foliation of Minkowski spacetime, 
we here collect all the variables for which an evolution equation 
in $t$ has been obtained:
the actual world variables
    $A(\,.\,,\SPvec{s})$, 
    $\SPvec{A}(\,.\,,\SPvec{s})$, 
    $\SPvec{B}(\,.\,,\SPvec{s})$,
    $\SPvec{D}(\,.\,,\SPvec{s})$, 
    $j(\,.\,,\SPvec{s})$, 
    $\SPvec{R}(\,.\,)$,
as well as 
    $\Upsilon(\,.\,,\SPvec{s})$ 
and 
    $\partial\Upsilon(\,.\,,\SPvec{s})$,
and the Hamilton--Jacobi variables
    ${A}^\sharp(\,.\,,\SPvec{s},\SPvec{S})$, 
    $\SPvec{A}^\sharp(\,.\,,\SPvec{s},\SPvec{S})$, 
    $\SPvec{D}^\sharp(\,.\,,\SPvec{s},\SPvec{S})$, 
    $\Phi(\,.\,,\SPvec{S})$.
  However, the information content in the listed variables is partly redundant.
  Indeed, in the configuration space setting we already cleaned up this redundancy,
but there is a reason why we left the redundancy of the actual world variables. 
  Namely, which variables most conveniently represent the physical evolutionary degrees of 
freedom,\footnote{There are different 
    conventions in the literature as to what constitutes a degree of freedom. 
    Traditioned in Newtonian mechanics one normally refers to the independent 
    second-order (in time) equations as the \emph{dynamical} degrees of freedom 
    (e.g., a Newtonian point particle moving along a line and satisfying Newton's 
    equation of motion with a linear restoring force (a harmonic oscillator) counts 
    as a one-degree-of-freedom dynamical system.) 
    However, we here mean the number of data needed to specify the \emph{state} 
    of the physical system on any leaf of the standard foliation, which is then evolved
    uniquely in time by independent first-order \emph{evolution} equations.}
depends on whether source-free evolutions or those with point charge sources are 
considered.

\smallskip
%%%%%%%%%%%%%%%%%%%%%%%%%%%%%%%%%%%%%%%%%%%%%%%%%%%%%%%%%%%%%%%%%%%%
%%%%%%%%%%%%%%%%%%%%%%%%%%%%%%%%%%%%%%%%%%%%%%%%%%%%%%%%%%%%%%%%%%%%
	\subsubsection{Source-free evolutions}
%%%%%%%%%%%%%%%%%%%%%%%%%%%%%%%%%%%%%%%%%%%%%%%%%%%%%%%%%%%%%%%%%%%%
%%%%%%%%%%%%%%%%%%%%%%%%%%%%%%%%%%%%%%%%%%%%%%%%%%%%%%%%%%%%%%%%%%%%

   In the absence of any point charge sources, the only evolutionary variables are 
the solenoidal vector fields $\SPvec{s}\mapsto\SPvec{B}(\,.\,,\SPvec{s})$ and 
$\SPvec{s}\mapsto\SPvec{D}(\,.\,,\SPvec{s})$.
  These are in fact canonically conjugate variables for the first-order 
field evolution equations \refeq{eq:FOLIeB} and \refeq{eq:FOLIeD} 
without source terms, which constitute a Hamiltonian dynamical system; cf.
    \cite{BiBiONE}.
  It is not too difficult to show that sufficiently regular, finite energy 
solenoidal initial data at $t=0$ launch a unique evolution locally in time. 
  The much harder problem of global existence is widely open; see however
          \cite{lindblad} 
for global existence results for a related scalar equation.\footnote{After the original writing of these lines,
                                                                     a proof of the global well-posedness
								     of the source-free Maxwell--Born--Infeld
								     field equations for small initial data patterned
								     after Lindblad's paper 
								     \cite{lindblad} 
								     has been announced 
								     \cite{ChaeHuh}.}
  Note that the qualifier `solenoidal' for the initial data implies that the 
constraint equations \refeq{eq:FOLIcB} and \refeq{eq:FOLIcD} are satisfied
for all $t$, as follows by taking the divergence of the
evolution equations \refeq{eq:FOLIeB} and \refeq{eq:FOLIeD}.
  Note furthermore that no gauge freedom needs to be taken care of.

\smallskip
%%%%%%%%%%%%%%%%%%%%%%%%%%%%%%%%%%%%%%%%%%%%%%%%%%%%%%%%%%%%%%%%%%%%
%%%%%%%%%%%%%%%%%%%%%%%%%%%%%%%%%%%%%%%%%%%%%%%%%%%%%%%%%%%%%%%%%%%%
	\subsubsection{Evolutions with point charge sources}
%%%%%%%%%%%%%%%%%%%%%%%%%%%%%%%%%%%%%%%%%%%%%%%%%%%%%%%%%%%%%%%%%%%%
%%%%%%%%%%%%%%%%%%%%%%%%%%%%%%%%%%%%%%%%%%%%%%%%%%%%%%%%%%%%%%%%%%%%

  The situation is considerably more involved when point charges are present.
  The actual electromagnetic fields on Minkowski spacetime, $\SPvec{B}$ and 
$\SPvec{D}$, lose the status they enjoyed in the charge-free situation.
  Indeed, to begin with, we sort  out the redundant field quantities on actual 
Minkowski spacetime.
  In particular, while the electromagnetic potentials $A$ and $\SPvec{A}$ now
enter, rather than as the constraint equation for $\SPvec{A}$ given 
$\SPvec{B}$, \refeq{eq:FOLIcAmagn} ought to be read as defining $\SPvec{B}$
given $\SPvec{A}$. 
  The constraint equation \refeq{eq:FOLIcB} for $\SPvec{B}$ 
is then automatically satisfied; furthermore, by taking the curl of the evolution equation 
\refeq{eq:FOLIeAmagn} for $\SPvec{A}$ we see 
that the evolution equation \refeq{eq:FOLIeB} for $\SPvec{B}$ is 
also satisfied if the one for $\SPvec{A}$, \refeq{eq:FOLIeAmagn}, is. 
  Hence, $\SPvec{B}$ can be eliminated from the list of independent field
variables on Minkowski spacetime.\footnote{We remark that with prescribed point charge sources, 
                                           $\SPvec{A}$ and $\SPvec{D}$ form a canonically conjugate 
					   pair for a Hamiltonian field-dynamical system, 
					   in which the constraint of Coulomb's 
					   law is incorporated at the expense of a 
					   Lagrange parameter field, $A$. 
					   Of course, technically one can work with 
                                           $A$, $\SPvec{A}$ and $\SPvec{D}$ 
					   also in the absence of point charges, 
                                           but in that case $A$ and $\SPvec{A}$ 
					   would then be considered to be purely auxiliary
					   fields, the primary fields still being $\SPvec{B}$ and $\SPvec{D}$.}
 Working with $\SPvec{A}$  and $\SPvec{D}$, plus $A$, now gives us some gauge freedom 
to choose from.
   By imposing the Lorenz--Lorentz gauge, which yields the evolution equation  \refeq{eq:FOLIeAelec} 
for $A$, we have already removed some of the gauge freedom, but we can still impose a convenient 
gauge constraint on part of the initial conditions for $\Phi, A,\SPvec{A}$ as long
as it is compatible with the physical constraints on  $\Phi, A,\SPvec{A}$, and to the
extent that it  can be accommodated uniquely by suitable initial 
data $\Upsilon(0,\SPvec{s})$ and $\partial\Upsilon(0,\SPvec{s})$ for
the wave equation $\Box\Upsilon =0$.
  For instance, although perhaps not the most convenient thing to do, 
it is always possible to demand $A(0,\SPvec{s}) = 0$ and
$\nabla\cdot\SPvec{A}(0,\SPvec{s}) =0$ which fixes $\Upsilon$ up to 
a harmonic function, still do be disposed of.
   Yet another evolution equation is redundant, namely \refeq{eq:FOLIchargecontinuityEQ}, 
for the definitions \refeq{eq:FOLIpointSOURCESrho} and  \refeq{eq:FOLIpointSOURCESjjj}
of $j$ and $\SPvec{j}$ guarantee that the continuity equation 
\refeq{eq:FOLIchargecontinuityEQ} is satisfied for all times.
 This leaves us with the following list of a-priori independent 
actual fields on Minkowski spacetime:
$\SPvec{D}(\,.\,,\SPvec{s})$,
the electric displacement vector at $\SPvec{s}$ away from all $\SPvec{r}_k(\,.\,)$;
$\SPvec{A}(\,.\,,\SPvec{s})$, 
the magnetic vector potential at $\SPvec{s}$ away from all $\SPvec{r}_k(\,.\,)$;
and  $A(\,.\,,\SPvec{s})$, 
the electric potential at $\SPvec{s}$ away from all $\SPvec{r}_k(\,.\,)$. 
 However, to compute those fields we need to know the motion of the actual 
particles configuration, and to compute that one, we actually need to work with 
a whole parameter family of such $\SPvec{D}$, $\SPvec{A}$, and  $A$ fields, 
namely their corresponding $^\sharp$field cousins, so that the
$\SPvec{D}$, $\SPvec{A}$, and  $A$ fields themselves are not anymore 
independent degrees of freedom either.
 Thus, the evolutionary degrees of freedom comprise:

\noindent 
--- the $^\sharp$fields variables

$\bullet$ 
$\SPvec{D}^\sharp(\,.\,,\SPvec{s},\SPvec{S})$,
the generalized electric displacement field;

$\bullet$ 
$\SPvec{A}^\sharp(\,.\,,\SPvec{s},\SPvec{S})$,
the generalized magnetic potential;

$\bullet$ 
${A}^\sharp(\,.\,,\SPvec{s},\SPvec{S})$,
the generalized electric potential; 

\noindent
--- the Hamilton--Jacobi field variable

$\bullet$ 
$\Phi(\,.\,,\SPvec{S})$, 
the phase function at $\SPvec{S}$, coincidence points excluded;

\noindent
--- and finally, the $N$-particles' variable

$\bullet$ 
$\SPvec{R}(\,.\,)=\big(\SPvec{r}_1(\,.\,),...,\SPvec{r}_N(\,.\,)\big)$, the actual configuration space point;

\smallskip
\noindent
all constrained by the constraints equations. 
 In regard to our gauge and the space and time decomposition, these variables constitute the 
classical electromagnetic state $\Omega^{\mathrm{cl}}$ with point charges.
 The set of classical states is denoted $\Gamma^{\mathrm{cl}}$.

   The Cauchy problem for $\Omega^{\mathrm{cl}}(t)$ is to solve
its system of evolution equations supplemented at initial time, say $t=0$,
by the data $\Omega^{\mathrm{cl}}(0)$, which in case of 
$A^\sharp(0,\SPvec{s},\SPvec{S})$, 
$\SPvec{A}^\sharp(0,\SPvec{s},\SPvec{S})$, 
and $\SPvec{D}^\sharp(0,\SPvec{s},\SPvec{S})$,
we demand to satisfy asymptotic vanishing conditions as $|{\SPvec{s}}|\to\infty$
for each fixed $\SPvec{S}$.
 The initial data for the $^\sharp$variables and for $\Phi$ have to be chosen such that,
if the initial data $\SPvec{R}(0)$ for the actual configuration are substituted for the 
generic configuration $\SPvec{S}$, then the $^\sharp$fields restricted to $\RR^3$ become 
just the actual initial fields, viz. 
$A^\sharp(0,\SPvec{s},\SPvec{R}(0))=A(0,\SPvec{s})$, 
$\SPvec{A}^\sharp(0,\SPvec{s},\SPvec{R}(0))= \SPvec{A}(0,\SPvec{s})$, 
and 
$\SPvec{D}^\sharp(0,\SPvec{s},\SPvec{R}(0))=\SPvec{D}(0,\SPvec{s})$,
and the covariant $k$-gradient of $\Phi$ (divided by the obvious 
relativistic square root term) gives the initial velocity of the $k$-th particle. 
 Otherwise the  Hamilton--Jacobi part (inclusive the $^\sharp$fields) does not
depend on the actual variables, and in this sense poses an autonomous problem;
the actual configuration space trajectory $t\mapsto\SPvec{R}(t)$ is solved for
subsequently by integrating  the $N$ guiding equations obtained from the 
velocity field generated by $\Phi$. 
  Finally $\SPvec{R}(t)$ is substituted for $\SPvec{S}$ in the computed $^\sharp$fields
to get the actual fields $A$, $\SPvec{A}$ (giving $\SPvec{B}$), 
and $\SPvec{D}$ which satisfy the Maxwell--Born--Infeld field equations with
actual point charge and current densities according to the motion described by 
$\SPvec{R}(t)$. 

 We claim that in sharp contrast to the ill-defined Lorentz electrodynamics with 
point charges, our Cauchy problem can be set up consistently  for (classical-)physically 
generic situations, such that the Hamilton--Jacobi PDE and the $^\sharp$field equations 
launch $\Phi$ and the $^\sharp$fields differentiably into the immediate future of $t=0$, 
and the guiding equation launches $\SPvec{R}$ into its immediate future.
 More to the point:
 
\begin{Prop}
\label{thmCAUCHYlocal} 
The Cauchy problem for $\Omega^{\mathrm{cl}}$ is well-defined.
\end{Prop}

  Whether the Cauchy problem is also well-posed is a subtler issue. 
  Recall that local well-posedness means that for all initial states ${\Omega}^{\mathrm{cl}}(0)$
in some sufficiently small open subset of $\Gamma^{\mathrm{cl}}$ there exists a $T>0$ independent of 
the specific initial state but dependent on the open subset, such that there is a unique evolution
$t\mapsto {\Omega}^{\mathrm{cl}}(t)\in C^0((0,T),\Gamma^{\mathrm{cl}})$, 
	  satisfying $\lim_{t\downarrow{0}}\Omega^{\mathrm{cl}}(t) = \Omega^{\mathrm{cl}}(0)$.
 This should be straightforward to prove for a large class of physically relevant data.
 Global well-posedness on the other hand is a truly hard issue; in fact, we do not expect
that the Cauchy problem will be generically globally well-posed. 
  However, as is well-known from the Hamilton--Jacobi PDE in classical non-relativistic
mechanics, the evolution may break down also because $\Phi$ may lose its single-valuedness.
 This type of breakdown is akin to coordinate-induced singularities in general
relativity that do not correspond to singularities in the curvature of spacetime, and it
has to be distinguished from a physical breakdown in finite time.  
 Hard analysis will illuminate the situation in future work. 
  
  We will illustrate the Cauchy problem explicitly with two examples of single-particle dynamics:
the simplest, static one, which already is too much to handle for the Newtonian law of motion
with total electromagnetic fields, and the next simplest one, two oppositely charged point 
particles initially at rest, one of which is treated in Born--Oppenheimer approximation. 
 However, this we defer to section 6, after we have collected some pertinent technical materials. 

%%%%%%%%%%%%%%%%%%%%%%%%%%%%%%%%%%%%%%%%%%%%%%%%%%%%%%%%%%%%%%%%%%%%
%%%%%%%%%%%%%%%%%%%%%%%%%%%%%%%%%%%%%%%%%%%%%%%%%%%%%%%%%%%%%%%%%%%%
	\subsubsection{Permutations and $t$-synchronized natural configurations}
%%%%%%%%%%%%%%%%%%%%%%%%%%%%%%%%%%%%%%%%%%%%%%%%%%%%%%%%%%%%%%%%%%%%
%%%%%%%%%%%%%%%%%%%%%%%%%%%%%%%%%%%%%%%%%%%%%%%%%%%%%%%%%%%%%%%%%%%%

  Since the labeling of the charges within each species of point charges 
is ambiguous, the configuration space of $N$ ordered space-points,
${\times}_{k\in\cN} {\RR}^{3}$, is actually unnatural, though convenient.
  Following 
                 \cite{duerretalC},
the natural configuration space for $N$ identical particles is the space of finite 
subsets of $\RR^3$ with cardinality $N$, i.e. $\RR^{3N}_{\neq} / S_{N}$,
denoted ${}^N\RR^3$. 
 Here, $S_N$ is the symmetric group of $N$ elements.
  The natural configuration space ${}^N\RR^3$ is not simply connected and quite 
naturally leads to the distinction of bosonic and fermionic wave functions in 
quantum mechanics 
                  \cite{duerretalC}.
  Curiously, the \emph{Pauli principle} for bosons
can be implemented {already at the classical level} by more or less similar
reasoning. 
  Thus, realizing $\RR^{3N}_{\neq}$ as the universal covering space for ${}^N\RR^3$, 
in the genuinely electromagnetic setting of the theory in which all point charges
represent electrons, we can impose the $t$-synchronized `bosonic' permutation 
relations on $\RR^{3N}_{\neq}$,
\beq
  \Phi(t,...,\SPvec{s}_k,...,\SPvec{s}_l,...)  
-
 \Phi(t,...,\SPvec{s}_l,...,\SPvec{s}_k,...)  
=
{2\pi Z}
\label{eq:GEObosonPERM}
\eeq
where $Z\in\ZZ$; in fact, one may simply have
\beq
  \Phi(t,...,\SPvec{s}_k,...,\SPvec{s}_l,...)  
-
 \Phi(t,...,\SPvec{s}_l,...,\SPvec{s}_k,...)  
=
{0}.
\label{eq:GEObosonPERMzero}
\eeq
 This symmetry postulate at time $t$ needs to be preserved under the evolution,
which requires that the $^\sharp$fields are symmetric under permutations as 
well. 
  We remark that imposing the Pauli principle 
does not interfere with the gauge freedom: If 
$\Phi$ satisfies the permutation relations \refeq{eq:GEOfermionPERM}
or \refeq{eq:GEObosonPERM}, then so does 
$\Phi\! +\! \alpha \sum_{k\in\cN} \pmk\Upsilon_k$, where $\Upsilon_k$ 
is short for $\Upsilon$ evaluated at $t,\SPvec{s}_k$.

  For fermions it is tempting to conjecture that
\beq
 \Phi(t,...,\SPvec{s}_k,...,\SPvec{s}_l,...)  
-
 \Phi(t,...,\SPvec{s}_l,...,\SPvec{s}_k,...)  
=
(2Z+1){\pi}\,
\label{eq:GEOfermionPERM}
\eeq
where $Z\in\ZZ$.
 However, as James Taylor kindly pointed out to me, this possibility is
spoiled through a subtle topological aspect of a Fermi bundle; we return 
to fermions in the sequel to this paper.

%\newpage

%%%%%%%%%%%%%%%%%%%%%%%%%%%%%%%%%%%%%%%%%%%%%%%%%%%%%%%%%%%%%%%%%%%%
%%%%%%%%%%%%%%%%%%%%%%%%%%%%%%%%%%%%%%%%%%%%%%%%%%%%%%%%%%%%%%%%%%%%
%%%%%%%%%%%%%%%%%%%%%%%%%%%%%%%%%%%%%%%%%%%%%%%%%%%%%%%%%%%%%%%%%%%%
	\subsection{The conservation laws}
%%%%%%%%%%%%%%%%%%%%%%%%%%%%%%%%%%%%%%%%%%%%%%%%%%%%%%%%%%%%%%%%%%%%
%%%%%%%%%%%%%%%%%%%%%%%%%%%%%%%%%%%%%%%%%%%%%%%%%%%%%%%%%%%%%%%%%%%%
%%%%%%%%%%%%%%%%%%%%%%%%%%%%%%%%%%%%%%%%%%%%%%%%%%%%%%%%%%%%%%%%%%%%

  Interestingly, a rigorous study of the conservation laws associated with the gauge invariance and the 
spacetime symmetries of the theory, i.e. the conservation of charge, energy, momentum, angular momentum, 
and moment of energy-momentum, shows that they hold  essentially in the forms anticipated in 
		\cite{BiBiBiBiBOOK}
by symbolic manipulations that pretend the validity of Newton's law of motion with a formal 
Lorentz force.
  The conservation laws reduce to the forms proposed by Born and Infeld 
		\cite{BornInfeldC}
only in the absence of charges.

%%%%%%%%%%%%%%%%%%%%%%%%%%%%%%%%%%%%%%%%%%%%%%%%%%
%%%%%%%%%%%%%%%%%%%%%%%%%%%%%%%%%%%%%%%%%%%%%%%%%%
	\subsubsection{Fields with point charge sources}
%%%%%%%%%%%%%%%%%%%%%%%%%%%%%%%%%%%%%%%%%%%%%%%%%%
%%%%%%%%%%%%%%%%%%%%%%%%%%%%%%%%%%%%%%%%%%%%%%%%%%

 In the presence of point sources the functionals of total
charge,
	energy, 
		momentum, 
			 angular momentum, 
and
				moment of  energy-momentum 
take the following form.

  The functional of the actual total electric charge is given by
\beq
\cQ\left(\Omega^{\mathrm{cl}}\right)  
:= 
\cQ_{\mathrm{field}}\left({\SPvec{D}}\right)  
\,,
\label{Qfunc}
\eeq
where
\beq
\cQ_{\mathrm{field}}\left({\SPvec{D}}\right)  
:= 
\frac{1}{4\pi} \int_{\RR^3} {\nabla}\cdot{{\SPvec{D}}}
\, \dvol({\SPvec{s}}) 
\,
\label{QfuncF}
\eeq
is the functional that counts the signed field defects. 

The functional of the actual total energy is given by
\beq
\cE\left(\Omega^{\mathrm{cl}}\right) 
=
\cE_{\mathrm{field}}\left({\SPvec{B}},{\SPvec{D}}\right) 
+ \sum_{k\in\cN} 
 \sqrt{1 +\abs{{\nabla}_k \Phi(t,{\SPvec{R}})  
-\pmk \alpha{\SPvec{A}_k}\left(t,{\SPvec{R}}\right)}^2}
\label{eq:HfuncTOT}
\,,
\eeq
where 
\beq
\cE_{\mathrm{field}}
\left({\SPvec{B}},{\SPvec{D}}\right) 
= 
\frac{1}{4\pi}
\frac{\alpha}{\beta^4}
\int_{\RR^3}\left(
 \sqrt{   1 
	+ \beta^4(|{\SPvec{B}}|^2 + |{\SPvec{D}}|^2) 
	+ \beta^8|{\SPvec{B}}\times {\SPvec{D}}|^2
       } 
- 1\right)
\dvol({\SPvec{s}})
\,
\label{eq:HfuncFIELDS}
\eeq
is the functional of the field energy.  
  At the same time, the actual field-energy functional is the 
field Hamiltonian for the conjugate field variables ${\SPvec{B}}$
and ${\SPvec{D}}$; to emphasize this we shall sometimes use the notation
$\cH_{\mathrm{field}}\left({\SPvec{B}},{\SPvec{D}}\right)$ 
instead of $\cE_{\mathrm{field}}\left({\SPvec{B}},{\SPvec{D}}\right)$.

	The functional of the actual total momentum of electromagnetic field 
plus defects reads
\beq
   \cP\left(\Omega^{\mathrm{cl}}\right) 
=
\cP_{\mathrm{field}}\left({\SPvec{B}},{\SPvec{D}}\right)  
+ 
\sum_{k\in\cN} 
 \left(\nabla_k\Phi\left(t,{\SPvec{R}}\right) 
-\pmk \alpha{\SPvec{A}_k}\left(t,{\SPvec{R}}\right)\right)
\, , 
\label{PfuncTOT}
\eeq
where 
\beq
\cP_{\mathrm{field}}\left({\SPvec{B}},{\SPvec{D}}\right) 
= 
	\frac{\alpha}{4\pi} \int_{\RR^3} {{\SPvec{D}}}\times{{\SPvec{B}}}
\, \dvol({\SPvec{s}}) 
\, 
\label{PfuncFIELDS}
\eeq
is the functional of the  field momentum.

The functional of the actual total angular momentum is given by
\beq
\cJ\left(\Omega^{\mathrm{cl}}\right) 
=
\cJ_{\mathrm{field}}\left({\SPvec{B}},{\SPvec{D}}\right) 
 + 
\sum_{k\in\cN} 
{\SPvec{r}}_k\times
 \left(\nabla_k\Phi\left(t,{\SPvec{R}}\right) 
-\pmk \alpha{\SPvec{A}_k}\left(t,{\SPvec{R}}\right)\right)
\, , \label{JfuncTOT}
\eeq
where 
\beq
\cJ_{\mathrm{field}}\left({\SPvec{B}},{\SPvec{D}}\right) 
= 
\frac{\alpha}{4\pi}
\int_{\RR^3}{{\SPvec{s}}}\times({{\SPvec{D}}}\times{{\SPvec{B}}})\,\dvol({\SPvec{s}}) 
\, , \label{JfuncFIELDS}
\eeq
is the functional of the  field angular momentum.

  The functional of the actual moment of the total energy is given by
\beq
\cM\left(\Omega^{\mathrm{cl}}\right) 
=
\cM_{\mathrm{field}}\left({\SPvec{B}},{\SPvec{D}}\right) 
+ \sum_{k\in\cN} 
 \sqrt{1 +\abs{{\nabla}_k \Phi(t,{\SPvec{R}})  
-\pmk \alpha{\SPvec{A}_k}\left(t,{\SPvec{R}}\right)}^2}
\, {\SPvec{r}}_k
\label{MofHfuncTOT}
\,,
\eeq
where
\beq
\cM_{\mathrm{field}}\left({\SPvec{B}},{\SPvec{D}}\right) 
= 
\frac{1}{4\pi}
\frac{\alpha}{\beta^4}
\int_{\RR^3}
\left(
 \sqrt{   1 
	+ \beta^4(|{\SPvec{B}}|^2 + |{\SPvec{D}}|^2) 
	+ \beta^8|{\SPvec{B}}\times {\SPvec{D}}|^2
       } 
	- 1
\right)\!{\SPvec{s}}\,
\dvol({\SPvec{s}})
\label{MofHfuncFIELDS}
\eeq
is the functional of the moment of the field energy.

 The conservation laws in the presence of point sources are 
collected in 

\begin{Prop} \label{propClawsMBI} 
	\textit{For $t\in (0,T)$, let 
		$t\mapsto \Omega^{\mathrm{cl}}(t)\in \Gamma^{\mathrm{cl}}$ 
		be a regular solution of the Cauchy problem with point
		sources, satisfying $\lim_{t\downarrow 0}  \Omega^{\mathrm{cl}}(t) = \Omega^{\mathrm{cl}}(0)$
		for prescribed initial data $\Omega^{\mathrm{cl}}(0)\in\Gamma^{\mathrm{cl}}$.
		Then the conventional physical conservation laws for
		total charge, energy, momentum, and angular momentum
		are satisfied, i.e. }
\bea
\cQ\left(\Omega^{\mathrm{cl}}(t)\right) 
\!\!\!  \!\!\!  \!\!\! && 
=  
Q
\,,
\label{QfuncCOM}
\\
\cE\left(\Omega^{\mathrm{cl}}(t)\right) 
\!\!\!  \!\!\!  \!\!\! &&
 = 
E
\,,
\label{EPSconstE}
\\
 \cP\left(\Omega^{\mathrm{cl}}(t)\right) 
\!\!\!  \!\!\!  \!\!\! && 
= 
{\SPvec{P}}
\,,
\label{EPSconstP}
\\
\cJ\left(\Omega^{\mathrm{cl}}(t)\right) 
\!\!\!  \!\!\!  \!\!\! && 
 = 
{\SPvec{J}}
\,,
\label{EPSconstJ}
\eea
\textit{with  $Q$, $E$, ${\SPvec{P}}$, and ${\SPvec{J}}$
independent of time.}

	\textit{Furthermore, the moment of total energy and the total
momentum are related by}
\beq
\cM\left(\Omega^{\mathrm{cl}}(t)\right) 
 -
t\cP\left(\Omega^{\mathrm{cl}}(t)\right) 
 = {\SPvec{M}}
\,,
\label{MofHminusPtGLEICHconstX}
\eeq
	\textit{with ${\SPvec{M}} = \cM\left(\Omega^{\mathrm{cl}}(0)\right)$ independent of time.}
\end{Prop}

\noindent
	\textit{Sketch of proof of Proposition \ref{propClawsMBI}.} 
	The law of the charge conservation holds because 
the continuity equation for the point charge and current 
densities holds in lieu of their definitions, and 
since ${\nabla}\cdot{{\SPvec{D}}}= 4{\pi}j$ by Coulomb's law \refeq{eq:FOLIcD}.

  To prove the other conservation laws, regularize $A$ and $\SPvec{A}$ by convolution
with a differentiable compactly supported probability density having
spherical symmetry in foliation space.
  The conservation laws for the regularized Hamilton--Jacobi dynamics can then be 
proved by differentiation in a similar straightforward fashion as done in 
        \cite{KiePLA}
for the old Abraham--Lorentz model.
 Since the limit as the regularizer concentrates on a point is absolute in the
Hamilton--Jacobi formulation, not merely conditional as in the Newtonian formulation,
the conservation laws for the point charge model follow.
\QED
\smallskip

  One can easily show that the conserved quantities 
	total energy, 
		total momentum, 
			total angular momentum, 
and the 
			 moment of total energy-momentum 
are the generators of the time translations, space translations, space rotations, 
and spacetime boosts.
 The total charge generates the gauge transformations.
 As a general reference to symmetries and conservation laws, see
      \cite{SudarshanBOOK}.

%%%%%%%%%%%%%%%%%%%%%%%%%%%%%%%%%%%%%%%%%%%%%%%%%%%%%%%%%%%%%%%%%%%%
%%%%%%%%%%%%%%%%%%%%%%%%%%%%%%%%%%%%%%%%%%%%%%%%%%%%%%%%%%%%%%%%%%%%
%%%%%%%%%%%%%%%%%%%%%%%%%%%%%%%%%%%%%%%%%%%%%%%%%%%%%%%%%%%%%%%%%%%%
	\subsubsection{Source-free fields}
%%%%%%%%%%%%%%%%%%%%%%%%%%%%%%%%%%%%%%%%%%%%%%%%%%%%%%%%%%%%%%%%%%%%
%%%%%%%%%%%%%%%%%%%%%%%%%%%%%%%%%%%%%%%%%%%%%%%%%%%%%%%%%%%%%%%%%%%%
%%%%%%%%%%%%%%%%%%%%%%%%%%%%%%%%%%%%%%%%%%%%%%%%%%%%%%%%%%%%%%%%%%%%

  In the absence of charges, when the evolution equations are reduced 
to the Maxwell--Born--Infeld field equations without source terms,
the conservation laws  of proposition \ref{propClawsMBI} reduce to their
obvious corollaries.
  In addition, as is the case with the vacuum Maxwell equations,
the source-free Maxwell--Born--Infeld field equations and their Hamiltonian $\cH_{\mathrm{field}}$ 
are also invariant under electric-magnetic duality transformations 
(generalized Hodge duality rotations; ``$\gamma$ transformations'' in 
                \cite{ErwinEiHBiD}), 
which implies a conservation law for the sum of the electric and magnetic field helicities, 
					 cf. \cite{BiBiONE}.
 These are defined as
\beq
	\cY_{\mathrm{field}}\left({\SPvec{B}}\right) 
=
{{\frac{1}{8\pi}}}\int_{\RR^3\times\RR^3}
\!\!
\frac{
{\SPvec{B}}(\hat{\SPvec{s}}) \cdot{\nabla}\times{{\SPvec{B}}}(\check{\SPvec{s}})}
	{|\hat{\SPvec{s}}-\check{\SPvec{s}}| }
\dvol(\hat{\SPvec{s}}) \,\dvol(\check{\SPvec{s}}) 
\eeq
and
\beq
	\cY_{\mathrm{field}}\left({\SPvec{D}}\right) 
=
{{\frac{1}{8\pi}}}\int_{\RR^3\times\RR^3}
\!\!
\frac{
{\SPvec{D}}(\hat{\SPvec{s}})\cdot{\nabla}\times{{\SPvec{D}}}(\check{\SPvec{s}})}
	{|\hat{\SPvec{s}}-\check{\SPvec{s}}|}
\dvol(\hat{\SPvec{s}}) \,\dvol(\check{\SPvec{s}}) .
\eeq

We summarize the conservation laws for the source-free evolution. 

\begin{Prop} \label{propClawsMBIvac} 
	\textit{For\, $t\in (0,T)$, let the pair of maps
		$t\mapsto {\SPvec{B}}(t,\,.\,)$ 
		and 
		$t\mapsto{\SPvec{D}}(t,\,.\,)$ 
		be a classical solution of the source-free
		Maxwell--Born--Infeld field equations, satisfying
$\lim_{t\downarrow 0}  {\SPvec{B}}(t,\,.\,) = {\SPvec{B}}(0,\,.\,)$
		and 
$\lim_{t\downarrow 0} {\SPvec{D}}(t,\,.\,) = {\SPvec{D}}(0,\,.\,)$
		for the	prescribed initial data 
		${\SPvec{B}}(0,\,.\,)$ 
		and
		${\SPvec{D}}(0,\,.\,)$.
		Then the conventional physical conservation laws for
		the charge, field energy, field momentum, and field angular momentum
		are satisfied, i.e. }
\bea
\cQ_{\mathrm{field}}\Bigl({\SPvec{D}}(t,.)\Bigr)  = 0
\,,
\qquad
&&  
\label{EPSconstQfields}
\\
\cH_{\mathrm{field}}\Bigl({\SPvec{B}}(t,.),{\SPvec{D}}(t,.)\Bigr) = E_{\mathrm{field}}
\,,
&&
\label{EPSconstEfields}
\\
 \cP_{\mathrm{field}}\Bigl({\SPvec{B}}(t,.),{\SPvec{D}}(t,.)\Bigr) = {\SPvec{P}_{\mathrm{field}}}
\,,
&&
\label{EPSconstPfields}
\\
\cJ_{\mathrm{field}}\Bigl({\SPvec{B}}(t,.),{\SPvec{D}}(t,.)\Bigr) = {\SPvec{J}_{\mathrm{field}}}
\,,
&&
\label{EPSconstJfields}
\eea
\textit{with $E_{\mathrm{field}}$, ${\SPvec{P}_{\mathrm{field}}}$, and ${\SPvec{J}_{\mathrm{field}}}$
independent of time.}

	\textit{In addition, the sum of electric and magnetic field helicities is conserved, i.e.}
\beq
\cY_{\mathrm{field}}\Bigl({\SPvec{B}}(t,.)\Bigr) 
+
\cY_{\mathrm{field}}\Bigl({\SPvec{D}}(t,.)\Bigr) 
= 
Y_{\mathrm{field}}
\,,
\label{BiBiconstYfields}
\eeq
	\textit{with $Y_{\mathrm{field}}$ independent of time.}

	\textit{Also, the moment of the field energy and the 
field momentum are related by}
\beq
\bigl(\cM_{\mathrm{field}}
-
t\cP_{\mathrm{field}}\bigr) \Bigl({\SPvec{B}}(t,.),{\SPvec{D}}(t,.)\Bigr) 
= 
{\SPvec{M}_{\mathrm{field}}}
\,,
\label{MofHminusPtGLEICHconstXfields}
\eeq
	\textit{with ${\SPvec{M}_{\mathrm{field}}}$ independent of time.}
\end{Prop}

\noindent
	\textit{Sketch of proof of Proposition \ref{propClawsMBIvac}.}
	Proposition \ref{propClawsMBIvac} is largely
a special case of our proof of Proposition \ref{propClawsMBI}, except
for the proof of the conservation law for the electromagnetic field helicity,
which can be done as yet another exercise in vector analysis.  

   An alternate proof, using the generators of the various symmetries 
associated with the conservation laws (except helicity), can be found in 
		\cite{BiBiBiBiBOOK}, 
pp. 92--95; the corresponding proof of the conservation law for the 
electromagnetic field helicity (though it is not called by that name) can 
be found in 
		\cite{BiBiONE}, 
pp. 41/42.
\QED

  We remark that in the two limiting cases $\beta\to 0$ and $\beta\to \infty$,
both of which yield field equations that are invariant under the full conformal 
group on Minkowski space,
further conservation laws of the source-free field dynamics emerge
		\cite{BiBiONE}.
 These additional conservation laws concern electric and magnetic field
helicities separately, as well as an electromagnetic cross-field helicity, 
\beq
	\cX_{\mathrm{field}}\left({\SPvec{B}},{\SPvec{D}}\right) 
=
{{\frac{1}{4\pi}}}\int_{\RR^3\times\RR^3}
\!\!
\frac{
{\SPvec{B}}(\hat{\SPvec{s}}) \cdot{\nabla}\times{{\SPvec{D}}}(\check{\SPvec{s}})}
	{|\hat{\SPvec{s}}-\check{\SPvec{s}}| }
\! 
\dvol(\hat{\SPvec{s}}) \,\dvol(\check{\SPvec{s}}) .
\eeq
%%%%%%%%%%%%%%%%%%%%%%%%%%%%%%%%%%%%%%%%%%%%%%%%%%%%%%%%%%%%%%%%%%%%
%%%%%%%%%%%%%%%%%%%%%%%%%%%%%%%%%%%%%%%%%%%%%%%%
%%%%%%%%%%%%%%%%%%%%%%%%%%%%%%%%%%%%%%%%%%%%%%%%
%%%%%%%%%%%%%%%%%%%%%%%%%%%%%%%%%%%%%%%%%%%%%%%%%%%%%%%%%%%%%%%%%%%%
     \section{Special types of solutions of the Maxwell--Born--Infeld field equations with point sources}
%%%%%%%%%%%%%%%%%%%%%%%%%%%%%%%%%%%%%%%%%%%%%%%%%%%%%%%%%%%%%%%%%%%%
%%%%%%%%%%%%%%%%%%%%%%%%%%%%%%%%%%%%%%%%%%%%%%%%
%%%%%%%%%%%%%%%%%%%%%%%%%%%%%%%%%%%%%%%%%%%%%%%%
%%%%%%%%%%%%%%%%%%%%%%%%%%%%%%%%%%%%%%%%%%%%%%%%%%%%%%%%%%%%%%%%%%%%
   We now list some physically relevant facts about 
special solutions of the electromagnetic 
Maxwell--Born--Infeld field equations with 
$\beta\in(0,\infty)$.

%%%%%%%%%%%%%%%%%%%%%%%%%%%%%%%%%%%%%%%%%%%%%%%%%%%%%%%%%%%%%%%%%%%%%%%%%%%%%%%%%%%%%%%%%%%%%%%%
%%%%%%%%%%%%%%%%%%%%%%%%%%%%%%%%%%%%%%%%%%%%%%%%
	\subsection{On the existence and uniqueness of electrostatic solutions}
%%%%%%%%%%%%%%%%%%%%%%%%%%%%%%%%%%%%%%%%%%%%%%%%
%%%%%%%%%%%%%%%%%%%%%%%%%%%%%%%%%%%%%%%%%%%%%%%%%%%%%%%%%%%%%%%%%%%%%%%%%%%%%%%%%%%%%%%%%%%%%%%%

  Everything that is rigorously known about solutions with point 
sources refers to solutions which are static in some Lorentz frame
after at most a Lorentz boost.
  Therefore it suffices to discuss the static versions.

\smallskip
%%%%%%%%%%%%%%%%%%%%%%%%%%%%%%%%%%%%%%%%%%%%%%%%%%%%%%%%%%%%%%%%%%%%%%%%%%%%%%%%%%%%%%%%%%%%%%%%
%%%%%%%%%%%%%%%%%%%%%%%%%%%%%%%%%%%%%%%%%%%%%%%%
	\subsubsection{The electric field of $N$ unit point charges}
%%%%%%%%%%%%%%%%%%%%%%%%%%%%%%%%%%%%%%%%%%%%%%%%
%%%%%%%%%%%%%%%%%%%%%%%%%%%%%%%%%%%%%%%%%%%%%%%%%%%%%%%%%%%%%%%%%%%%%%%%%%%%%%%%%%%%%%%%%%%%%%%%

 For an electrostatic solution, $\SPvec{B} = \SPvec{0} = \SPvec{H}$, while 
the electrostatic field ${\SPvec{s}}\mapsto {\SPvec{E}}$ and the electric 
displacement field ${\SPvec{s}}\mapsto{\SPvec{D}}$ satisfy the electrostatic 
Maxwell equations
\beq
	 {\nabla}\times {\SPvec{E}}
=
        {\SPvec{0}}
\, ,
\qquad\qquad
\label{eq:FOLIeBstatic} 
\eeq
and
\beq
        {\nabla}\cdot {\SPvec{D}}
=
        4 \pi \sum_{k\in\cN}\pmk\delta_{{\SPvec{s}}_{k}}
\,
\label{eq:FOLIcDstatic}
\eeq
in an aether governed by the electrostatic Born and Infeld law
\beq
{\SPvec{E}} 
= 
\frac{
{\SPvec{D}}
}{
\sqrt{    1 + \beta^4|{\SPvec{D}}|^2
      }}
\label{eq:FOLIeqEofDstatic}
\, .
\eeq

\begin{Prop} 
\label{staticNcharges}
\textit{A unique finite-energy solution to the static Maxwell--Born--Infeld field equations 
\refeq{eq:FOLIeBstatic}, \refeq{eq:FOLIcDstatic}, \refeq{eq:FOLIeqEofDstatic}
with $N\in\NN$ fixed unit point charges as sources 
exists whenever there exists a maximal space-like slice in Minkowski space
with $N$ null-like point defects at the locations of the charges.}
\end{Prop} 

\noindent
\textit{Sketch of proof of Proposition \ref{staticNcharges}.}
 To have finite energy $\cE\left(\Omega^{\mathrm{cl}}_{\mathrm{static}}\right)  = \widehat\cE\left({\SPvec{D}}\right)$,
\beq
\widehat\cE\left({\SPvec{D}}\right) 
= 
N + \frac{1}{4\pi}
\frac{\alpha}{\beta^4}
\int_{\RR^3}\left(
 \sqrt{   1 
	+ \beta^4|{\SPvec{D}}|^2
       } 
- 1\right)
\dvol({\SPvec{s}})
\,,
\label{eq:HfuncFIELDSelectroSTATIC}
\eeq
we need to impose the asymptotic  condition that
${\SPvec{D}}({\SPvec{s}})\to {\SPvec{0}}$ as $|{\SPvec{s}}|\to\infty$, 
which is inherited by ${\SPvec{E}}$. 
   Then \refeq{eq:FOLIeBstatic} is satisfied identically if there exists an electrostatic
scalar potential ${\SPvec{s}}\mapsto{A}$, vanishing for $|{\SPvec{s}}|\to\infty$, such that 
\beq
{\SPvec{E}} ({\SPvec{s}})
=
-\nabla A({\SPvec{s}})
\eeq
for all ${\SPvec{s}}\neq {\SPvec{s}}_k$, $k\in\cN$.
  We next invert the electrostatic Born and Infeld law of the aether 
\refeq{eq:FOLIeqEofDstatic} to expresses $\SPvec{D}$ explicitly in 
terms of $\nabla{A}$, viz.
\beq
\SPvec{D}
= 
- 
\frac{ \nabla{A} }{ \sqrt{ 1 - \beta^4| \nabla{A}|^2 }}
\label{eq:FOLIeqAofZstaticINV} 
\,. 
\eeq 
 Taking now the divergence of \refeq{eq:FOLIeqAofZstaticINV} and noting 
\refeq{eq:FOLIcDstatic} gives
\beq
-\nabla \cdot
\frac{ \nabla{A} }{ \sqrt{ 1 - \beta^4| \nabla{A}|^2 }}
= 
   4 \pi \sum_{k\in\cN}\pmk\delta_{{\SPvec{s}}_{k}}
\label{eq:FOLIeqAstatic}
\, ,
\eeq 
together with the asymptotic condition that ${A}({\SPvec{s}})\to 0$ 
for $|{\SPvec{s}}|\to\infty$.

 All the hard work has been absorbed in the single, nonlinear, second-order
partial differential equation \refeq{eq:FOLIeqAstatic} for $A$.
 But \refeq{eq:FOLIeqAstatic} is the Euler--Lagrange equation for
\beq
\int_{\RR^3}
           \left(\sqrt{ 1 - \beta^4|{\nabla{A}}|^2 } - 1 \right)
\dvol({\SPvec{s}})
+ 
{4\pi}{\beta^4}
\sum_{k\in\cN}\pmk{A}(\SPvec{s}_k)\quad =\quad 
\mathrm{maximum}
\,,
\label{eq:HYPERSURFACEfuncSTATIC}
\eeq
for $A\in C^{0,1}_0(\RR^3)\cap C^1(\RR^3\backslash\{\SPvec{s}_k\}_{k=1}^{N})$
with Lip$(A)=\beta^{-2}$.
 With $\beta^2A$ interpreted as the time function $T$ of the subsection 3.1.4,
with lapse function $\ell = 1$, see \refeq{eq:lapseMETRIC}, the
variational principle \refeq{eq:HYPERSURFACEfuncSTATIC} characterizes a 
maximal space-like slice in Minkowski space with $N$ null-like point defects 
at the locations of the charges, and \refeq{eq:FOLIeqAstatic} states that the
mean curvature of that space-like slice vanishes except at the defects.
  This concludes the existence part of the proposition. 
  We remark that theorem 5.4, and Remark 1 thereafter, of
                \cite{bartnik}
seems to guarantee the existence of the solutions of \refeq{eq:HYPERSURFACEfuncSTATIC};
see also \cite{KlyMik} and \cite{Kly} for results in certain domains with boundary.

 As for the uniqueness of an electrostatic solution, it suffices to discuss 
uniqueness of an asymptotically (for $|\SPvec{s}|\to\infty$) vanishing solution 
$A$ of \refeq{eq:FOLIeqAstatic} for any given configuration of the $N$ charges.
 The argument is based on convexity, entirely standard, and a little shorter 
than Pryce's proof     
   \cite{PryceB}.
 Thus assume that two distinct solutions $A_1\not\equiv A_0$ of 
\refeq{eq:FOLIeqAstatic} exist for a given configuration of the $N$ charges.
 We subtract \refeq{eq:FOLIeqAstatic} for $A_0$ from \refeq{eq:FOLIeqAstatic} for $A_1$,
multiply by $A_1-A_0$, integrate over $\RR^3$, use integration by parts and obtain
the identity
\beq
0 =  \int_{\RR^3} \nabla(A_1-A_0)\cdot 
\int_0^1\frac{\dd}{\dd{u}}\frac{\nabla A_u}{\sqrt{ 1 - \beta^4|{\nabla{A_u}}|^2 }}\,\dd{u}\,\dvol(\SPvec{s})
\eeq
where $A_u = u A_1 + (1-u)A_0$. 
 Now exchange the volume and the $u$ integrations, carry out the indicated $u$-differentiation of the integrand, 
and obtain a manifestly positive definite integral on the r.h.s. which vanishes iff $A_1-A_0\equiv 0$. \QED

%%%%%%%%%%%%%%%%%%%%%%%%%%%%%%%%%%%%%%%%%%%%%%%%%%%%%%%%%%%%%%%%%%%%%%%%%%%%%%%%%%%%%%%%%%%%%%%%
%%%%%%%%%%%%%%%%%%%%%%%%%%%%%%%%%%%%%%%%%%%%%%%%
	\subsection{Special electrostatic solutions}
%%%%%%%%%%%%%%%%%%%%%%%%%%%%%%%%%%%%%%%%%%%%%%%%
%%%%%%%%%%%%%%%%%%%%%%%%%%%%%%%%%%%%%%%%%%%%%%%%%%%%%%%%%%%%%%%%%%%%%%%%%%%%%%%%%%%%%%%%%%%%%%%%

%%%%%%%%%%%%%%%%%%%%%%%%%%%%%%%%%%%%%%%%%%%%%%%%
%%%%%%%%%%%%%%%%%%%%%%%%%%%%%%%%%%%%%%%%%%%%%%%%
\subsubsection{Born's solution (The electric field of a single unit point charge)}
%%%%%%%%%%%%%%%%%%%%%%%%%%%%%%%%%%%%%%%%%%%%%%%%
%%%%%%%%%%%%%%%%%%%%%%%%%%%%%%%%%%%%%%%%%%%%%%%%

 The static Maxwell--Born--Infeld equations with a single 
point charge as source can be solved explicitly, as already announced in
		\cite{BornA}
and elaborated on further in
		\cite{BornInfeldA, BornB, BornInfeldB}.
	We identify the origin of space with the location of the charge at rest.
	The electric displacement field of Born's solution is 
identical to the Coulomb field of a point charge,
\beq
{\SPvec{D}}_{\mathrm{Born}}^\plumi\bigl(\SPvec{s}\bigr) 
=
{\SPvec{D}}_{\mathrm{Coulomb}}^\plumi\bigl(\SPvec{s}\bigr) 
\equiv
\plumi \frac{\SPvec{s}}{|\SPvec{s}|^3}
\,,
\eeq
where $\plumi$ indicates a positive or negative electron.
   The associated electric field ${\SPvec{E}}_{\mathrm{Born}}^\plumi$
is bounded, but undefined at the origin.
   For $\SPvec{s}\neq \SPvec{0}$, 
${\SPvec{E}}_{\mathrm{Born}}^\plumi\bigl(\SPvec{s}\bigr) = -\nabla{A}_{\mathrm{Born}}^\plumi({\SPvec{s}})$,
where
\beq
A_{\mathrm{Born}}^\plumi({\SPvec{s}}) 
=  
\plumi \frac{1}{\beta}
\int_{|{\small\SPvec{s}}|/\beta}^\infty \frac{\dd{x}}{\sqrt{1+ x^4}} 
\,.
\label{eq:BornsElectricPot}
\eeq
 This spherically symmetric electrostatic potential is asymptotic to Coulomb's 
potential, $A_{\mathrm{Born}}^\plumi({\SPvec{s}}) \sim \plumi |\SPvec{s}|^{-1}$ for $|\SPvec{s}|\gg\beta$,
and $\lim_{|\SPvec{s}|\to{0}}A_{\mathrm{Born}}^\plumi(\SPvec{s}) = A_{\mathrm{Born}}^\plumi(\SPvec{0})<\infty$.

 In
                \cite{PryceB}
the uniqueness of \refeq{eq:BornsElectricPot}
under the condition of finite field energy is shown,
by an argument similar in spirit to the one we used above.
  Ecker 
                   \cite{ecker}
showed that, interpreted as the time function of a maximal space-like slice with defect,
Born's solution is the unique solution to \refeq{eq:HYPERSURFACEfuncSTATIC}
among all asymptotically flat space-like slices with any single isolated singularity; 
note that the singularity of Born's solution is a 
so-called (light)cone singularity in the space-like slice interpretation.

\smallskip
%%%%%%%%%%%%%%%%%%%%%%%%%%%%%%%%%%%%%%%%%%%%%%%%%%%%%%%%%%%%%%%%%%%%%%%%%%%%%%%%%%%%%%%%%%%%%%%%
%%%%%%%%%%%%%%%%%%%%%%%%%%%%%%%%%%%%%%%%%%%%%%%%%%%%%%%%%%%%%%%%%%%%%%%%%%%%%%%%%%%%%%%%%%%%%%%%
	\subsubsection{Hoppe's solution (The electric field of an infinite crystal of unit point charges)}
%%%%%%%%%%%%%%%%%%%%%%%%%%%%%%%%%%%%%%%%%%%%%%%%%%%%%%%%%%%%%%%%%%%%%%%%%%%%%%%%%%%%%%%%%%%%%%%%
%%%%%%%%%%%%%%%%%%%%%%%%%%%%%%%%%%%%%%%%%%%%%%%%%%%%%%%%%%%%%%%%%%%%%%%%%%%%%%%%%%%%%%%%%%%%%%%%

  If we allow $N \to \infty$ and relax the condition of finite energy, another
exact many-body solution of \refeq{eq:FOLIeqAstatic} for $\beta\in(0,\infty)$ becomes
available, which was discovered by Hoppe 
             \cite{HoppeA, HoppeB}.
  It is defined implicitly by 
\beq
\wp\left(A_{\mathrm{Hoppe}}(s^1,s^2,s^3)\right) = \wp(s^1)\wp(s^2)\wp(s^3)
\eeq
where $\wp$ is the elliptic Weierstrass 
function.\footnote{Recall that 
            $\zeta\mapsto\wp(\zeta)$ satisfies the nonlinear ODE
            $
	    {\wp^{\prime}}^2 = 4 \wp^3 - g_2\wp - g_3
	    $
	    with invariants
	    $
	    g_2 
	    = 
	    60
	    \sum_{(n,m)\in \ZZ^2_*} 
	    (n2\omega + m2\omega^\prime)^{-4}
	    $
	    and
	    $
            g_3
	    = 
	    140
	    \sum_{(n,m)\in \ZZ^2_*} (n2\omega + m2\omega^\prime)^{-6}
	    $,
	    where $\ZZ^2_*=\ZZ^2\backslash (0,0)$, and $\omega$ and $\omega^\prime$ 
	    are the real half-periods (note the bizarre notation is standard). 
	    $\wp$ has a pole of order two at the origin; more precisely,
	    $\wp(\zeta) - \zeta^{-2}$ is analytic in a neighborhood of the origin.}
    This solution describes an infinite `electron-positron crystal' with NaCl symmetry. 
    For more, see  the review 
         \cite{GibbonsA}.
%%%%%%%%%%%%%%%%%%%%%%%%%%%%%%%%%%%%%%%%%%%%%%%%%%%%%%%%%%%%%%%%%%%%
%%%%%%%%%%%%%%%%%%%%%%%%%%%%%%%%%%%%%%%%%%%%%%%%%%%%%%%%%%%%%%%%%%%%
%%%%%%%%%%%%%%%%%%%%%%%%%%%%%%%%%%%%%%%%%%%%%%%%%%%%%%%%%%%%%%%%%%%%
%%%%%%%%%%%%%%%%%%%%%%%%%%%%%%%%%%%%%%%%%%%%%%%%%%%%%%%%%%%%%%%%%%%%
  \section{Illustrations of the electromagnetic Cauchy problem with point charges}
%%%%%%%%%%%%%%%%%%%%%%%%%%%%%%%%%%%%%%%%%%%%%%%%%%%%%%%%%%%%%%%%%%%%
%%%%%%%%%%%%%%%%%%%%%%%%%%%%%%%%%%%%%%%%%%%%%%%%%%%%%%%%%%%%%%%%%%%%
%%%%%%%%%%%%%%%%%%%%%%%%%%%%%%%%%%%%%%%%%%%%%%%%%%%%%%%%%%%%%%%%%%%%
%%%%%%%%%%%%%%%%%%%%%%%%%%%%%%%%%%%%%%%%%%%%%%%%%%%%%%%%%%%%%%%%%%%%
 We finally have collected all the relevant pieces of information that we
need for explicitly illustrating the setup of the electromagnetic Cauchy problem 
for fields and point charges. 
 It suffices to consider fields coupled to a single electron.

%%%%%%%%%%%%%%%%%%%%%%%%%%%%%%%%%%%%%%%%%%%%%%%%%%%%%%%%%%%%%%%%%%%%
%%%%%%%%%%%%%%%%%%%%%%%%%%%%%%%%%%%%%%%%%%%%%%%%%%%%%%%%%%%%%%%%%%%%
  \subsection{A single electron at rest surrounded by its own electrostatic field}
%%%%%%%%%%%%%%%%%%%%%%%%%%%%%%%%%%%%%%%%%%%%%%%%%%%%%%%%%%%%%%%%%%%%
%%%%%%%%%%%%%%%%%%%%%%%%%%%%%%%%%%%%%%%%%%%%%%%%%%%%%%%%%%%%%%%%%%%%

 We begin with the simplest possible case: an electron initially at rest somewhere
in space, surrounded only by its own electrostatic field. 
 Even this simplest one-particle case is already too much to handle for the Newtonian law of motion
with the total Lorentz force, while for our Hamilton--Jacobi law of motion it is trivial.

 Clearly, the initial $\SPvec{A}$ vanishes identically, so we can take 
the initial $\SPvec{A}^\sharp$ to vanish identically too.
 Also, the initial $\SPvec{D}$ is just the Coulomb field of the point charge,
with the corresponding electrostatic potential field in space given by 
Born's solution
       \refeq{eq:BornsElectricPot}
if the electron is at the origin of space, and by a
translate of it if the electron is elsewhere. 
 This fixes the initial $\SPvec{D}^\sharp$ and ${A}^\sharp$; the 
latter is the corresponding translate in space by $\SPvec{s}_1$ of
        \refeq{eq:BornsElectricPot}.
 Lastly, since the particle is initially at rest no matter where it is,
we take $\SPvec{V}(0,\SPvec{s}_1) \equiv \SPvec{0}$, and
$\Phi(0,\SPvec{s}_1) \equiv 0$.

 It is now easy to see that these data imply that 
$A^\sharp(0, \SPvec{s}_1, \SPvec{s}_1) 
= A_{\mathrm{Born}}^{(-)}({\SPvec{0}})$ for all $\SPvec{s}_1$.
 Therefore the Hamilton--Jacobi PDE for $\Phi$ initially reads
$\partial \Phi =K$, a constant function in $\RR^3_1$, while 
$\partial \SPvec{X}^\sharp =\SPvec{0}$ for all $^\sharp$fields $\SPvec{X}^\sharp$.
 Hence, the $^\sharp$fields initial data form a stationary solution 
of the $^\sharp$field equations, while the Hamilton--Jacobi equation
is solved by $\Phi(t,\SPvec{s}_1) = Kt$. 
 This gives $\SPvec{V}(t,\SPvec{s}_1) \equiv \SPvec{0}$ for all time, 
which implies that the actual particle position and, hence, that the 
actual electromagnetic fields and their potentials on space, 
all retain their initial data for all time, as they should. 
 
 We take the opportunity to re-enforce what we emphasized earlier in section 3,
namely that $\Phi$ should not be interpreted as a field on physical space. 
  Indeed, if instead of for a field $\Phi$ on configuration space we had misinterpreted 
the Hamilton--Jacobi PDE as a scalar equation for a field $\phi$ on physical space, say 
with the particle initially at the origin, then initially and near the origin of space 
that field equation would read qualitatively like $\partial \phi = b - |\SPvec{s}|$. 
 Now $\phi$ too would immediately develop a kink, as function of space, at the 
location of the point charge, and the guiding equation would become as meaningless 
as the Newtonian equation of motion. 
 The same remark applies verbatim to the upgraded test particles Hamilton--Jacobi 
theory.
 
%%%%%%%%%%%%%%%%%%%%%%%%%%%%%%%%%%%%%%%%%%%%%%%%%%%%%%%%%%%%%%%%%%%%
%%%%%%%%%%%%%%%%%%%%%%%%%%%%%%%%%%%%%%%%%%%%%%%%%%%%%%%%%%%%%%%%%%%%
  \subsection{A single electron and an oppositely charged, infinitely massive nucleus, both 
   initially at rest in the total electrostatic field}
%%%%%%%%%%%%%%%%%%%%%%%%%%%%%%%%%%%%%%%%%%%%%%%%%%%%%%%%%%%%%%%%%%%%
%%%%%%%%%%%%%%%%%%%%%%%%%%%%%%%%%%%%%%%%%%%%%%%%%%%%%%%%%%%%%%%%%%%%

 To have another, less trivial example, consider a single negative electron 
initially at rest at $\SPvec{r}_1(0)= \SPvec{r}_0$ and an infinitely massive positive 
unit point charge at rest at the origin of space, so that we are allowed 
to restrict our discussion of configuration space to the single electron space. 
 The initial $\Phi(0,\SPvec{s}_1)$ on the single electron configuration space
is a constant, $\Phi_0$ say.
  The initial electromagnetic fields are electrostatic, the field
generated by the two point charges, for which existence we once again 
invoke Bartnik's results (see above).
 The time derivative of these initial fields in space vanishes initially. 
 Again, we take the initial magnetic potential to vanish identically, and
do the same for its magnetic $^\sharp$cousin.
 Unfortunately, to set up the initial value problem for the $^\sharp$fields
\emph{explicitly}, we need to compute the initial total electrostatic 
displacement field and potential on space - configuration space, 
$\SPvec{D}^\sharp(0,\SPvec{s},\SPvec{s}_1)$ and 
$A^\sharp(0,\SPvec{s},\SPvec{s}_1)$, for which we need to solve the 
electrostatic Maxwell--Born--Infeld equations with two-point sources, 
and which no one has been able to do, as far as we know.
  However, to set up the initial value problem for $\Phi(t,\SPvec{s}_1)$,
we only need to know the initial  
$A_1(0,\SPvec{s}_1) 
= A^\sharp(0,\SPvec{s}_1,\SPvec{s}_1)$, 
for which we were able to find an explicit integral representation.
 Clearly, $A_1(0,\SPvec{s}_1)$ is here a nontrivial function of 
$|\SPvec{s}_1|$.
 More precisely, $A_1(0,\SPvec{s}_1)$ is real analytic as function of 
$|\SPvec{s}_1|\in(0,\infty)$, (extended to $|\SPvec{s}_1|\in\RR$) 
its Taylor series about the origin has finite radius of convergence and begins as
\beq
A_1(0,\SPvec{s}_1)
=
-\textstyle{ \frac{1}{ 2 \beta}\left[
            \frac{|\SPvec{s}_1|}{\beta} -O\left(\frac{|\SPvec{s}_1|^5}{\beta^5}\right)
                              \right]}
\,,
\label{eq:AatSoneINTEGRALformulaSMALLsONE}
\eeq
and the asymptotic expansion for large $|\SPvec{s}_1|$ reads
\beq
A_1(0,\SPvec{s}_1) 
=
 A_{\mathrm{Born}}^{(-)}({\SPvec{0}}) 
+
\textstyle{\frac{1}{|\SPvec{s}_1|}}\left[1 - U\left(\textstyle{\frac{\beta}{|\SPvec{s}_1|}}\right)\right]
\,,
\label{eq:AatSoneASYMPTOTICS}
\eeq
with 
$|U(\beta/|\SPvec{s}_1|)|< C {\beta}/{|\SPvec{s}_1|}$ 
for large $|\SPvec{s}_1|$.
 This result is just part of a more detailed theorem which we need
for our discussion of the hydrogen spectrum, and for which reason
we defer its rigorous proof to our follow-up paper on the partially quantized theory
        \cite{KiePapII}.
 Here we just note that our results show that the 
Hamilton--Jacobi PDE is well-defined initially with these data $A_1(0,\SPvec{s}_1)$ and $\Phi_0$.
 Different from the first, trivial example, the Hamilton--Jacobi PDE now launches a solution 
$\Phi(t,\SPvec{s}_1)$ which immediately develops a spatial dependence on the position in 
configuration space. 
 Its non-trivial configuration space gradient (viz., velocity field on configuration space)
is well-defined everywhere except at the origin, where $\Phi(t,\SPvec{s}_1)$ develops a {kink}. 
 Thus, since the point electron cannot be initially at rest exactly on top of the nuclear point charge,
for coincidence points are removed from configuration space,
in its immediate future it begins to move according to its guiding equation. 
  Better than that, our asymptotic expansion of $A_1(0,\SPvec{s}_1)$ also shows that to leading order 
after the irrelevant constant term $A_{\mathrm{Born}}^{(-)}({\SPvec{0}})$ the negative point 
electron sees the familiar electrostatic Coulomb potential of the infinitely massive
positive point charge at the origin. 
 Thus, not only does the point electron begin to move; if it is initially far enough
from the origin, then in leading order of the asymptotic expansion it begins to move 
precisely according to the familiar Newtonian law of motion in a \emph{given}
 attractive Coulomb field, as it should.

 As to the question of global well-posedness vs. finite-time blow-up, we remark that
it is to be expected that the evolution of the point electron will run into the nuclear point
charge in finite time, which is the shorter the closer to the nucleus the point electron started 
out of rest initially.
 Once it has hit the nucleus, the motion of the point electron cannot be continued 
uniquely beyond this dynamical singularity, but until this happens we expect the evolution 
to be regular. 
  In any event, we have not tried to prove this yet. 

 We also remark that the initial $\beta$-correction is tiny unless the two charges are sufficiently 
close together. 
 How close is `sufficiently close' will be of relevance for our next section, the 
assessment of Born's calculation of $\beta$. 
%\newpage
%%%%%%%%%%%%%%%%%%%%%%%%%%%%%%%%%%%%%%%%%%%%%%%%%%%%%%%%%%%%%%%%%%%%
%%%%%%%%%%%%%%%%%%%%%%%%%%%%%%%%%%%%%%%%%%%%%%%%%%%%%%%%%%%%%%%%%%%%
%%%%%%%%%%%%%%%%%%%%%%%%%%%%%%%%%%%%%%%%%%%%%%%%%%%%%%%%%%%%%%%%%%%%
%%%%%%%%%%%%%%%%%%%%%%%%%%%%%%%%%%%%%%%%%%%%%%%%%%%%%%%%%%%%%%%%%%%%
  \section{The values of the universal constants $\alpha$ and $\beta$ \\
             (preliminary assessment)}
%%%%%%%%%%%%%%%%%%%%%%%%%%%%%%%%%%%%%%%%%%%%%%%%%%%%%%%%%%%%%%%%%%%%
%%%%%%%%%%%%%%%%%%%%%%%%%%%%%%%%%%%%%%%%%%%%%%%%%%%%%%%%%%%%%%%%%%%%
%%%%%%%%%%%%%%%%%%%%%%%%%%%%%%%%%%%%%%%%%%%%%%%%%%%%%%%%%%%%%%%%%%%%
%%%%%%%%%%%%%%%%%%%%%%%%%%%%%%%%%%%%%%%%%%%%%%%%%%%%%%%%%%%%%%%%%%%%
  We are finally ready to vindicate our identification of 
the universal constant $\alpha$ in our guiding laws with 
Sommerfeld's fine structure constant, 
as indeed done in \refeq{eq:SOMMERFELDconstant}.
 However, we have collected barely enough material to identify $\beta$
correctly.
 We begin with the tentative determination of $\alpha/\beta$.
 
%%%%%%%%%%%%%%%%%%%%%%%%%%%%%%%%%%%%%%%%%%%%%%%%%%%%%%%%%%%%%%%%%%%%
%%%%%%%%%%%%%%%%%%%%%%%%%%%%%%%%%%%%%%%%%%%%%%%%%%%%%%%%%%%%%%%%%%%%
    \subsection{Born's determination of $\alpha/\beta$}
%%%%%%%%%%%%%%%%%%%%%%%%%%%%%%%%%%%%%%%%%%%%%%%%%%%%%%%%%%%%%%%%%%%%
%%%%%%%%%%%%%%%%%%%%%%%%%%%%%%%%%%%%%%%%%%%%%%%%%%%%%%%%%%%%%%%%%%%%

 Inspired by the idea of the later 19th century that the electron's 
inertia a.k.a. mass has a purely electromagnetic origin, Born argued 
that the electrostatic energy of the spherically symmetric Born alias 
Coulomb field $\SPvec{D}_{\mathrm{Born}}$ of the electronic point 
charge at rest be identified with the empirical rest energy of the 
electron
    \cite{BornA}.
 Since  no other fields are present than the Coulomb field, we have 
${\SPvec{B}} = {\SPvec{0}}$ and ${\SPvec{D}} = {\SPvec{D}}_{\mathrm{Born}}^{\plumi}$, 
where the superscript $\plumi$ again indicates the sign of the electron's charge.
 The electrostatic field energy of Born's solution is therefore given by
\beq
\cH_{\mathrm{field}}\bigl({\SPvec{0}},{\SPvec{D}}_{\mathrm{Born}}^{\plumi}\bigr) 
=
\frac{1}{4\pi}\frac{\alpha}{\beta^4}
\int_{\RR^3} \Big(\sqrt{1 +\beta^4|\SPvec{D}_{\mathrm{Born}}^{\plumi}|^2} -1\Big)\dvol(\SPvec{s})
,
\eeq
which evaluates independently of the sign of the charge to 
\beq
\cH_{\mathrm{field}}\bigl({\SPvec{0}},{\SPvec{D}}_{\mathrm{Born}}^{\plumi}\bigr) 
= 
\frac{\alpha}{\beta} \frac{1}{6} \Beta\left({\textstyle{\frac{1}{4},\frac{1}{4}}}\right) 
\,,
\eeq
where $\Beta(p,q)$ is Euler's Beta function.
 In our dimensionless formulation, the empirical rest energy $\me c^2$ of the electron is 
the unit of energy. 
Hence, setting 
\beq
\cH_{\mathrm{field}}\bigl({\SPvec{0}},{\SPvec{D}}_{\mathrm{Born}}^\plumi\bigr) = 1
\,,
\eeq
we find  Born's result\footnote{As 
     mentioned before, Born used a different, dimensional 
     notation; our $\beta^2\propto a$ of Born.}
    \cite{BornA}
\beq
\frac{\beta}{\alpha}\Big|_{_{\mathrm{Born}}}
=
     \frac{1}{6} 
\Beta\left({\textstyle{\frac{1}{4},\frac{1}{4}}}\right)
\approx 
1.2361
\label{eq:BORNconstantCOMPUTED}
\,.
\eeq
 Remark that with Born's value for $\beta$,
the value at the origin of the electrostatic potential of Born's solution 
\refeq{eq:BornsElectricPot} is given by
\beq
\alpha A_{\mathrm{Born}}^\plumi({\SPvec{0}}) 
=  
 \plumi \frac{3}{2}
\,.
\label{eq:BornsElectricPotATnull}
\eeq

%\newpage
%%%%%%%%%%%%%%%%%%%%%%%%%%%%%%%%%%%%%%%%%%%%%%%%%%%%%%%%%%%%%%%%%%%%
\textit{Comments on Born's calculation of $\alpha/\beta$}
%%%%%%%%%%%%%%%%%%%%%%%%%%%%%%%%%%%%%%%%%%%%%%%%%%%%%%%%%%%%%%%%%%%%
 
 The derivation of formula \refeq{eq:BORNconstantCOMPUTED} is not as 
unproblematic  as it pretends to be. 

  Born's thoughts about the purely electromagnetic origin of the electron's inertia 
were reinforced by his and Infeld's conviction that the field energy functional 
$\cH_{\mathrm{field}}\bigl({\SPvec{B}},{\SPvec{D}}\bigr)$ was the conserved total energy
quantity, which our energy conservation theorem shows not to be true in the presence of
     point charges.
 Moreover, the qualitative content of the law of energy conservation
does not change if an arbitrary  constant is added to the expression 
on the r.h.s. of \refeq{eq:HfuncTOT}, but clearly its quantitative 
content does.
 One might want to argue on behalf of the mathematical integrity of
\refeq{eq:HfuncTOT} that the adding of any nontrivial  constant,
other than perhaps $-1$ per single 
electron,\footnote{As 
          for adding $-1$ per electron to the r.h.s. of \refeq{eq:HfuncTOT}, 
	  this gives the alternate total energy functional 
\beq
\widetilde\cE\left(\Omega^{\mathrm{cl}}\right) 
=
\cH_{\mathrm{field}}\left({\SPvec{B}},{\SPvec{D}}\right) 
+ \sum_{k\in\cN} \left(
 \sqrt{1 +\abs{{\nabla}_k \Phi(t,{\SPvec{R}})  
-\pmk \alpha{\SPvec{A}_k}\left(t,{\SPvec{R}}\right)}^2}
 -1\right)
\label{eq:HfuncTOTminusONE}
\,.
\eeq
           For a system containing only a single electron at rest, the energy 
	   functional $\widetilde\cE(\Omega^{\mathrm{cl}}_0)$ coincides with the field-energy 
	   functional $\cH_{\mathrm{field}}\bigl({\SPvec{B}},{\SPvec{D}}\bigr)$.
	   The identification of the electron's empirical rest energy with 
	   either the electrostatic field energy or the total energy of the
	   static single-electron state  then give the same result, which
	   would seem like an attractive way out of the dilemma.}
would be totally perverse unless compelling reasons for such an additive 
constant would be found elsewhere, but this is not an entirely convincing 
way of reasoning. 
 In any event, Born's calculation of the value of $\alpha/\beta$ 
suddenly seems rather arbitrary.
 An entirely unambiguous identification of $\alpha/\beta$ or even $\beta$
can only be made on the basis of truly dynamical considerations.

 The first clues can be obtained by taking the other conservation laws 
into account, which already provide some additional pieces of information
of the underlying dynamical system.
 In particular, the usually-little-attention-paid-to law of the linear 
motion of the moment of energy \refeq{MofHminusPtGLEICHconstX} suggests
that the total energy as identified in  \refeq{eq:HfuncTOT} is singled
out among all other possibilities that differ from \refeq{eq:HfuncTOT} 
by an additive constant, whether per electron or in total.
 Once we have accepted \refeq{eq:HfuncTOT} as the likely candidate for
the correct total energy, the next question is whether it is still a 
reasonable working hypothesis to identify the electron's rest energy ($=1$) 
with the electrostatic field energy, or whether we should identify it with 
the total energy of the static single-electron state, or neither of the two.

 With the help of elementary  considerations we can immediately dispose
of the option of identifying the total energy \refeq{eq:HfuncTOT} 
of the electrostatic single-electron 
state with the electron's rest energy ($=1$), for this leads to the conclusion 
$\alpha\beta^{-4} =0$. 
 However, $\alpha =0$ means the particles feel no influence of the 
electromagnetic fields, so we need to have $\alpha  \neq 0$. 
 The only other possibility, then, is to let $\beta \to \infty$,
which gives us the ultra Born--Infeld laws 
\refeq{eq:FOLIeqEofBDultra} and
\refeq{eq:FOLIeqHofBDultra}.
 However, the ultra Born--Infeld field equations allow electrostatic 
solutions with arbitrarily many point charges placed arbitrarily in 
space, and the total energy of any such static $N$-electrons state 
is then always simply $N$, which is clearly incorrect. 
 Hence, the alternate identification of the electron rest energy
with the total energy \refeq{eq:HfuncTOT} of the static single-electron state
is not feasible.

 Incidentally, we have actually disposed of either of the possibilities 
$\alpha =0$ and $\beta=\infty$.
 As we will see next, this in turn suffices to determine $\alpha$, and from 
there to get some upper estimate on $\beta$, which comes close to the value given 
by Born's result. 
 While this gives some a-posteriori confidence in the viability of Born's formula
\refeq{eq:BORNconstantCOMPUTED}, 
we will re-assess the $\beta$ value in our follow-up paper
\cite{KiePapII}
on the spinless quantum theory, and  once again
when spin and photon are incorporated into the theory. 

%%%%%%%%%%%%%%%%%%%%%%%%%%%%%%%%%%%%%%%%%%%%%%%%%%%%%%%%%%%%%%%%%%%%
%%%%%%%%%%%%%%%%%%%%%%%%%%%%%%%%%%%%%%%%%%%%%%%%%%%%%%%%%%%%%%%%%%%%
\subsection{Identification of $\alpha$ with Sommerfeld's fine structure constant}
%%%%%%%%%%%%%%%%%%%%%%%%%%%%%%%%%%%%%%%%%%%%%%%%%%%%%%%%%%%%%%%%%%%%
%%%%%%%%%%%%%%%%%%%%%%%%%%%%%%%%%%%%%%%%%%%%%%%%%%%%%%%%%%%%%%%%%%%%

 We now vindicate our identification $\alpha = e^2/\hbar{c}$ by 
showing formally that in the limit of radiation-reaction-free gentle motions, 
our guiding equation for a single  electron reduces to the  correct law of motion.
  Incidentally, the second example of our subsection illustrating the Cauchy problem
already vindicates our claim about $\alpha$ for the special two-body Coulomb problem
with one particle infinitely massive. 
 We here adapt this line of reasoning to the more general situation depicted below.
 Another, more rigorous vindication will be supplied in our follow-up paper 
         \cite{KiePapII}
on the quantum theory.

 We consider a Lorentz frame in which a single (negative) point electron is deflected 
(in Born--Oppenheimer approximation) in the Coulomb field $\SPvec{E}_{\mathrm{Coulomb}}$ of an infinitely 
massive point charge at rest, and perhaps also by an electromagnetic radiation field with very low 
intensity and very long wavelength which impinges on the electron.
 For the notion of Coulomb field $\SPvec{E}_{\mathrm{Coulomb}}$ to apply, we need to have
$|\SPvec{s}| \gg \beta$, as a perusal of \refeq{eq:BornsElectricPot} reveals; similarly, 
`long wavelength $\lambda$' is defined as $\lambda \gg\beta$, while `low intensity' 
is defined as $\beta^2(|\SPvec{B}|+|\SPvec{D}|) \ll 1$. 
 These definitions of smallness make sense as long as $\beta\in(0,\infty)$.
 Recall that $\beta =0$ yields just the ill-defined Lorentz electrodynamics with point charges,
while $\beta = \infty$ yields the fully conformally invariant ultra Maxwell--Born--Infeld
field equations with point charge sources, which we have just disposed of; hence, we do have 
$\beta\in(0,\infty)$ and can proceed unimpeded.
 
 When a particle moves at high speeds, but is only gently accelerated by the 
radiation field and the static Coulomb field, then according to the established 
physics, the asymptotically correct evolutionary 
law of the point electron is Newton's law of radiation-reaction-free motion, which
equates the rate of change of kinematical particle momentum to a Lorentz force in 
which only the external (i.e. incoming radiation and static Coulomb) fields
enter, while the kinematical particle momentum 
and particle velocity are related by Einstein's relativistic formula. 
  It is an easy exercise to verify that the familiar textbook 
formulas, when converted from conventional dimensional Gaussian units into our dimensionless 
units, become
\bea
	\bulldif{\SPvec{r}}(t)
\!\!\!&=&\!\!\!
    \frac{{\SPvec{p}}(t)}{ \sqrt{1+ |{\SPvec{p}}(t))|^2}}
\label{eq:EINSTEINvOFp}
\\
        \bulldif{\SPvec{p}}(t)
\!\!\!&=&\!\!\!
         - \frac{e^2}{\hbar{c}}
       \Bigl( {\SPvec{E}}^{\mathrm{ext}}(t,{\SPvec{r}}(t)) 
+ \bulldif{\SPvec{r}}(t) \times {\SPvec{B}}^{\mathrm{ext}}(t,{\SPvec{r}}(t))\Bigr)
\label{eq:NEwTONlawOFmotion}
\, ,
\eea
in which the dimensionless Sommerfeld fine structure constant $e^2/\hbar{c}$ features 
prominently as the only universal physical constant. 
  We next argue, non-rigorously, that in the approximation in which radiation-reaction 
and non-linear field response is neglected, our Hamilton--Jacobi guiding laws reduce  
precisely to 
\refeq{eq:EINSTEINvOFp}
and
\refeq{eq:NEwTONlawOFmotion}.

 We define the external electric and magnetic fields 
${\SPvec{E}}^{\mathrm{ext}}(t,{\SPvec{s}})$ 
and ${\SPvec{B}}^{\mathrm{ext}}(t,{\SPvec{s}})$ as
solutions to the Maxwell--Born--Infeld field equations 
in $\RR_+\times\RR^3$ with just the infinitely massive 
point source present.
 By solving the familiar partial differential equations 
\refeq{eq:FOLIeAmagn}, \refeq{eq:FOLIcAmagn}, and \refeq{eq:FOLIeAelec}
for appropriate initial conditions, and assuming vanishing conditions at 
spatial infinity, we obtain electric and magnetic potentials 
$A^{\mathrm{ext}}(t,{\SPvec{s}})$ and $\SPvec{A}^{\mathrm{ext}}(t,{\SPvec{s}})$ 
in the Lorentz--Lorenz gauge which generate the external fields 
${\SPvec{E}}^{\mathrm{ext}}(t,{\SPvec{s}})$ 
and 
${\SPvec{B}}^{\mathrm{ext}}(t,{\SPvec{s}})$.
 Of course, the Lorentz--Lorenz gauge is also used 
for the electric and magnetic potentials of the 
total electric and magnetic fields, assuming the total
potentials vanish at spatial infinity as well.
 On configuration space, we write 
${A}_1 = {A}_1^\prime + A^{\mathrm{ext}}$
and
${\SPvec{A}_1} = {\SPvec{A}}_1^\prime + {\SPvec{A}}^{\mathrm{ext}}$.
 In the radiation-reaction-free approximation, and neglecting also the nonlinear modifications 
of the weak external fields at the location of the electron, we have
(in $1$-form notation, for brevity), 
\beq
{\AQ}_1^\prime
\approx {A}_{\mathrm{Born}}^{(-)}(\SPvec{0}) {\uQ}
\label{eq:convectedA}
\eeq
with 
\beq
A_{\mathrm{Born}}^{(-)}({\SPvec{0}}) 
=
- \viertel \Beta\left({\textstyle{\frac{1}{4},\frac{1}{4}}}\right) \beta^{-1} 
\,.
\eeq
 We may assume that in a local spacetime neighborhood of short truncated point histories,
different initial $\uQ_0$ will define a vector field on configurational spacetime.
 In this neighborhood we then gauge-transform this convected electromagnetic 
potential away with the help of the gauge potential (0-form)
$\Upsilon(t,\SPvec{s})  = -A_{\mathrm{Born}}^{(-)}({\SPvec{0}}) \int_{\Eta_1} {\uQ}$,
understood as a function of the endpoint of the integration along the truncated history.
 Thus, for such small times and regions in configuration space, 
after the radiation-reaction-free approximation \refeq{eq:convectedA}
has been made, our Hamilton--Jacobi partial differential equation 
\beq 
{\partial} \Phi(t,{\SPvec{s}_1}) 
= 
-  {\sqrt{1 +{| {\nabla}_1 \Phi(t,{\SPvec{s}_1}) + \alpha
({\SPvec{A}}_1^\prime + {\SPvec{A}}_1^{\mathrm{ext}})
\left(t,{\SPvec{s}_1}\right)|}^2} } + \alpha 
({{A}}_1^\prime + A^{\mathrm{ext}})(t,{\SPvec{s}_1}) 
\label{eq:HamJacPDEsplit}
\eeq
and the Hamilton--Jacobi guiding equation 
\beq
	\frac{\dd\SPvec{s}_1}{\dd{t}}
=
\frac{{\nabla}_1 \Phi(t,{\SPvec{s}_1}) +  \alpha({\SPvec{A}}_1^\prime + {\SPvec{A}}^{\mathrm{ext}})
\left(t,{\SPvec{s}_1}\right)}
  {\sqrt{1 +{|{\nabla}_1 \Phi(t,{\SPvec{s}_1}) + 
\alpha({\SPvec{A}}_1^\prime + {\SPvec{A}}^{\mathrm{ext}})\left(t,{\SPvec{s}_1}\right)|}{}^2} }
\,,
\label{eq:HamJacGuideEQsplit}
\eeq
are gauge equivalent to the Hamilton--Jacobi equations
\beq 
{\partial} {\Phi}(t,{\SPvec{s}_1}) 
= 
-  {\sqrt{1 +{| {\nabla}_1 {\Phi}(t,{\SPvec{s}_1}) 
+ \alpha  {\SPvec{A}}^{\mathrm{ext}}\left(t,{\SPvec{s}_1}\right)|}^2} } 
+ \alpha   A^{\mathrm{ext}}(t,{\SPvec{s}_1}) 
\,,
\eeq
\beq
	\frac{\dd\SPvec{s}_1}{\dd{t}}
=
\frac{{\nabla}_1 \Phi(t,{\SPvec{s}_1}) +  \alpha {\SPvec{A}}^{\mathrm{ext}}\left(t,{\SPvec{s}_1}\right)}
  {\sqrt{1 +{|{\nabla}_1 \Phi(t,{\SPvec{s}_1}) + 
\alpha{\SPvec{A}}^{\mathrm{ext}}\left(t,{\SPvec{s}_1}\right)|}{}^2} }
\,.
\eeq
 These equations are just the test particle Hamilton--Jacobi equations, 
well-known to be equivalent to 
\refeq{eq:EINSTEINvOFp} and \refeq{eq:NEwTONlawOFmotion}, with $\alpha$ in 
place of $e^2/\hbar{c}$, which was to be shown.
%%%%%%%%%%%%%%%%%%%%%%%%%%%%%%%%%%%%%%%%%%%%%%%%%%%%%%%%%%%%%%%%%%%%
%%%%%%%%%%%%%%%%%%%%%%%%%%%%%%%%%%%%%%%%%%%%%%%%%%%%%%%%%%%%%%%%%%%%
\subsection{An upper estimate for $\beta$}
%%%%%%%%%%%%%%%%%%%%%%%%%%%%%%%%%%%%%%%%%%%%%%%%%%%%%%%%%%%%%%%%%%%%
%%%%%%%%%%%%%%%%%%%%%%%%%%%%%%%%%%%%%%%%%%%%%%%%%%%%%%%%%%%%%%%%%%%%
 As the discussion in the previous subsection shows, only the information that
$\beta\in(0,\infty)$, but not the precise value of the parameter $\beta$, enters 
in the derivation of the classical law of radiation-reaction-free motion of a 
point charge. 
 It is clear that $\beta$ will eventually enter in some higher-order-of-$\alpha$ 
corrections to this leading order law, but some early investigations by Schr\"odinger 
   \cite{ErwinDUBLINb}
indicate that to first order in radiation-reaction 
correction\footnote{In 
           our dimensionless units, this is an order $\alpha^2$
           correction to the r.h.s. of \refeq{eq:NEwTONlawOFmotion}.}
the value of $\beta$ still does not play any r\^{o}le.
 While Schr\"odinger's calculations are not based on a truly consistent 
dynamical model, we may have some confidence in his result because of
the known fact that the first radiation-reaction correction to the 
equations of radiation-reaction-free motion \refeq{eq:EINSTEINvOFp},
\refeq{eq:NEwTONlawOFmotion} is independent of any specific assumptions 
about the charge structure of the electron when computed from the classical 
electron theory 
	   \cite{lorentzENCYCLOPb, abrahamBOOK, lorentzBOOKb, spohnBOOK},
and also independent of the regularization in Dirac's re-calculation of this
term for point electrons 
            \cite{DiracA}.
 Although not a substitute for a rigorous discussion of radiation-reaction, 
this indicates that to see $\beta$ enter explicitly in the asymptotic expansion 
of the equation of motion may require going to such high order in $\alpha$
that quantum physical corrections may be of equal importance.
 An assessment based on the hydrogen 
spectrum will be supplied in our next paper.
 
 Yet, some quantitative information on $\beta$ becomes available by rule of thumb
if we ask for the empirical range of validity of the equations \refeq{eq:EINSTEINvOFp},
\refeq{eq:NEwTONlawOFmotion}, which is quite impressive. 
 Since for the derivation of \refeq{eq:EINSTEINvOFp}, \refeq{eq:NEwTONlawOFmotion}, 
we had to assume a low intensity and long wavelength of the incoming radiation, 
and a sufficiently large impact parameter for the scattering at the Coulomb
potential, all of which terms are defined against $\beta$, the range of validity 
of  \refeq{eq:EINSTEINvOFp}, \refeq{eq:NEwTONlawOFmotion} sets some rough
upper bound on $\beta$. 
 In particular, from Coulomb scattering experiments with electrons we may have 
confidence in  \refeq{eq:EINSTEINvOFp}, \refeq{eq:NEwTONlawOFmotion} for length scales
roughly down to the classical electron radius, which is a factor $\alpha$ smaller 
than the Compton wave length of the electron, our reference unit of length.
  This suggests an upper estimate on $\beta$ roughly equal to $\alpha$, so
that the $\beta$ value found by Born remains viable, for now.

 We end by noting that the possibility $\beta =\alpha$ is currently as viable 
as Born's proposal \refeq{eq:BORNconstant}, and so is Euler--Kockel's early QED result
$\beta^4 = (\#/45\pi)\alpha^3$, with $\#\in[4,7]$, giving $\beta/\alpha \approx 1.4 - 1.6$.
 However, future estimates based on the quantum theory may well rule out either of these
possibilities. 
%%%%%%%%%%%%%%%%%%%%%%%%%%%%%%%%%%%%%%%%%%%%%%%%%%%%%%%%%%%%%%%%%%%%
%%%%%%%%%%%%%%%%%%%%%%%%%%%%%%%%%%%%%%%%%%%%%%%%%%%%%%%%%%%%%%%%%%%%
%%%%%%%%%%%%%%%%%%%%%%%%%%%%%%%%%%%%%%%%%%%%%%%%%%%%%%%%%%%%%%%%%%%%
%%%%%%%%%%%%%%%%%%%%%%%%%%%%%%%%%%%%%%%%%%%%%%%%%%%%%%%%%%%%%%%%%%%%
	\section{Summary and Outlook}
%%%%%%%%%%%%%%%%%%%%%%%%%%%%%%%%%%%%%%%%%%%%%%%%%%%%%%%%%%%%%%%%%%%%
%%%%%%%%%%%%%%%%%%%%%%%%%%%%%%%%%%%%%%%%%%%%%%%%%%%%%%%%%%%%%%%%%%%%
%%%%%%%%%%%%%%%%%%%%%%%%%%%%%%%%%%%%%%%%%%%%%%%%%%%%%%%%%%%%%%%%%%%%
%%%%%%%%%%%%%%%%%%%%%%%%%%%%%%%%%%%%%%%%%%%%%%%%%%%%%%%%%%%%%%%%%%%%
	In this paper we achieved the first consistent implementation 
of the notion of the point charge into the classical theory of electromagnetism.
 No regularization and no renormalization is called for. 
 The actual electromagnetic fields are solutions of the Maxwell--Born--Infeld field equations 
with point charges sources, which move according to a relativistic guiding law
of Hamilton--Jacobi type. 
  The guiding field is generated by the solution of a
relativistic Hamilton--Jacobi partial differential equation which is coupled 
self-consistently to the continuous potentials of generalized 
electromagnetic fields that live on space $\times$ configuration space. 
 When the actual configuration is substituted for the generic one, then 
these generalized fields reduce to the actual electromagnetic fields on 
actual space.
 The formalism works because in the Maxwell--Born--Infeld theory of the
electromagnetic fields, singularities associated with the point charges 
feature merely as mild defects in the electromagnetic potentials.
 It will not work for the fields of the older Maxwell--Lorentz theory 
with point charges.
 Curiously, since the (total) electromagnetic Maxwell--Born--Infeld fields
are ill-defined at the actual positions of the point charges, the various
generalized fields on space $\times$ configuration space that enter the
Hamilton--Jacobi formalism cannot be eliminated, and this compels us 
to regard those fields not as mere mathematical auxiliary constructs,
but as fields that enjoy a certain physical status of their own. 
 This introduces a new element of physics into classical electrodynamics
which is akin to the wave function in quantum physics.

 Most of the paper is concerned with the consistent dynamical coupling
of point charges to the classical electromagnetic Maxwell--Born--Infeld 
field equations, in an appendix we also address the question whether 
charge-free solitons with finite energy exist.
 There we prove rigorously that such soliton solutions of the Maxwell--Born--Infeld 
field equations cannot exist if their proper electric and magnetic field strengths 
remain below a huge threshold. 
 Put differently, if such solitons exist, their peak field strengths must be
enormous.

 While in this paper we have worked out in detail only the special-relativistic 
theory, in an appendix on the action principle we note that the 
extension to a general-relativistic electromagnetic theory with point charges 
is feasible in which spacetime is not flat and static but curved and dynamical, 
and in which the electromagnetic stress-energy-momentum couples gravitationally 
to the spacetime as a source of its curvature.
 Interestingly enough, we anticipate that the point charges 
will appear as naked singularities of the spacetime;
yet our formulation also yields a general relativistic
Hamilton--Jacobi guiding law of motion for these singularities,
thus indicating that the theory will contain enough information
to remove the evolutionary ambiguities associated with the
occurrence of naked singularities in general relativity. 
  Technically, the nonlinear mathematical structure of Einstein's field 
equations poses formidable challenges to the rigorous implementation of 
this picture, but these will be taken up. 
 It might be helpful to treat singularities as defects in the smoothness 
of spacetime, similar in spirit to the work 
       \cite{HeinzleSteinbauer}
on the Schwarzschild metric, rather than punctures of spacetime.
 We plan to return to these issues at a later time.

 Our special-relativistic theory should describe the physics of positive and negative 
point electrons and their radiation fields in the classical regime where 
spin effects and the photonic nature of the electromagnetic fields can be 
neglected.
 In this spirit, we have given the first assessment of the correctness of Born's 
value for his aether constant $\beta$, which enters the theory through the Born--Infeld 
laws of the aether. 
 Born calculated the value of $\beta$ by arguing that the empirical electron 
rest mass ${\me}$ $({\times}c^{2})$ be identified with the electrostatic energy 
of his spherically symmetric electrostatic solution for a single point charge.
 We showed that Born's argument, based as it is on his dynamically incomplete
formulation of the theory, is inconclusive; yet  Born's value for $\beta$ 
remains viable for now, in the sense that it does not seem to conflict with 
any established classical electromagnetic effects.
 In particular, we showed for the two simplest examples of the initial value problem 
for a point charge that in leading order the conventional physical wisdom is reproduced.
 Our investigation led us to conclude that the $\beta$-induced corrections to these 
known effects in the classical domain of electromagnetism are presumably so 
small that they are masked by quantum effects. 
 In other words, the definitive calculation of $\beta$ can presumably be done 
only after the full quantization of our theory.
 A partial quantization of our theory, without spin and the photon, 
will be presented in 
         \cite{KiePapII},
the follow-up paper to this one. 
 The incorporation of spin and the photon will be taken up subsequently.
 We add that, even though spin has not been incorporated 
at this classical level, our formalism has allowed us to implement 
the Pauli principle for many `bosonic electrons.'
 We also add that only minor modifications of the theory are needed to accommodate 
the electromagnetic effects of other, non-genuinely electromagnetic
particles, representing nuclei with or without magnetic moment and form factor. 
	This requires putting in by hand the parameters 
$z$ for the charge number,  $\kappa$  for the ratio of 
the electron's to the nucleus' rest mass, and, if desired, 
a smeared-out spinning charge distribution.

\noindent
\textbf{Acknowledgments:} This work began in early 1992 when the author held
a German-Dartmouth distinguished visiting professorship at Dartmouth College. 
 It was supported in the past two years by NSF grant DMS-0103808.
 I am indebted to many individuals, but I am most grateful to 
S. Goldstein and H. Spohn for many invaluable scientific 
discussions about electromagnetism and quantum theory.
 I also thank S. Chanillo for Moser's theorem, and J. Taylor for his
insights into the Fermi bundle.
 I owe very special thanks to S. Goldstein and, especially, to R. Tumulka 
for their helpful comments and penetrating criticisms of an earlier version 
of this paper, which prompted me to improve and clarify the presentation.
 My sincere thanks go also to the five  referees for their favorable
reactions to this non-mainstream paper and their helpful suggestions.
 I thank T. Dorlas for sending me copies of Schr\"odinger's 
Dublin papers, and Y. Brenier and I. Bia{\l}ynicki-Birula for bringing
their more recent works to my attention after the first version of this 
paper was circulated.
\newpage
%%%%%%%%%%%%%%%%%%%%%%%%%%%%%%%%%%%%%%%%%%%%%%%%%%%%%%%%%%%%%%%%%%%%
%%%%%%%%%%%%%%%%%%%%%%%%%%%%%%%%%%%%%%%%%%%%%%%%%%%%%%%%%%%%%%%%%%%%
%%%%%%%%%%%%%%%%%%%%%%%%%%%%%%%%%%%%%%%%%%%%%%%%%%%%%%%%%%%%%%%%%%%%
%%%%%%%%%%%%%%%%%%%%%%%%%%%%%%%%%%%%%%%%%%%%%%%%%%%%%%%%%%%%%%%%%%%%
%%%%%%%%%%%%%%%%%%%%%%%%%%%%%%%%%%%%%%%%%%%%%%%%%%%%%%%%%%%%%%%%%%%%
%%%%%%%%%%%%%%%%%%%%%%%%%%%%%%%%%%%%%%%%%%%%%%%%%%%%%%%%%%%%%%%%%%%%

\section{Appendix}

%%%%%%%%%%%%%%%%%%%%%%%%%%%%%%%%%%%%%%%%%%%%%%%%%%%%%%%%%%%%%%%%%%%%
%%%%%%%%%%%%%%%%%%%%%%%%%%%%%%%%%%%%%%%%%%%%%%%%
%%%%%%%%%%%%%%%%%%%%%%%%%%%%%%%%%%%%%%%%%%%%%%%%
%%%%%%%%%%%%%%%%%%%%%%%%%%%%%%%%%%%%%%%%%%%%%%%%%%%%%%%%%%%%%%%%%%%%
     \subsection{On source-free solutions of the Maxwell--Born--Infeld field equations}
%%%%%%%%%%%%%%%%%%%%%%%%%%%%%%%%%%%%%%%%%%%%%%%%%%%%%%%%%%%%%%%%%%%%
%%%%%%%%%%%%%%%%%%%%%%%%%%%%%%%%%%%%%%%%%%%%%%%%
%%%%%%%%%%%%%%%%%%%%%%%%%%%%%%%%%%%%%%%%%%%%%%%%
%%%%%%%%%%%%%%%%%%%%%%%%%%%%%%%%%%%%%%%%%%%%%%%%%%%%%%%%%%%%%%%%%%%%

 In this appendix we collect some interesting results about solutions 
of the Maxwell--Born--Infeld field equations without sources
(in particular without point sources) which, while pertinent to
the content of the present paper, are somewhat beside its main thrust.
 We also prove a new no-soliton result. 

  All the known source-free solutions are electromagnetic waves of some sort.

%%%%%%%%%%%%%%%%%%%%%%%%%%%%%%%%%%%%%%%%%%%%%%%%
%%%%%%%%%%%%%%%%%%%%%%%%%%%%%%%%%%%%%%%%%%%%%%%%%%%%%%%%%%%%%%%%%
             {\subsubsection{Monochromatic plane waves}}
%%%%%%%%%%%%%%%%%%%%%%%%%%%%%%%%%%%%%%%%%%%%%%%%%%%%%%%%%%%%%%%%%
%%%%%%%%%%%%%%%%%%%%%%%%%%%%%%%%%%%%%%%%%%%%%%%%

 As observed already by Born
                     (\cite{BornB}, p.434)
and Schr\"odinger
                     (\cite{ErwinEiHBiD}, p.474),
any monochromatic plane wave solution of Maxwell's field equations in vacuum also solves the source-free 
Maxwell--Born--Infeld field equations.\footnote{Curiously, Born and Infeld, apparently overlooking the fact that
                                                the electromagnetic field strengths in a plane monochromatic 
						wave can be arbitrarily large, interpreted Born's field parameter 
						$b\, (\propto\beta^{-2})$ (originally Born used $a\, \equiv b^{-1}$)
						as an upper bound to the field strengths by alluding to some
						``principle of finiteness'' that Nature supposedly adheres to
			  			(p.427 in \cite{BornInfeldB}).
						It seems to have been Schr\"odinger who first pointed out
						that ``none of the field quantities has an insurmountable upper
						limit in this theory''  (p.82 in \cite{ErwinDUBLINa}).
						This simple fact to the contrary notwithstanding, $b$ continues 
						to be misinterpreted as an absolute upper bound to the electric and 
						magnetic field strengths even as recently as 2002, when some 
						``extended relativity'' 
						was proposed in which, beside ${c}$ as an absolute speed limit, 
						$eb/\me$ is interpreted as absolute bound on accelerations.}
	Indeed, in Maxwell's theory the electromagnetic vacuum 
fields satisfy ${\SPvec{E}}={\SPvec{D}}$ and ${\SPvec{B}}={\SPvec{H}}$, 
and for a monochromatic plane electromagnetic wave they satisfy 
also $|{\SPvec{E}}|=|{\SPvec{B}}|$ 
and ${\SPvec{E}}\cdot{{\SPvec{B}}} =0$. 
	It is now an easy exercise to show that such field solutions of
the vacuum Maxwell equations also satisfy the electromagnetic 
aether laws of Born and Infeld.\footnote{Schr\"odinger subsequently extended this result to a whole class
                                        of nonlinear electromagnetic field equations, see
                                        App.I in \cite{ErwinDUBLINc}.}

        The plane wave solutions themselves do not have a finite total
energy even when their  energy density is locally integrable.
	However, suitably cut off, plane wave solutions may provide useful 
approximations to solutions with finite total energy, and indeed are frequently 
used for this purpose.

\smallskip
%%%%%%%%%%%%%%%%%%%%%%%%%%%%%%%%%%%%%%%%%%%%%%%%
%%%%%%%%%%%%%%%%%%%%%%%%%%%%%%%%%%%%%%%%%%%%%%%%%%%%%%%%%%%%%%%%%
             {\subsubsection{Polychromatic plane waves with linear polarization}}
%%%%%%%%%%%%%%%%%%%%%%%%%%%%%%%%%%%%%%%%%%%%%%%%%%%%%%%%%%%%%%%%%
%%%%%%%%%%%%%%%%%%%%%%%%%%%%%%%%%%%%%%%%%%%%%%%%

  Unlike in Maxwell's linear electromagnetic vacuum theory, 
arbitrary linear superpositions of monochromatic plane waves 
will in general not furnish a solution of the Maxwell--Born--Infeld
field equations, but certain linearly polarized polychromatic plane waves
will do. 
  More specifically, by at most an $SO(3)$ rotation of our coordinate system, 
we may assume that the plane wave propagates in the $\SPvec{e}_z$ direction,
that $\SPvec{B}$ points along (or against) the $\SPvec{e}_x$ direction, while 
$\SPvec{E}$ points along (or against) the $\SPvec{e}_y$ direction. 
  Let $p:\RR\to\RR$ by any differentiable function. 
  Then 
\bea
&&
 \SPvec{B}^\pm(t,z) = p(z\pm t)\SPvec{e}_x =  \SPvec{H}^\pm(t,z) 
\,,
\\
&&
 \SPvec{D}^\pm(t,z) = -p(z\pm t)\SPvec{e}_y =  \SPvec{E}^\pm(t,z)
\eea
with either sign solves the source-free Maxwell--Born--Infeld field equations.
  In particular, one can choose $p$ to have pulse shape.

  While the superposition of one left- and one right-propagating fixed pulse shape
will in general not solve the Maxwell--Born--Infeld equations, this can provide
interesting asymptotic conditions for studies of the nonlinear pulse interactions.
  Such studies have been carried out in
		\cite{IgnatovPoponin, GibbonsB};
see also the earlier review by Gibbons
		\cite{GibbonsA}.
  Two such pulses resemble two colliding solitons.\footnote{Meanwhile, the complete integrability of the 
                                             dynamical equations for plane simply periodic waves was proven in 
					     \cite{Brenier}.
					     Interestingly, the equations are globally well-posed if and only if the data
					     satisfy a smallness condition.}

\smallskip
%%%%%%%%%%%%%%%%%%%%%%%%%%%%%%%%%%%%%%%%%%%%%%%%
%%%%%%%%%%%%%%%%%%%%%%%%%%%%%%%%%%%%%%%%%%%%%%%%%%%%%%%%%%%%%%%%%
             {\subsubsection{Bichromatic plane waves with circular polarizations}}
%%%%%%%%%%%%%%%%%%%%%%%%%%%%%%%%%%%%%%%%%%%%%%%%%%%%%%%%%%%%%%%%%
%%%%%%%%%%%%%%%%%%%%%%%%%%%%%%%%%%%%%%%%%%%%%%%%

  A bichromatic plane wave solution with circular polarizations was discovered
by Schr\"odinger
		\cite{ErwinDUBLINc},
see 
		\cite{GibbonsA}
for a recent discussion.
  Schr\"odinger's solution was perhaps the first hint at some 
hidden continuous symmetries of the source-free Maxwell--Born--Infeld 
equations that may lead to further conservation laws, features known 
from completely integrable systems.

\medskip
%%%%%%%%%%%%%%%%%%%%%%%%%%%%%%%%%%%%%%%%%%%%%%%%
%%%%%%%%%%%%%%%%%%%%%%%%%%%%%%%%%%%%%%%%%%%%%%%%
             {\subsubsection{On finite energy field solitons 
                	     traveling at speeds less than light}}
%%%%%%%%%%%%%%%%%%%%%%%%%%%%%%%%%%%%%%%%%%%%%%%%
%%%%%%%%%%%%%%%%%%%%%%%%%%%%%%%%%%%%%%%%%%%%%%%%

  Since the source-free Maxwell--Born--Infeld field equations feature 
infinite-energy solutions with  planar symmetry that display nonlinear 
soliton-like dynamics, naturally one wonders whether  stable \emph{finite-energy} 
solutions exist that could belong in the category particle-like soliton. 
 In the following we provide some negative answers the proofs of which are very 
different from the well-known proof of the ``no-solitons theorem'' of Derrick. 

  Since any soliton which travels at a speed less than the speed of light
can always be boosted into a Lorentz frame in which it is at rest, the 
question of the existence of such solitons can be reduced to the question
whether static finite-energy solutions of the source-free Maxwell--Born--Infeld
solutions exist.
   Such an inquiry was carried out by Y. Yang
          \cite{Yisong, YisongBOOK},
though only for Born's first field model 
 \cite{BornA, BornB, BornD}
in which the $O(\beta^8)$ terms under the square root are missing from the Lagrangian.
  Yang 
  \cite{Yisong, YisongBOOK}
showed by Moser's application of Harnack's inequality 
              \cite{Moser}
that the only static electromagnetic entire solutions vanishing at infinity as $O(1/|\SPvec{s}|)$ 
are $\SPvec{B} = \SPvec{0}$ \emph{and} ${\SPvec{D}} = \SPvec{0}$.
 Yang also showed, by entirely different arguments which are rather similar to the proofs of 
the geometric Bernstein type theorems on Minkowski spacetime\footnote{This uses the same differential-geometric
                                                                      analogy which figured also in our discussion on 
								      static solutions with point sources alias maximal
								      space-like slices in Minkowski spacetime.}
by Calabi
                 \cite{Calabi}
and by Cheng and Yau
		\cite{ChengYau},
that the only entire solutions of the source-free Maxwell--Born--Infeld field equations having finite energy
which are electrostatic are given by $\SPvec{D} = \SPvec{0}$,
and those which are magnetostatic are given by  ${\SPvec{B}} =\SPvec{0}$.
 Since for purely electrostatic or purely magnetostatic solutions the field equations of
Born's first model coincide with the Maxwell--Born--Infeld field equations, we can take over
Yang's Bernstein results.

\bigskip
\begin{Prop} 
\label{YisongA} 
	\textit{Let $(\SPvec{B},\SPvec{D})$ be a static electromagnetic 
entire solution of the source-free Maxwell--Born--Infeld field equations 
for $\beta <\infty$. 
  If $\cH_{\mathrm{field}}(\SPvec{B},\SPvec{D})<\infty$ (finite energy),
then if $\SPvec{B} = \SPvec{0}$ one also has ${\SPvec{D}} = \SPvec{0}$,
and vice versa.
}
\end{Prop}

  Unfortunately, Yang's Harnack--Moser result does not apply to the Maxwell--Born--Infeld 
field equations.
  However, we found that under an additional smallness condition,  Moser's application of 
Harnack's inequality 
              \cite{Moser}
can be adapted  to the Maxwell--Born--Infeld field equations.
 Thus we have

\bigskip
\begin{Prop} 
\label{Harnack} 
	\textit{Let $(\SPvec{B},\SPvec{D})$ be a static electromagnetic 
entire solution of the source-free Maxwell--Born--Infeld field equations 
for $\beta <\infty$, which decay to zero as $|\SPvec{s}|\to\infty$, say like 
$O(1/|\SPvec{s}|)$, which implies 
$\cH_{\mathrm{field}}(\SPvec{B},\SPvec{D})<\infty$ (finite energy).
 Assume that there is a (positive) $\epsilon\ll 1$
such that the associated fields $(\SPvec{E},\SPvec{H})$ satisfy the bounds
$\beta^4\abs{\SPvec{E}}^2 \leq 1 -\epsilon$
and
$\beta^4\abs{\SPvec{H}}^2 \leq 1 -\epsilon$,
uniformly on $\RR^3$.
 Then $\SPvec{B} = \SPvec{0}$ and ${\SPvec{D}} = \SPvec{0}$.}
\end{Prop}

\noindent
\textit{Sketch of proof of Proposition \ref{Harnack}.}
 We first note that from the  Born and Infeld aether laws 
\refeq{eq:FOLIeqEofBD} 
and 
\refeq{eq:FOLIeqHofBD} 
it follows that when 
$(\SPvec{B},\SPvec{D})$ go to zero at infinity, then so do
$(\SPvec{E},\SPvec{H})$, at the same rate.
 We next invert 
\refeq{eq:FOLIeqEofBD} 
and 
\refeq{eq:FOLIeqHofBD} 
to expresses $\SPvec{B}$ and $\SPvec{D}$ explicitly in terms of $\SPvec{E}$ and $\SPvec{H}$, 
\newpage

\bea
&&
{\SPvec{B}} 
= 
\frac{
{\SPvec{H}} + \beta^4{\SPvec{E}}\times({\SPvec{E}}\times{\SPvec{H}})
}{
\sqrt{    1 
	- \beta^4(|{\SPvec{E}}|^2 + |{\SPvec{H}}|^2) 
	+ \beta^8|{\SPvec{E}}\times {\SPvec{H}}|^2
      }}
\label{eq:FOLIeqBofEH}
\\
&&
{\SPvec{D}} 
 = 
\frac{
{\SPvec{E}} +\beta^4{\SPvec{H}}\times({\SPvec{H}}\times {\SPvec{E}})
     }{
\sqrt{	  1 
	- \beta^4(|{\SPvec{E}}|^2 + |{\SPvec{H}}|^2) 
	+ \beta^8|{\SPvec{E}}\times {\SPvec{H}}|^2
      }}
\label{eq:FOLIeqDofEH}
\, ,
\eea
for $\beta\in (0,\infty)$. 
  The fields ${\SPvec{B}}$ and ${\SPvec{D}}$ must satisfy the vanishing divergence equations
$\nabla\cdot{\SPvec{B}}= 0$ and $\nabla\cdot{\SPvec{D}}= 0$, and
since $\nabla\times{\SPvec{E}} = {\SPvec{0}}$ and $\nabla\times{\SPvec{H}} = {\SPvec{0}}$ implies
that there exist scalar fields $f$ and $g$ such that ${\SPvec{E}} = \nabla{f}$ and ${\SPvec{H}} = \nabla{g}$,
the problem of finding entire static electromagnetic solutions of the Maxwell--Born--Infeld field
equations reduces to solving a coupled system of two scalar PDEs of divergence form, namely
\beq
\nabla\cdot\left(
\frac{
{\nabla{g}} + \beta^4{\nabla{f}}\times({\nabla{f}}\times{\nabla{g}})
}{
\sqrt{    1 
	- \beta^4(|{\nabla{f}}|^2 + |{\nabla{g}}|^2) 
	+ \beta^8|{\nabla{f}}\times {\nabla{g}}|^2
      }}\right)
=0
\, ,
\label{eq:DIVfEQ}
\eeq
and the same equation with $f$ and $g$ interchanged.
 The numerator between the big parentheses can be rewritten as 
\beq
{\nabla{g}} + \beta^4{\nabla{f}}\times({\nabla{f}}\times{\nabla{g}})
=
\left((1-\beta^4|\nabla{f}|^2)\Id +\beta^4\nabla{f}\otimes\nabla{f}\right)
\cdot\nabla{g}
\, ,
\label{eq:MATRIXinDIVeq}
\eeq
so \refeq{eq:DIVfEQ} reads 
$\nabla\cdot\bigl( M\cdot\nabla{g}\bigr)=0$, with $M(\nabla{f},\nabla{g})$ a symmetric
matrix. 
 The eigenvalues of $M$ are easily seen to be $m_1=\beta^4/\sqrt{\dots}$ and
$m_2=(1- \beta^4|\nabla{f}|^2)/\sqrt{\dots}$, where $\sqrt{\dots}$ denotes the denominator
between the big parentheses in \refeq{eq:DIVfEQ}.
 Clearly, $\sqrt{\dots}
\leq \sqrt{ 1 + \beta^8|{\nabla{f}}\times {\nabla{g}}|^2}$, and
for entire solutions for which $(\nabla{f},\nabla{g})\to (\SPvec{0}, \SPvec{0})$ at 
infinity, we have $\sqrt{ 1 + \beta^8|{\nabla{f}}\times {\nabla{g}}|^2}<C$ ($C$ a 
generic, positive constant), hence $m_1 \geq C >0$ uniformly on $\RR^3$.
 The same estimate for $\sqrt{\dots}$ combined with the bound 
$1- \beta^4|\nabla{f}|^2 \geq \epsilon$, which holds by hypothesis, shows that also
$m_2 \geq C >0$  uniformly on $\RR^3$.
 Thus, \refeq{eq:DIVfEQ} is strictly elliptic, and $M$ symmetric, hence by Moser's
application of Harnack's inequality 
               \cite{Moser}
we conclude that $\nabla{g}=\SPvec{0}$.
 The same argument holds for \refeq{eq:DIVfEQ}'s twin equation, hence also
$\nabla{f}=\SPvec{0}$. \QED

 Thus, nontrivial static electromagnetic entire solutions with finite-energy 
with both $\SPvec{B} \neq \SPvec{0}$ and ${\SPvec{D}} \neq \SPvec{0}$ can at most
exist if the magnitude of the associated fields $\SPvec{E}$ and $\SPvec{H}$
exceeds a certain value.
  The possibility of finite-energy solitons which travel precisely at the speed of light
is another open question.

\newpage

%%%%%%%%%%%%%%%%%%%%%%%%%%%%%%%%%%%%%%%%%%%%%%%%%%%%%%%%%%%%%%%%
%%%%%%%%%%%%%%%%%%%%%%%%%%%%%%%%%%%%%%%%%%%%%%%%%%%%%%%%%%%%%%%%
%%%%%%%%%%%%%%%%%%%%%%%%%%%%%%%%%%%%%%%%%%%%%%%%%%%%%%%%%%%%%%%%
                \subsection{The action principle}
%%%%%%%%%%%%%%%%%%%%%%%%%%%%%%%%%%%%%%%%%%%%%%%%%%%%%%%%%%%%%%%%
%%%%%%%%%%%%%%%%%%%%%%%%%%%%%%%%%%%%%%%%%%%%%%%%%%%%%%%%%%%%%%%%
%%%%%%%%%%%%%%%%%%%%%%%%%%%%%%%%%%%%%%%%%%%%%%%%%%%%%%%%%%%%%%%%

  Any decent relativistic physical theory satisfies an action principle
              \cite{christodoulou}.
  The geometrical equations of our classical theory are no exception.
  The action $\cA$ is defined as integral over a four-dimensional 
spacetime domain $\Xi$ sandwiched between two disjoint space-like slices, 
called the past and future boundaries of $\Xi$, thus
$
        {\cA} =  \int_{\Xi} { {L}}
$,
where ${{L}}$ is a four-form, called Lagrangian `density.' 
 The quotes here indicate that the point particle 
terms are not true densities but Dirac measures; having
pointed this out we will from now on simply speak of Lagrangian
density.
 We remark that the Dirac measure-valued four-forms can be recast 
as regular one-forms to be integrated along those truncated histories 
which are cut off by the future and past boundaries 
of $\Xi$, but to work out the Euler--Lagrange equations one then has to
convert back to four-forms. 
   In any event, the Lagrangian density naturally splits into a sum of two 
terms, one associated with the time-like line defects $\Eta_k$ of the 
electromagnetic potential, the other with the differentiable part 
that gives the electromagnetic field $\FQ$ of the electromagnetic 
spacetime between those defects, which gives the familiar
\begin{equation}
{{L}} =  {{L}}_{\mathrm{particle}} + {{L}}_{\mathrm{field}}
\, .
\label{eq:LAGRANGIANdensity}
\end{equation}

%%%%%%%%%%%%%%%%%%%%%%%%%%%%%%%%%%%%%%%%%%%%%%%%%%%%%%%%%%%%%%%%
%%%%%%%%%%%%%%%%%%%%%%%%%%%%%%%%%%%%%%%%%%%%%%%%%%%%%%%%%%%%%%%%
                \subsubsection{The Lagrangian density in $\AQ$ and $\uQ_k$ variables}
%%%%%%%%%%%%%%%%%%%%%%%%%%%%%%%%%%%%%%%%%%%%%%%%%%%%%%%%%%%%%%%%
%%%%%%%%%%%%%%%%%%%%%%%%%%%%%%%%%%%%%%%%%%%%%%%%%%%%%%%%%%%%%%%%

 The term ${L}_{\mathrm{field}}$ is the Born--Infeld Lagrangian density,
which involves only the electromagnetic curvature $\dQ\AQ =\FQ$ 
in $\Xi\backslash\bigcup_k\Eta_k$
(recall that $\FQ$ is defined by $\dQ\AQ =\FQ$),
\begin{equation}
  {L}_{\mathrm{field}}(\varpi)
\equiv \, 
	 {\textstyle{\frac{1}{4\pi}}} 
	 {\phantom{\Big|}}^\star\!
\left({\textstyle{ \frac{\alpha}{\beta^4}}}
    - {\textstyle{ \frac{\alpha}{\beta^4}}}\sqrt{\detg \big(\gQ + \beta^2 \dQ\AQ(\varpi)\big)}\, 
\right)
\prod_{k\in\cN}\chi_{_{\Xi\backslash{\Eta_k}}}(\varpi)
\, ,
\label{eq:GEOaetherLAWborninfeldDETlagrangian}
\label{eq:Sf}
\end{equation}
where $\detg$ means determinant w.r.t. the metric $\gQ$
        (for rank-two tensors on $\MM^4$ we have $\detg =- \det$).
    The determinant can be expanded as follows,
\begin{equation}
\detg \big(\gQ + \beta^2 \dQ\AQ\big)
=
  1 
- \beta^4\, \Hodge(\FQ\wedge\Hodge\FQ \big)
- \beta^8\, \left(\Hodge\bigl(\FQ\wedge\FQ \bigr)\right)^2
\, .
\label{eq:detEXPAND}
\end{equation}
   In the limit of a weak electromagnetic curvature, the Born--Infeld Lagrangian density reduces 
to the familiar Lagrangian density of the vacuum Maxwell fields
         \cite{WheeleretalBOOK},
\begin{equation}
        {{L}}_{\mathrm{field}} 
        \sim
        \, -  {\textstyle{\frac{1}{8\pi}}}\, \alpha\, \FQ\wedge \Hodge\FQ\,
{\textstyle{\prod_{k}}}\chi_{_{\Xi\backslash{\Eta_k}}}
\, \qquad (\mathrm{weak\ field\ limit}).
\label{eq:SfSCHWARZSCHILD}
\end{equation}

  The Lagrangian density ${{L}}_{\mathrm{particle}}$ involves only 
the line defects of the electromagnetic connection $\AQ$.
  It consists of a sum of linear and quadratic terms in the $\uQ_k$s, 
\bea
{{L}}_{\mathrm{particle}} (\varpi)
  \equiv   && \!\!\!\!\!\!\!\!
        \alpha \MTWvec{A} \wedge 
	\sum_{k\in\cN} \int_{-\infty}^{+\infty}
		\pmk\Hodge\uQ_k(\tau)\delta_{\eta_k(\tau)}\big(\varpi\big)\,\dd\tau
\nonumber
\\
&&\!\!\!\!\!\!\!\!\!
+\, \haelfte \sum_{k\in\cN} 
\int_{-\infty}^{+\infty}
  \uQ_k(\tau)\wedge{}\Hodge\uQ_k(\tau)\delta_{\eta_k(\tau)}\big(\varpi\big)\,\dd\tau
\,.
\label{eq:Ldefect}
\eea
  The term quadratic in the $\uQ_k$s is the four-form version of a familiar expression
          \cite{ThirringBOOKa}, 
the proper-time integral along point histories.\footnote{Notice that the quadratic term in the $\uQ_k$s is the only term 
                                                     which is not proportional to $\alpha$; hence, formally we can 
						     `switch off' electromagnetic influences on the point histories by 
						     letting $\alpha \downarrow{0}$, retaining only the quadratic term in 
						     the $\uQ_k$s. We thereby obtain the familiar Lagrangian for non-interacting 
						     particles which satisfy Galileo's law of inertia, viz. geodesic motion in $\MM^4$.}
 The term linear in the $\uQ_k$s can be written more concisely as $\alpha \MTWvec{A} \wedge \JQ $
and reveals itself as the familiar ``minimal coupling'' term recast as a four-form on $\MM^4$, cf. 
          \cite{ThirringBOOKa}.
 Note that because of the Dirac $\delta$ function we can alternatively pull $\AQ$ under the  $\tau$-integral 
and  switch to $\widetilde\AQ$. 

  The \emph{relativistic principle of `least' action} demands that $\cA$ be 
{extremal} w.r.t. independent variations of $\AQ$ and the $\uQ_k$. 
  The variations w.r.t. $\AQ$ are standard and yield the Maxwell--Born--Infeld
field equations with point sources. 
  Notice that $\alpha$ does not figure in these variations.
  The variations w.r.t. $\uQ_k$ are constrained by the fact that each $\uQ_k$ is a 
Minkowski-velocity co-vector, i.e. dual to the unit tangent vector at the respective history, 
and as such satisfying $\Hodge(\uQ_k\wedge\Hodge\uQ_k) = 1$.
 These constraints are taken into account by adding to the Lagrangian density the term
\beq
{L}_{\mathrm{particle}}^\Phi (\varpi)
=
- \sum_{k\in\cN}\int\limits_{-\infty}^{+\infty}
		{}\dQ_k\widetilde\Phi(\eta_1(\tau),...,\eta_N(\tau))
\wedge\Hodge\uQ_k(\tau)
\delta_{\eta_k(\tau)}\big(\varpi\big)\,\dd\tau
.
\label{eq:LdefectPhi}
\eeq
 Note that after integration over $\Xi$ each summand indeed vanishes, 
independently of the gauge, by virtue of the same reason why each summand in  \refeq{eq:GEOptchargecurrent} 
separately satisfies the law of charge conservation \refeq{eq:GEOconservechargeLAW}.
  Unconstrained variation of $\cA$ w.r.t. $\uQ_k$ now gives the guiding laws \refeq{eq:GEOhamjacLAW}.
  The constraint $\Hodge\left(\uQ_k\wedge\Hodge\uQ_k\right) = 1$ 
(equivalent to $\uQ_k\cdot\uQ_k = -1$) applied to the guiding law then gives equation \refeq{eq:GEOhamjacPDE}
for each $\dQ_k\widetilde\Phi$. 
 We note that $\widetilde\Phi$ plays a r\^{o}le close to a familiar Lagrange parameter, which is what it would
be if $\AQ$ were of class $C^1$.
  Here $\widetilde\Phi$ turns out to have some life of its own. 

%%%%%%%%%%%%%%%%%%%%%%%%%%%%%%%%%%%%%%%%%%%%%%%%%%%%%%%%%%%%%%%%
%%%%%%%%%%%%%%%%%%%%%%%%%%%%%%%%%%%%%%%%%%%%%%%%%%%%%%%%%%%%%%%%
      \subsubsection{The Lagrangian density in $\AQ$  and $\Phi$ variables}
%%%%%%%%%%%%%%%%%%%%%%%%%%%%%%%%%%%%%%%%%%%%%%%%%%%%%%%%%%%%%%%%
%%%%%%%%%%%%%%%%%%%%%%%%%%%%%%%%%%%%%%%%%%%%%%%%%%%%%%%%%%%%%%%%

 We note that we may switch from the $\uQ_k$s to $\widetilde\Phi$ as
variational degrees of freedom.
 Namely, similarly to defining $\FQ$ via $\FQ = \dQ\AQ$, we may simply \emph{define} $\uQ_k$ via
$\uQ_k = \dQ_k\widetilde\Phi - \pmk \alpha \widetilde\AQ_k$
and rewrite the action principle with $\AQ$ and $\widetilde\Phi$ as variables. 
 However, different from the definition of $\FQ$, which automatically implies the Faraday--Maxwell
law $\dQ\FQ=\MTWvec{0}$, the definition of $\uQ_k$ does not automatically imply that $\uQ_k$ is a Minkowski 
velocity co-vector, so that this piece of information has to be incorporated for the variations. 
 The total Lagrangian density then reads
\bea
        {{L}} (\varpi)
=&&\!\!\!\!\!\!\!\!
 {\textstyle{\frac{1}{4\pi}}} 
	 {\phantom{\Big|}}^\star\!
\left({\textstyle{ \frac{\alpha}{\beta^4}}}
    - {\textstyle{ \frac{\alpha}{\beta^4}}}\sqrt{\detg \big(\gQ + \beta^2 \dQ\AQ(\varpi)\big)}\, 
\right)
\prod_{k\in\cN}\chi_{_{\Xi\backslash{\Eta_k}}}(\varpi)
\label{eq:LtotalAPhi}
\\
&&\!\!\!\!\!\!\!\!\!\!
- \haelfte \int_{-\infty}^{\infty}\sum_{k\in\cN} 
\!\!\left(\!
\left(\dQ_k\widetilde\Phi - \pmk \alpha \widetilde\AQ_k\right)
\wedge
{}^\star\!
\left(\dQ_k\widetilde\Phi- \pmk \alpha \widetilde\AQ_k\right)
\! +\! {}^\star{1}
\right)\!\!\Big|_{\{\varpi_n=\eta_n(\tau)\}}\Big.
\delta_{\eta_k(\tau)}\big(\varpi\big)\,\dd\tau
\nonumber
\eea

 The action principle inherits a change from ${z}_k\in\{-1,1\}$
to $z_k\in\ZZ$; to incorporate the $\kappa_k$, replace $z_k\to \kappa_kz_k$
and $\sum_{k\in\cN}\to \sum_{k\in\cN}\kappa_k^{-1}$ 
in \refeq{eq:Ldefect}, \refeq{eq:LdefectPhi}, \refeq{eq:LtotalAPhi}.

%%%%%%%%%%%%%%%%%%%%%%%%%%%%%%%%%%%%%%%%%%%%%%%%%%%%%%%%%%%%%%%%%%%%
%%%%%%%%%%%%%%%%%%%%%%%%%%%%%%%%%%%%%%%%%%%%%%%%%%%%%%%%%%%%%%%%%%%%
%%%%%%%%%%%%%%%%%%%%%%%%%%%%%%%%%%%%%%%%%%%%%%%%%%%%%%%%%%%%%%%%%%%%
  \subsection{Extensions to general-relativistic spacetimes}
	\label{sect:generalRELATIVITY}
%%%%%%%%%%%%%%%%%%%%%%%%%%%%%%%%%%%%%%%%%%%%%%%%%%%%%%%%%%%%%%%%%%%%
%%%%%%%%%%%%%%%%%%%%%%%%%%%%%%%%%%%%%%%%%%%%%%%%%%%%%%%%%%%%%%%%%%%%
%%%%%%%%%%%%%%%%%%%%%%%%%%%%%%%%%%%%%%%%%%%%%%%%%%%%%%%%%%%%%%%%%%%%
  
 The formal extension of our classical special-relativistic electromagnetic 
theory with point charges to a general-relativistic electromagnetic theory 
with point charges in which spacetime is no longer flat but dynamical, 
is perfectly straightforward. 
  All we need to do, besides allowing $\gQ$ to be a general Lorentz metric 
with signature $+2$ and interpreting all the $p$ forms and covariant derivatives 
accordingly, is to add the gravitational Lagrangian density of the Einstein--Hilbert
variational principle to ${L}$, 
\beq
 {L} =   {{L}}_{\mathrm{particle}} + {{L}}_{\mathrm{field}} +   {{L}}_{\mathrm{spacetime}} 
\label{eq:GEMlagrangian}
\eeq
with 
\beq
{{L}}_{\mathrm{spacetime}} 
=
\frac{1}{8\pi\gamma} {}^\star\trg \RQ
\label{eq:EINSTEINlagrangian}
\eeq
where $\RQ$ is the Ricci curvature tensor of the metric $\gQ$, and $\gamma$ is given
in \refeq{eq:gravitationalSOMMERFELDconstant}.
 Beside $\AQ$ and $\uQ_k$, also $\gQ$ is now a variable for the principle of least
action. 
 Variation w.r.t. $\gQ$ yields the Einstein equations of spacetime with the 
electromagnetic energy-momentum tensor as source of the spacetime curvature.
 The r\^{o}le of $\gamma$ is that of a coupling constant which calibrates the gravitational 
influence of the electromagnetism (and perhaps other non-spacetime sources) on the spacetime 
curvature.\footnote{Formally,  by letting $\gamma \downarrow{0+}$, we can switch off the gravitational influence of 
                    electromagnetism on the spacetime curvature and obtain the  Einstein equations for so-called vacuum 
		    spacetimes.
		    Beside the passive, flat Minkowski spacetime, other global solutions 
		    of the vacuum Einstein equations exist which are truly dynamical; see
		    \cite{christodoulouklainermanBOOK}.}

%\newpage
		%%%%%%%%   BIBLIOGRAPHY  %%%%%%%%%%%
\scriptsize{

}
\end{document}